\documentclass[3p, times, sort&compress]{elsarticle}
\usepackage[utf8]{inputenc}
\usepackage{textcomp}

\usepackage{hyperref}

\usepackage{graphicx}
\graphicspath{{./Figures/}}
\usepackage{natbib}
\usepackage{amsmath}
\usepackage[version=4]{mhchem}
\usepackage{siunitx}
\usepackage{longtable,tabularx}
\usepackage{textcase}
\usepackage{multirow}
\usepackage{xcolor}
\usepackage{threeparttable}
\usepackage{comment}

\journal{Acta Astronautica}

\definecolor{PerpendicularTrueDetection}{RGB}{255, 184, 77}
\definecolor{TiltedTrueDetection}{RGB}{113, 218, 113}
\definecolor{PerpendicularFalseDetection}{RGB}{128, 159, 255}
\definecolor{TiltedFalseDetection}{RGB}{255, 133, 102}

\newcommand\T{\rule{0pt}{2.6ex}}       %
\newcommand\B{\rule[-1.2ex]{0pt}{0pt}} %

\usepackage{bm}
\usepackage{subcaption}
\DeclareMathAlphabet{\mymathbb}{U}{BOONDOX-ds}{m}{n}

\newcommand{\SdR}[1]{\ensuremath{\mathcal{\MakeTextUppercase{#1}} = \left\{\MakeLowercase{#1},\,\MakeLowercase{\bm{#1}}_1,\,\MakeLowercase{\bm{#1}}_2,\,\MakeLowercase{\bm{#1}}_3\right\}}}
\newcommand{\sdr}[1]{\ensuremath{\mathbb{\MakeTextUppercase{#1}} = \left\{\MakeTextUppercase{#1},\,\MakeTextUppercase{\bm{#1}}_1,\,\MakeTextUppercase{\bm{#1}}_2\right\}}}

\title{Vision-Based Estimation of Small Body Rotational State \tnoteref{t2}}
\tnotetext[t1]{Preliminary results of this work were presented in ``Panicucci, P., Autonomous vision-based navigation and shape reconstruction of an unknown asteroid during approach phase, Ph.D. thesis, ISAE-SUPAERO, 30 March 2021. URL \url{http://www.theses.fr/2021ESAE0011}.}
\date{}

\author[1,2,3]{Paolo Panicucci\footnote[1]{Currently: Assistant Professor, Dartimento di Scienze e Tecnologie Aerospaziali, Politecnico di Milano}}
\address[1]{Complex System Engineering Department, ISAE-SUPAERO, 10 Avenue Edouard Belin, 31400, Toulouse, France}
\address[2]{Future Mission Engineering Department, CNES, 18 Avenue Edouard Belin, 31400, Toulouse, France}
\address[3]{Sensor Processing Chain Department, Airbus Defence \& Space, 31 Rue des Cosmonautes, 31400, Toulouse, France}

\author[3]{J\'er\'emy Lebreton}
\author[3]{Roland Brochard}

\author[1]{Emmanuel Zenou}

\author[2]{Michel Delpech}

\usepackage[normalem]{ulem}
\newcommand{\stkout}[1]{\ifmmode\text{\sout{\ensuremath{#1}}}\else\sout{#1}\fi}
\newcommand{\EPP}[1]{{\color{black}#1}}

\begin{document}

\begin{abstract}
The heterogeneity of the small body population complicates the  prediction of small body properties before the spacecraft's arrival. In the context of autonomous small body exploration, it is crucial to develop algorithms that estimate the small body characteristics before orbit insertion and close proximity operations. This paper develops a vision-based estimation of the small-body rotational state (i.e., the center of rotation and rotation axis direction) during the approach phase. In this mission phase, the spacecraft observes the rotating celestial body and tracks features in images. As feature tracks are the projection of the landmarks' circular movement, the possible rotation axes are computed. Then, the rotation axis solution is chosen among the possible candidates by exploiting feature motion and a heuristic approach. Finally, the center of rotation is estimated from the center of brightness.  The algorithm is tested on more than 800 test cases with two different asteroids (i.e., Bennu and Itokawa), three different lighting conditions, and more than 100 different rotation axis orientations. Each test case is composed of about 250 synthetic images of the asteroid which are used to track features and determine the rotational state. Results show that the error between the true rotation axis and its estimation is below $10^{\circ}$ for $80\%$ of the considered test cases, implying that the proposed algorithm is a suitable method for autonomous small body characterization.
\end{abstract}


\maketitle

\section{Introduction}
The heterogeneity of small body characteristics is a crucial factor to consider during mission preparation and in-orbit operations. In particular, the limited knowledge of small bodies' properties imposes numerous challenges during mission design and operations. Shape, rotational state, and geophysical characteristics can strongly change depending on the small body under study and it is hard to predict the exact small body properties before the encounter. When a small body mission is planned, an observation campaign is performed to bound the characteristics of the small body under study and to estimate the range of uncertainty of these quantities \cite{melosh2019finding}. In particular, preliminary shape, pole orientation, and rotation period can be deduced from light-curve inversion \cite{kaasalainen2001optimization, kaasalainen2001optimization2} and, if feasible, radar campaign \cite{ostro2002asteroid}. The rotational period has shown high agreement with the light-curve inversion procedure and it can be estimated from on-board during far approach \cite{castellini2015far, bandyonadhyay2019silhouette}. Rotation pole, shape, and gravity field estimation is a more complex problem and ground-based observations are often inaccurate.\newline
On the one hand, the rotation pole orientation is directly coupled, through the rotational equations, to the inertial tensor which is deduced from the shape. Even though small bodies, in particular asteroids, are mainly principal axis rotators \cite{scheeres2016orbital}, fine shape estimation remains deeply coupled with the rotation pole inertial orientation. Current techniques to solve this problem rely on stereophotoclinometry or its variations \cite{gaskell2006landmark, gaskell2008characterizing,capanna2013three}. On the other hand, the gravity field can be derived from the shape under the assumption of constant density \cite{werner1996exterior, panicucci2020uncertainties, bercovici2020analytical, werner1997spherical} or gathered by solving the orbit determination problem by processing long trajectory arcs \cite{mcmahon2018osiris, jorda2016global}. Unfortunately, these techniques strongly rely on human intervention to control the solution convergence and validate the output shape. Moreover, the required computational time makes it unsuitable for onboard applications. It is worth noting that all these techniques rely on communications with the ground implying radio signal delays and high costs.\newline
To overcome these limitations, vision-based systems have been gaining attention as a cost-effective and accurate solution because they can provide real-time information to the spacecraft with limited impact on system budgets. In this context, it is crucial to design autonomous vision-based algorithms capable of supporting small-body parameter estimation without ground communications. The rotation pole, the shape, and the gravity field are of primary importance to allow the GNC architecture to perform critical mission phases like orbit insertion and close proximity characterization \cite{scheeres2019autonomous}. These quantities are strongly correlated and their onboard estimation is difficult to be performed independently. The gravity field depends on the small body shape and its density distribution. Therefore it can be derived from the shape under the assumption of constant density \cite{werner1996exterior, panicucci2020uncertainties, bercovici2020analytical, werner1997spherical} or estimated during the localization procedure \cite{stacey2022robust}. The shape can be deduced by shape from motion algorithms \cite{panicucci2023shadow, bandyonadhyay2019silhouette} or by methods solving for the SLAM (Simultaneous Localization And Mapping) problem \cite{panicucci2021autonomous, dor2022astroslam, villa2022point}. The rotational state initial estimation is required for two main reasons. First, it is necessary to define a relative reference frame which implies the estimation of a point and three vectors. Second, the knowledge - even if coarse - of the small body rotational dynamics can help the convergence of the relative localization as proposed in \citet{panicucci2021autonomous}. The rotational state can be determined implicitly by solving the SLAM problem as the spacecraft poses, i.e., positions and orientations, are estimated in the small-body-fixed reference frame. Alternatively, it can be determined during close approach when the small body is resolved and image processing algorithms can extract information about the small body rotation in time. \citet{bandyonadhyay2019silhouette} proposes a method to determine the rotation state of the small body by simultaneously optimizing the shape and the rotation state. The proposed method accurately estimates the pole orientation and the small body shape, but it requires a Monte-Carlo optimization which increases the computational burden. \citet{bissonnette2015vision} studies the estimation of the pole orientation from matched features. This work studies the possibility of detecting ellipses in the image and reconstructing the landmark's circles movements in 3D. In recent year several algorithms have been developed in the spacecraft close proximity community to solve for the angular velocity and inertia tensor f an unknown target. The pioneering work of \citet{masutani1994motion} studied the dimensionless inertia estimation from frame-to-frame tracking by exploiting the analytical solution of the Euler equations. Despite the angular velocity was not the focus of the estimation, angular velocity data were derived from noiseless synthetic data and used to determine the inertial tensor of a torque-free angular motion. \citet{augenstein2011improved} proposes a Bayesian filtering approach to estimate the chaser transnational state, the target rotation state and its inertia parameter ratio. Despite the  . \citet{padial2012tumbling} extended \citet{augenstein2011improved}'s work by merging range measurements and vision data to solve the scale factor. Range measurements are also used in \citet{lichter2004state} where the proposed methodology determines the angular state and the inertia matrix of a uncooperative target from  range measurements filtered in a Kalman filter. Also \citet{hillenbrand2005motion} uses range measurements with iteratively-reweighted-least-squares to determine the target rotational state and its inertia parameters. Results show agreement when measurements were available, but the target rotational state diverges quite rapidly, implying a coarse estimation of the rotational parameters. A pureley vision-based approach was used by \citet{tweddle2015factor} by exploiting bundle adjustment in a graph-based SLAM framework to solve for the rotational and translational evolution of two uncooperative spacecrafts. Results show agreement with testing conducted on the ISS, leading also to a correct estimation of the target inertia ratios. A different approach is used in \citet{setterfield2018inertial} where the polhode is analyze to estimate and predict the target rotational state and parameters. The previous references mainly focus on rotational state and inertia ratio estimation for artificial satellite close proximity, where the target undergoes high and variable rotation rate, the observational period is short, and the observed spacecraft is an human-made object. On the contrary, the algorithm presented in this work studies small body applications which differs in several aspect to human-made uncooperative target. First the rotation axis is generally fixed and the rotational period is long. This implies long tracking windiws which are challenging for feature tracking and matching algorithms. Second, the observed body is composed of rocky and dusty terrain implying different performance on image processing performance.\newline
Therefore, in the context of making space exploration more autonomous and avoid to perform long-lasting ground-based characterization before launch, the current study aims to develop an algorithm that estimate the rotation state of an asteroid during the approach phase. The main motivation is to avoid relying to ground-based observations and to purely rely on images obtained during the preliminary characterization phase. It is worth to underline the paradigm shift between the this approach and the ground-based lightcurve inversion. Lightcurve inversion processes photometric data from ground-based observations which can span up to several years \cite{muinonen2020asteroid}. Data are validated and chosen by ground-based operators who plan carefully the observations schedule to improve the output accuracy and remove outliers caused by corrupted observation. The rotation pole ambiguity (see Sections~\ref{sec:rotationaxestilted} and \ref{sec:rotationaxesperp} for details) is solved by exploiting multiple observational geometry and long-lasting observations, even thought this is a complex task for low-inclination asteroid orbit \cite{muinonen2020asteroid}. Moreover, despite pole orientation estimation seems to provide valuable and reliable results for near-Earth asteroids, \citet{hanuvs2011study} shows that pole orientation estimation errors can be considerably higher (i.e., 10 degrees of standard deviation for the pole latitude and 5 degree for the pole longitude). Finally, the computational burden of the lightcurve inversion take often dozens of hours of computing time on a desktop computer \cite{muinonen2020asteroid}. On the contrary, the proposed paradigm wants to exploit short observation timeframe during in-situ exploration to determine the pole orientation. The proposed approach works with hour-lasting observations (i.e., the observation time is comparable with the asteroid rational period) by exploiting geometrical information from image processing and computer vision algorithms developed for autonomous and on-board applications.\newline
In this context, this paper develops an algorithm to autonomously determine the small body rotational state during close approach without the need of performing the localization task starting from Bissonnette et al. \cite{bissonnette2015vision}'s work. The proposed approach also estimates the small-body-fixed reference frame origin leading to the definition of the small-body-fixed reference frame. This task is crucial to be performed before the localization task because it enables changing the navigation from the inertial reference frame to the small-body-fixed reference frame. By tracking points extracted from small body images taken during the approach, it is possible to retrieve the circular movement of the landmark in the inertial frame. Indeed, the trajectory of each 3D landmark is a circle having as center a point on the rotation axis and lying on a plane defined by the same axis. By assuming known the rotational period, the feature circular movements provide an estimate of all the possible rotation axes. Multiple solutions are present because the camera cannot retrieve the movement of the feature along its boresight. The correct one is determined by using a heuristic approach. The algorithm is tested through numerical simulation with synthetic images in the loop for a wide range of observation geometries and two different asteroids, i.e., Bennu and Itokawa. Finally, numerical results are presented and the algorithm's performance and limitations are discussed.

\section{Notation}
In this paper the following notation is used:
\begin{itemize}
	\item 3D vectors are in lower case bold text, such as $\bm{r}$, and 2D vectors are in upper case bold, such as $\bm{R}$.
	\item Matrices are in plain text in brackets, such as $\left[A\right]$
	\item Vector initialization are performed with parenthesis, such as $\bm{b} = \left(\bm{a}^T\; \bm{a}^T\right)^T$
	\item \SdR{a} is a 3D reference frame centered in $a$ with axes $\bm{a}_1$, $\bm{a}_2$, and $\bm{a}_3$. All the reference frame are right-handed.
	\item \sdr{a} is  a 2D reference frame. This reference frame is centered in $A$ with axes $\bm{A}_1$ and $\bm{A}_2$.%
	\item The vector $\bm{r}$ expressed in the $\mathcal{S}$ reference frame is denoted ${}^{\mathcal{S}}\bm{r}$
		\item The homography matrix from $\mathbb{S}$ to $\mathbb{C}$ is $\left[\mathbb{C}\mathbb{S}\right]$.
	\item The vector $\bm{R}$ expressed in the $\mathbb{S}$ reference frame is denoted ${}^{\mathbb{S}}\bm{R}$
	\item The angular velocity of reference frame $\mathcal{B}$ with respect to reference frame $\mathcal{N}$ is labeled $\bm{\omega}_{\mathcal{B}/\mathcal{N}}$
\end{itemize}

\section{Problem Statement and Algorithm Overview}\label{sec:probstat}
\begin{figure}[!t]
	\centering
	\includegraphics[width=0.7\textwidth]{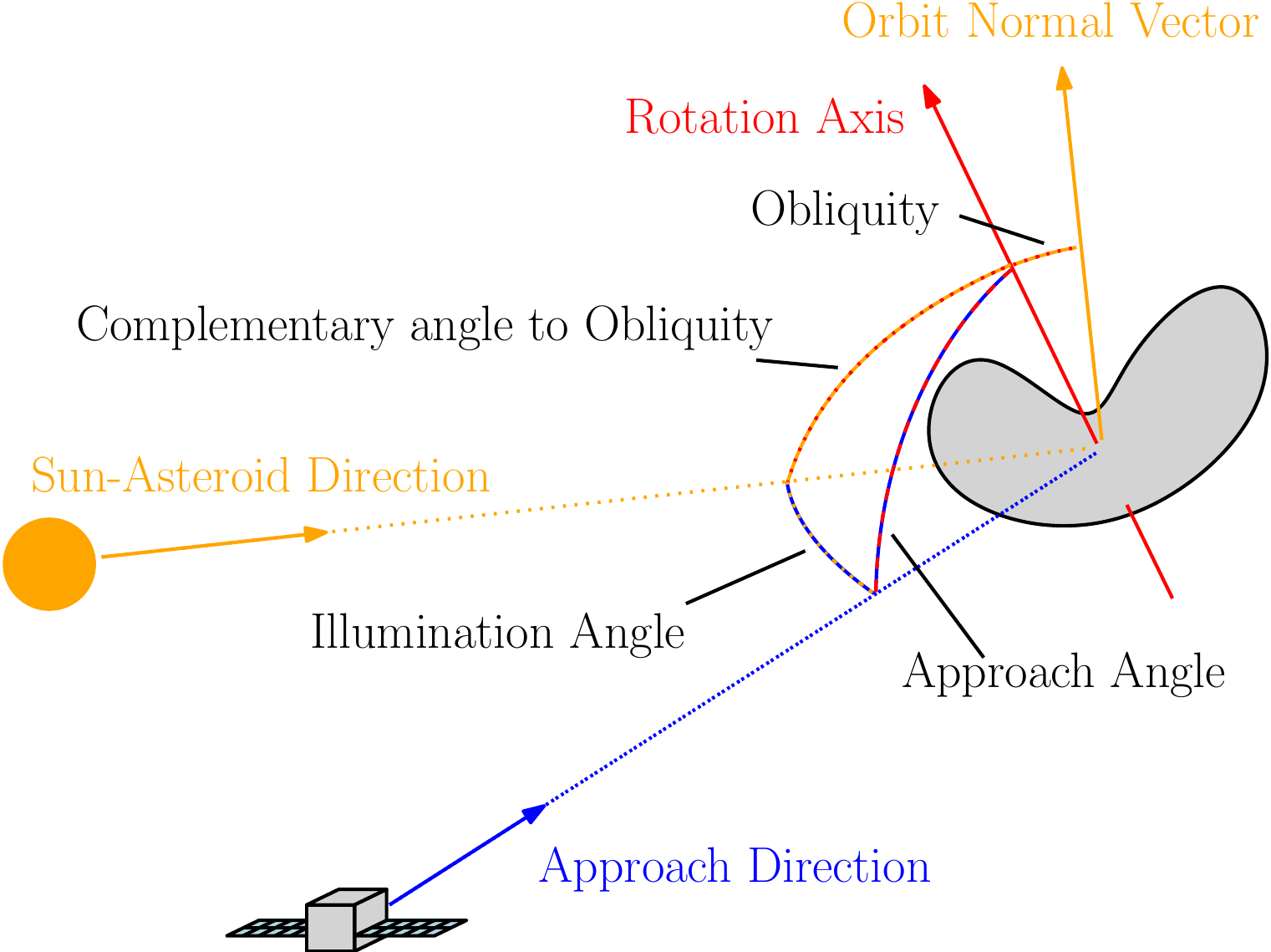}
	\caption{The observational geometry during small body approach.}
	\label{fig:RotEst_anglesapproach}
\end{figure}

During close approach to a small body, the spacecraft moves at low velocity with respect to its target and acquires images of the small body to perform characterization. \EPP{The proposed algorithm is conceived to be used during the early characterization phase during which the shape and the rotation state are coarsely estimated before performing orbit insertion \cite{scheeres1995navigation}.} Note that the small body gravity field is negligible in the dynamics \EPP{as the early characterization is performed during the approach}, thus the spacecraft motion is fully determined by the Sun gravity and deep-space perturbations. The spacecraft observes with an onboard camera the small body which is resolved in the image and rotates around its rotation axis which is assumed fixed in the inertial reference frame. \EPP{Note that the small body must be resolved to enable the possibility of tracking features on its surface. Moreover, to ensure image processing accuracy and precision during tracking, it is convenient to process images where the small body occupies a large portion of the field of view (e.g., hundreds of pixels), such that features can be correctly identified and followed.}\newline
The observation geometry is depicted in Fig.~\ref{fig:RotEst_anglesapproach} where the main geometrical entities are defined. In particular:
\begin{itemize}
	\item The approach angle is the angle between the small body rotation axis and the approach direction, i.e., the camera-small-body direction. 
	\item The illumination angle or phase angle is defined as the angle between the approach direction and the Sun-small-body direction.
	\item The obliquity, also known as the axial tilt, is the angle between the small-body rotation axis and the small-body orbital plane.
\end{itemize}
Note that the approach angle defines which hemisphere of the small body is observed during the approach.  Moreover, the illumination angle is an indicator of possible shadows due to the terminator line and self-shadowing. A more detailed analysis of shadow generation can be found in \citet{panicucci2023shadow}. Finally, the obliquity defines which hemisphere is illuminated at approach time, thus which is the observable part of the small body with a vision-based sensor. The combination of these three angles defines the appearance of the small body in the image, which infuences the performance of the image processing and vision-based system under study.
\newline
Note that the movement of the spacecraft in the small-body-fixed reference frame can be decomposed in the motion of the spacecraft in the inertial reference frame and the motion due to the small-body rotational dynamics. As the spacecraft's inertial trajectory is usually dynamically slower than the rotational dynamics of the small body, the motion of the latter in the images is mainly due to its rotation. When this is not the case as in fast fly bys, an estimation of the spacecraft inertial poses can be exploited to reconstruct the inertial epipolar geometry between the different poses and correct the images for the inertial spacecraft rototranslational motion.
From images of the rotating small body, it is possible to extract and follow 2D features which represent the projection of surface 3D landmarks. In the inertial reference frame landmarks, which are points anchored to the small body, rotate according to the small-body rotational dynamics. When considering a principal axis rotator, the landmark trajectory in the inertial space is circular. This movement is projected as a conic or as a single line in the camera frame (see Sec.~\ref{sec:circleprojection} for a more rigorous analysis). By fitting the conics from the observed small body feature tracks, it is possible to reconstruct the circle traced by the landmarks. This procedure provides a direct estimation of all possible solutions for the orientation of the small body rotation axis. As outlined in detail in Sec.~\ref{sec:rotstate}, some solutions can be pruned by exploiting the information about feature movement in the images. Others cannot be removed without information about the landmarks' movement in the boresight direction (which is not observable), thus a heuristic approach is exploited to determine the correct solution.\newline
The algorithm is composed as follows:
\begin{enumerate}
	\item The small-body-fixed reference frame origin is estimated from the first image to define the small-body-fixed reference frame origin as outlined in Sec.~\ref{sec:IP}.
	\item Images are processed sequentially to identify in the image the projection of the circle associated with the landmark movement. Several features are extracted which implies that several landmarks are observed.
	\item The algorithm decides whether to consider the conics as degenerated. This is a crucial algorithm step as the optimization procedure is different between the two cases (see Sec.~\ref{sec:rotstate} for more details).
	\item The conics are fitted and the possible solutions for the rotation axis orientation are identified. This process is outlined for the not-degenerate case in Sec.~\ref{sec:tiltedcase} and for the degenerate case in Sec.~\ref{sec:perpaxis}.
	\item The solution is identified by discarding unfeasible solutions. This process strongly depends on the degeneracy of the observed conics as outlined in Sec.~\ref{sec:tiltedcase} and  Sec.~\ref{sec:perpaxis}.
\end{enumerate}
The main assumptions are:
\begin{enumerate}
	\item The rotational period is known from light curves. This is a standard estimation from ground-based campaigns \cite{jorda2016global} or far-approach photometric studies \cite{castellini2015far, bandyonadhyay2019silhouette}.
	\item The distance between the small body and the probe is known for the first camera view. This is required to break the scale ambiguity and to place the small body reference frame at the time of the first camera view. This could be considered a strong assumption but a wrong estimation of the distance of the first camera view, or equivalently the scale factor, would simply induce an overall scaling of the solution. The small-body-spacecraft distance can be determined by a preliminary scale estimation through $\Delta$V ranging with circle or ellipsoid fitting  \cite{takahashi2021autonomous,wright2018optical}.
	\item The small body is observed for one rotation period to enable the spacecraft to observe long feature tracks needed to bound the rotation axis estimation.
\end{enumerate}
\EPP{Note that the knowledge of the distance between the small body and the probe only affects the determination of the small-body reference frame origin (see Section \ref{sec:origin}). The image processing pipeline developed in the following sections is not affected by the scale factor as the calculations are directly performed on image points, leading to be scale invariant. Indeed, the possible solution for the rotation axis are scale invariant as well because their calculation is computed from the eigenvectors which are known up to a scale factor (see Sections \ref{sec:rotationaxestilted} and \ref{sec:rotationaxesperp}). It is worth noting that the small-body reference frame origin is used to discard unfeasible solutions for the not-degenerate case (see Section \ref{sec:tiltedpruningandselection}), but a change in the scale - due to an error in the asteroid-spacecraft range - would affect only the small body overall size without any impact in the pruning process. In other words, the spacecraft would estimate points moving on a larger small body if the range is larger than the truth. Otherwise, the spacecraft would estimate points moving a smaller celestial body.}

\section{Determination of the Small-Body-Fixed Reference Frame Origin}\label{sec:origin}
In this section the estimation of an approximation of the rotation center is presented. From rigid body dynamics, the rotation axis passes through the small body barycenter, thus the rotation center is coincident with the barycenter. As the probe is not orbiting around the small body, the mass distribution, and thus the barycenter, can not be determined.  To do so, \citet{bandyonadhyay2019silhouette} proposes to determine the center of rotation as the average between a given image and the one with the asteroid rotated of 180 degrees. This approach can be useful for low illumination angle, where self-shadowing does not influence the center of brightness estimation. A different approach foresees the computation of the center of brightness and to apply a correction to estimate the projection of the center of mass \cite{bhaskaran1998autonomous, pugliatti2022data}. For the sake of simplicity. the center of brightness is computed only for the first image in this work and no correction is applied. Only the first image is used as the distance to the small body is considered known only for that instant. Note that the center of brightness is usually close to the center of mass only for low illumination angles, as the center of brightness can shift considerably when the illumination angle increases. In the framework of this work, no correction is applied for the sake of simplicity and future work should investigate the benefit of adding the correction within the proposed algorithmic pipeline.\newline
To compute the center of brightness, the first image is binarized by using Otsu's method \cite{otsu1979threshold} to identify the bright pixels. By labeling $\chi_1$ the bright pixel map in the first image, the center of brightness ${}^{\mathbb{C}_1}\bm{R}_{\textrm{CB}}$ expressed in the first image reference frame $\mathbb{C}_1$ is computed as
\begin{equation}
{}^{\mathbb{C}_1}\bm{R}_{\textrm{CB}} = \left(\frac{\sum_{P\in\chi_1}^{\,}\bm{P}_x}{\left|\chi_1\right|},\, \frac{\sum_{P\in\chi_1}^{\,}\bm{P}_y}{\left|\chi_1\right|}\right)^T
\end{equation}
where $\bm{P}$ is an arbitrary pixel in $\chi_1$, the subscripts $x$ and $y$ label the $x$ and $y$ coordinates in the image, and $\left|\chi_1\right|$ is the area of $\chi_1$.\newline
By knowing the distance from the first camera pose to the small body, the rotation center $\bm{r}_{\text{RC}}$ is computed by backprojecting in the 3D space the center of brightness:
\begin{equation}
{}^{\mathcal{N}}\bm{r}_{\text{RC}} = {}^{\mathcal{N}}\bm{r}_{\text{SC}_1} + \frac{d_{\rm SC_1/SB}}{\left|\left|\left[K\right]^{-1} {}_h^{\mathbb{C}}\bm{R}_{\textrm{CB}}\right|\right|} \left[NC_1\right] \left[K\right]^{-1} {}_h^{\mathbb{C}_1}\bm{R}_{\textrm{CB}}
\end{equation}
where$\left[K\right]$ is the camera calibration matrix, $\bm{r}_{\text{SC}_1}$ is the spacecraft position at the first image time, $d_{\rm SC_1/SB}$ is the small-body-spacecraft distance at the first image time, and $\left[NC_1\right]$ is the rotation matrix from the camera frame $\mathcal{C}_1$ at the first image time to the inertial frame $\mathcal{N}$. Note that the camera calibration matrix can be computed from camera characteristics as follows:
\begin{equation}
	\left[K\right] = \begin{bmatrix}
	\frac{f}{s_x} & 0 & C_x\\ 0 & \frac{f}{s_y} & C_y\\ 0 & 0& 1
	\end{bmatrix}
\end{equation}
where $f$ if the camera focal length, $s_x$ and $s_y$ are the pixel physical size in $x$ and $y$ components, and  $\bm{C} = \left(C_x, C_y\right)^T$ is the camera center in pixels.\newline
The computation of ${}^{\mathcal{N}}\bm{r}_{\text{RC}}$ defines the origin of the small-body-fixed reference frame $\mathcal{B}$. In order to fully define $\mathcal{B}$, the angular velocity $\bm{\omega}_{\mathcal{B}/\mathcal{N}}$ of the small-body-fixed reference frame $\mathcal{B}$ with respect to the inertial reference frame $\mathcal{N}$ must be characterized. As the rotational period is known, only the unit vector of the rotational axis $\hat{\bm{\omega}}_{\mathcal{B}/\mathcal{N}}$ must be estimated. 

\section{Feature Extraction and Tracking}\label{sec:IP}
\subsection{Feature-Based Image Processing}
The first step of the algorithm is to process images to detect landmarks' movement. From subsequent images, information about the relative motion of the scene can be recovered. The apparent motion of the scene and objects in the image is called optical flow, which is caused by both the observer's and the object's motion. The optical flow estimation provides information about the relative movement between the observer and the scene.\newline 
A standard procedure to perform this task is to use a feature-based image processing algorithm. Features are usually 2D entities that are characterized by a location and a compact description of the feature information. As a consequence, two tasks are performed to identify a feature \cite{szeliski2022computer}: feature detection and feature description. Feature extraction deals with the problem of determining the feature locations in the image. Many algorithms (e.g., SURF \cite{bay2006surf}, BRISK \cite{leutenegger2011brisk}, ORB \cite{rublee2011orb}, SIFT \cite{lowe2004distinctive}) are available to perform this task. The main difference is the type of structure (e.g., corners, edges, or blobs) they try to detect and the required computational effort. 
Feature description is designed to build a compact and informative descriptor of the information contained in the feature location neighborhood designed to be non-redundant and optimized for feature association. Feature detection algorithms often have an associated descriptor (e.g., SURF \cite{bay2006surf}, BRISK \cite{leutenegger2011brisk}, and KAZE \cite{alcantarilla2012kaze}), but any combination of feature description and detection is possible.\newline
Once features are detected from different images, it is necessary to determine their path  to gather the optical flow. This is a complex procedure as feature appearance can significantly change from different points of view and illumination conditions. Moreover, features can be shadowed during their motion which implies their loss in the image. This task can be fulfilled with two different approaches \cite{szeliski2022computer}: feature matching and feature tracking. On the one hand, feature matching exploits feature description to perform feature association between different images. A naive and widely-used method to perform this task is brute force matching where features in one image are associated according to the closest descriptor in the other image. As feature matching is based on feature description, it is usually exploited when images are generated from different observation geometries and when the motion is rapidly changing in time. On the other hand, feature tracking determines feature association by searching in the neighborhood of the feature location. A classical approach to perform feature tracking is Kanade-Lucas-Tomasi (KLT) algorithm \cite{lucas1981iterative,tomasi1992shape}.
The KLT tracker exploits the spatial gradient of image intensity to find the feature  in the next image by Newton-Raphson \EPP{gradient} descent algorithm.\newline
In this work SURF features are used to identify salient points in small body images. They are chosen as they are extracted faster than other features and are robust to camera rotation and translation. Feature association is then performed with the KLT algorithm as it does not require the use of time-consuming descriptor algorithms.

\subsection{Kanade-Lucas-Tomasi Optical Flow Tracker}
In this section, an overview of the KLT algorithm is provided as depicted in \citet{lucas1981iterative} and \citet{tomasi1992shape}. Let ${I}$ and ${J}$ be two different grayscale images. The aim is to find the displacement $\bm{D}$ of the feature $\bm{P}$ from ${I}$ to ${J}$ by minimizing the error between the two images. The error function $\varepsilon$ between the two images is defined by computing the squared difference in intensity in a region of interest $R$ which is normally considered a user-defined rectangle or square. As a consequence:
\begin{equation}
	\varepsilon = \sum_R \left({I}\left(\bm{P} + \bm{D}\right) - {J}\left(\bm{P}\right)\right)^2
\end{equation}
Image ${I}$ can be linearly approximated:
\begin{equation}
	\varepsilon \simeq \sum_R \left({I}\left(\bm{P}\right) + \left[\frac{\partial{I}}{\partial\bm{D}}\right]\bm{D} - {J}\left(\bm{P}\right)\right)^2
\end{equation}
where the spatial partial derivative $\left[\dfrac{\partial{I}}{\partial\bm{D}}\right]$ can be computed by finite differences. The error can be thus minimized:
\begin{equation}
	\begin{aligned}
		\left[\frac{\partial\varepsilon}{\partial\bm{D}}\right] \simeq& \frac{\partial}{\partial\bm{D}} \sum_R \left({I}\left(\bm{P}\right) + \left[\frac{\partial{I}}{\partial\bm{D}}\right]\bm{D} - {J}\left(\bm{P}\right)\right)^2 =\\=& 2\sum_R \left({I}\left(\bm{P}\right) + \left[\frac{\partial{I}}{\partial\bm{D}}\right]\bm{D} - {J}\left(\bm{P}\right)\right)^T\left[\frac{\partial{I}}{\partial\bm{D}}\right] = 0
	\end{aligned}
\end{equation} 
The displacement $\bm{D}$ is analytically computed:
\begin{equation}
	\bm{D} = -\left(\sum_R\left[\frac{\partial{I}}{\partial\bm{D}}\right]^T\left[\frac{\partial{I}}{\partial\bm{D}}\right]\right)^{-1}\left(\sum_R \left[\frac{\partial{I}}{\partial\bm{D}}\right]^T\left({I}\left(\bm{P}\right) - {J}\left(\bm{P}\right)\right)\right)
\end{equation}
In the KLT algorithm, not all features are tracked as only a subset is informative for the tracker. Good features are defined internally from the tracking algorithm as the ones having the highest eigenvalues of the matrix $\left(\sum_R\left[\frac{\partial{I}}{\partial\bm{D}}\right]^T\left[\frac{\partial{I}}{\partial\bm{D}}\right]\right)$ \cite{shi1994good}. This ensures that feature tracking is a well-conditioned process and that the algorithm is numerically stable. Moreover, to increase algorithm performance and to exploit different detail levels in the image a pyramidal scheme is often used~\cite{bouguet2001pyramidal}.\newline
In the present work the feature extraction and tracking are performed with Airbus Defence \& Space software Themis \cite{duteis2019genevis}. Themis is an enhanced space-certified KLT tracker which compensates for features rotation, scaling, and translation. Moreover, it implements a pyramidal scheme and homography filtering to improve the algorithm's overall performance. The use of Themis is justified by the fact that it has been optimized for space application, and it has shown its embeddability in space-certified processors LEON4 \cite{duteis2019genevis}.\newline
The output of Themis is a series of feature tracks of different length. These tracks are exploited to fit the conic obtained from the circle projection in the image and to determine the rotation axis orientation.

\section{Estimation of Small Body Rotation Axis}\label{sec:rotstate}
\subsection{Projection of the circle}\label{sec:circleprojection}
Before entering into the details of the algorithm fitting the conics generated by the projection of the tracked landmarks, it is worth analyzing how a 3D circle projects into a camera.\newline
Let  $\mathcal{P}$ the 3D reference centered in the circle origin and whose third axis oriented as the circle plane normal $\bm{n}_{\textrm{cl}}$. The circle plane is the plane where the circle lies and where the first two axes of $\mathcal{P}$ lie. Moreover, let $\mathbb{P}$ be the 2D reference frame centered in circle center projected on the circle plane and whose axes are oriented as the projection on the the circle plane of the first two axes of $\mathcal{P}$. In  $\mathbb{P}$, the equation of the circle can be expressed as follows:
\begin{equation}
		{}_h^{\mathbb{P}} \bm{P}^T 
	\begin{bmatrix}
		1 & 0 & 0\\ 0& 1 & 0\\ 0& 0& -R^2
	\end{bmatrix}
		{}_h^{\mathbb{P}} \bm{P} = {}_h^{\mathbb{P}} \bm{P}^T 
		\left[C\right]
		{}_h^{\mathbb{P}} \bm{P} = 0
\end{equation}
where $R$ is the circle radius and ${}_h^{\mathbb{P}} \bm{P}^T $ is a 2D homogeneous point expressed in  $\mathbb{P}$ and belonging to the circle plane defined by $\bm{n}_{\textrm{cl}}$. More details about homogeneous coordinates can be found in \citet{hartley2003multiple}. By denoting $\mathbb{C}$  the image plane reference, the mapping from the circle plane to the image plane is defined by the following homography \cite{hartley2003multiple}:
\begin{equation}
	\left[\mathbb{C}\mathbb{P}\right] = \left[K\right]\left[{}^{\mathcal{C}}\bm{p}_1\; {}^{\mathcal{C}}\bm{p}_2\; {}^{\mathcal{C}}\bm{r}_{\textrm{cl}} \right]
\end{equation}
where ${}^{\mathcal{C}}\bm{p}_i$  is the $i$th axis of $\mathcal{P}$ expressed in $\mathcal{C}$ and  ${}^{\mathcal{C}}\bm{r}_{\textrm{cl}}$ is the vector from the camera to the circle origin in $\mathcal{C}$. Note that ${}^{\mathcal{C}}\bm{p}_i$  is also the $i$th column of the rotation matrix $\left[CP\right]$ from $\mathcal{P}$ to  the camera frame $\mathcal{C}$.\newline
The circle as projected in the camera can be computed by noticing that:
\begin{equation}
	{}_h^{\mathbb{C}} \bm{P} = \left[\mathbb{C}\mathbb{P}\right]{}_h^{\mathbb{P}} \bm{P} 
\end{equation}
Thus:
\begin{equation}\label{eq:projectioncircle}
	{}_h^{\mathbb{C}} \bm{P}^T \left[\mathbb{C}\mathbb{P}\right]^T \left[C\right]\left[\mathbb{C}\mathbb{P}\right]{}_h^{\mathbb{P}} \bm{P} = {}_h^{\mathbb{C}} \bm{P}^T \left[E\right]{}_h^{\mathbb{C}} \bm{P} = 0
\end{equation}
where $\left[E\right]$ is the conic representing the projection of the circle in the image.
If $\left[E\right]$ is full rank, the conic represents an ellipse, an hyperbola, or a parabula. Otherwise, the conic degenerates to a line. As the circle is not a degenerate conics and its matrix has full rank, it is worth noticing that the conics degenerate to a line only if the homography matrix has not full rank. This happens when the camera boresight is perpendicular to the circle plane normal, i.e., $\bm{p}_3^T \bm{r}_{\textrm{cl}} = 0$.\newline
To prove that the projection is a single line, it is worth demonstrating that the projection of $\bm{p}_1$ and $\bm{p}_2$ in the image are parallel. For the sake of simplicity, let $\hat{\bm{r}}_{\textrm{cl}}$ be the unit vector pointing as ${\bm{r}}_{\textrm{cl}}$. Note that $\hat{\bm{r}}_{\textrm{cl}}$ is linearly dependent of $\bm{p}_1$ and $\bm{p}_2$, thus $\hat{\bm{r}}_{\textrm{cl}} = \hat{\bm{r}}_{\textrm{cl}}^T\bm{p}_1 \bm{p}_1  +\hat{\bm{r}}_{\textrm{cl}}^T\bm{p}_2 \bm{p}_2$. The vector perpendicular to the camera boresight unit vector are:
\begin{equation}
	\bm{e}_1 = \bm{p}_1 - \hat{\bm{r}}_{\textrm{cl}}^T\bm{p}_1 \hat{\bm{r}}_{\textrm{cl}}
\end{equation}
\begin{equation}
	\bm{e}_2 = \bm{p}_2 - \hat{\bm{r}}_{\textrm{cl}}^T\bm{p}_2 \hat{\bm{r}}_{\textrm{cl}}
\end{equation}
Note that these vectors belongs to the circle plane and to the image plane. To prove that these vectors are parallel, it must be proven that $\bm{e}_1 \times \bm{e}_2 = 0$. Thus:
\begin{equation}
\begin{aligned}
	\bm{e}_1 \times \bm{e}_2 &= \bm{p}_1 \times \bm{p}_2  - \hat{\bm{r}}_{\textrm{cl}}^T\bm{p}_2 \bm{p}_1 \times  \hat{\bm{r}}_{\textrm{cl}} - \hat{\bm{r}}_{\textrm{cl}}^T\bm{p}_1 \hat{\bm{r}}_{\textrm{cl}} \times \bm{p}_2 + \hat{\bm{r}}_{\textrm{cl}}^T\bm{p}_1 \, \hat{\bm{r}}_{\textrm{cl}}^T\bm{p}_2 \hat{\bm{r}}_{\textrm{cl}} \times \hat{\bm{r}}_{\textrm{cl}} = \\
	& = \bm{p}_3 - \left(\hat{\bm{r}}_{\textrm{cl}}^T\bm{p}_2\right)^2 \bm{p}_3 - \left(\hat{\bm{r}}_{\textrm{cl}}^T\bm{p}_1\right)^2 \bm{p}_3= \left(1 - \left(\hat{\bm{r}}_{\textrm{cl}}^T\bm{p}_2\right)^2 - \left(\hat{\bm{r}}_{\textrm{cl}}^T\bm{p}_1\right)^2\right)\bm{p}_3 =\\ 
	&= \left(1 -\left|\left|\hat{\bm{r}}_{\textrm{cl}} \right|\right|^2\right)\bm{p}_3 = 0
\end{aligned}
\end{equation}
This implies that the degenerate conic is composed of a single line. Therefore either the circle is projected as a full rank conic or to a rank-1 degenerate conic. No other situations are possible. From an operative perspective, this suggests that the algorithm must be able to cope with full-rank conic estimation and with single line fitting.\newline
When the conic is full rank, it is represented by a full rank matrix which is usually written as follows:
\begin{equation}\label{eq:ellipsematrix}
	\left[E\right] = \begin{bmatrix}
		A & \frac{B}{2} & \frac{D}{2}\\ \frac{B}{2}&  C & \frac{E}{2}\\ \frac{D}{2}& \frac{E}{2}& F
	\end{bmatrix}
\end{equation}
Otherwise, when the conic degenerates, it is represented by the following equation \cite{hartley2003multiple}
\begin{equation}
	\left[E\right] = \bm{M}\bm{L}^T  + \bm{L}\bm{M}^T
\end{equation}
where $\bm{M}$ and $\bm{L}$ are the two lines composing the degenerate conic. As the two lines coincide for the circle projection, $\bm{L} = \bm{M}$. Thus:
\begin{equation}
	\left[E\right] = 2\bm{L}\bm{L}^T
\end{equation}
which implies that Eq.~\ref{eq:projectioncircle} can be rewritten as the classical line equation ${}_h^{\mathbb{C}} \bm{P}^T\bm{L} = 0$.\newline
As multiple features are tracked by Themis, several 3D circles are projected in the image, so different ellipses or lines are present. Note that the 3D landmark movement is caused by the rotational dynamics of the small body. Therefore, landmarks move on 3D circles whose circle plane normal is the small body rotation axis. Thus, the 3D circles all have the same circle plane normal and have different origins lying on the rotation axis.

\subsection{Detection of Degenerate Solutions}\label{sec:conicfitting}
As shown in Sec.~\ref{sec:circleprojection}, the projection of the 3D circle, generated by the landmarks, is a series of ellipses when the camera boresight is not perpendicular to the circle plane normal. As the circle plane normal is the rotation axis direction for all the 3D circles, this situation happens when the rotation axis is tilted with respect to the camera boresight. Otherwise, the projection of the 3D circle degenerates into a sheaf of parallel lines. It is thus necessary to develop a method to detect whether the rotation axis is perpendicular to the camera boresight only using tracked features. To do so, it is proposed hereafter a detection algorithm to understand if the conic projection is degenerate.\newline
Recall that a point lying on an ellipse can be represented as follows:
\begin{equation}\label{eq:conic}
	A\,\xi^2 + B\,\xi\eta + C\,\eta^2 + D\,\xi + E\,\eta + F =0 \quad \text{with}\quad 4AC-B^2>0
\end{equation}
where $\left(\xi,\,\eta\right)$ are Cartesian coordinates obtained from the image coordinates through normalization with respect to the image size. The normalization is required to let the following steps of the algorithm be independent with respect to the image size.\newline
All the reprojected points of the same track must verify Eq.~\ref{eq:conic}. Thus:
\begin{equation}
	\begin{bmatrix}
		\xi_1^2 & \xi_1\eta_1 &\eta_1^2 &\xi_1 & \eta_1&1\\
		\vdots& \vdots &\vdots&\vdots& \vdots &\vdots \\
		\xi_i^2 & \xi_i\eta_i &\eta_i^2 &\xi_i & \eta_i&1\\
		\vdots& \vdots &\vdots&\vdots& \vdots &\vdots \\
		\xi_{N_j}^2 & \xi^{\,}_{N_j}\eta^{\,}_{N_j} &\eta_{N_j}^2 &\xi^{\,}_{N_j} & \eta^{\,}_{N_j} &1\\
	\end{bmatrix}\begin{pmatrix}
		A_j\\B_j\\C_j\\	D_j\\E_j\\F_j
	\end{pmatrix} = \left[\left[D_{2,j}\right]\; \left[D_{1,j}\right]\right]\begin{pmatrix}
		A_j\\B_j\\C_j\\	D_j\\E_j\\F_j
	\end{pmatrix}=\left[\mymathbb{0}_{N_j\times1}\right]
\end{equation}
where $N_j$ is the number of frames during which the 3D point is tracked for the $j$th conic, $\left[\mymathbb{0}_{n\times m}\right]$ is the zero $n\times m$ matrix, the $j$ subscript define the parameters associated with the $j$th conic, $\left[D_{2,j}\right]\in\mathbb{R}^{N_j\times 3}$ denotes the matrix associated with the second-order conic coefficients, and $\left[D_{1,j}\right]\in\mathbb{R}^{N_j\times 3}$ the matrix associated with the remaining-orders conics coefficients.\newline
Recall that a point lying on a line can be represented as follows:
\begin{equation}\label{eq:lineEq}
	D\,\xi + E\,\eta + F =0 \quad \text{with}\quad D^2+E^2 =1
\end{equation}
Moreover, the points of the $j$th line verify:
\begin{equation}\label{eq:line}
	\begin{bmatrix}
		\xi_1 & \eta_1&1\\
		\vdots& \vdots &\vdots \\
		\xi_i & \eta_i&1\\
		\vdots& \vdots & \vdots\\
		\xi^{\,}_{N_j} & \eta^{\,}_{N_j} &1\\
	\end{bmatrix}\begin{pmatrix}
		D_j\\E_j\\F_j
	\end{pmatrix} = \left[D_{1,j}\right]\begin{pmatrix}
		D_j\\E_j\\F_j
	\end{pmatrix}=\left[\mymathbb{0}_{N_j\times1}\right]
\end{equation}
As the lines of the matrix in Eq.~\ref{eq:line} are linearly dependent, the rank of the matrix $\left[D_{1,j}\right]$ is one. Due to tracking errors, the rank is not one. Nevertheless, being $\left[D_{1,j}\right]$ nearly singular,  its condition number is high. Note that this is only true when the conics are degenerate. As $\left[D_{1,j}\right]$ has $N_j$ lines, it is more convenient to work with $\left[S_{3,j}\right] = \left[D_{1,j}\right]^T\left[D_{1,j}\right] \in\mathbb{R}^{3\times 3}$.\newline
To understand if the family of conics degenerates, the following steps are applied:
\begin{enumerate}
	\item For each tracked feature, the track length is computed to store only the longest $M_{\text{curv}}$ tracks. This step is necessary to avoid processing a high number of tracks which increases the numerical burden. Moreover, it is worth noting that long-tracked features are the ones that provide more information about the rotational state. $M_{\text{curv}}$ is set to 50 in this work. The maximum number of curves number can be also set free and select a minimum tracking time to reject short tracks. The proposed approach is used in this work also to limit the computational burden.
	\item For each tracked feature, the conditioning number of $\left[S_{3,j}\right]$ is computed. If the tracks conditioning number is greater than a threshold $\gamma_{\text{CN}}$ for more than $\gamma_{\text{sheaf}}$, the conics are considered degenerate and processed as a sheaf of parallel lines. Otherwise, they are considered a family of ellipses. $\gamma_{\text{CN}}$ and $\gamma_{\text{sheaf}}$ are set to $10^5$ and 0.8 respectively in this work. See Section~\ref{sec:detectionresults} for comments about parameter selection.
\end{enumerate}
This procedure detects autonomously whether the rotation axis is perpendicular to the camera boresight. As the 3D circle projection generates different conics, the estimation procedure is not identical between the two cases and two algorithms must be developed. First, in Sec.~\ref{sec:tiltedcase} the general case of a tilted axis with respect to the camera boresight is presented. Then, in Sec.~\ref{sec:perpaxis} the algorithm when the two axes are perpendicular is explained.

\subsection{Rotation Axis Tilted with respect to Camera Boresight}\label{sec:tiltedcase}
\subsubsection{Ellipses Fitting}
\begin{figure}[!t]
	\centering
	\begin{subfigure}{0.4\textwidth}
		\includegraphics[width=\textwidth,trim={1.5cm 0cm 2.5cm 0cm},clip]{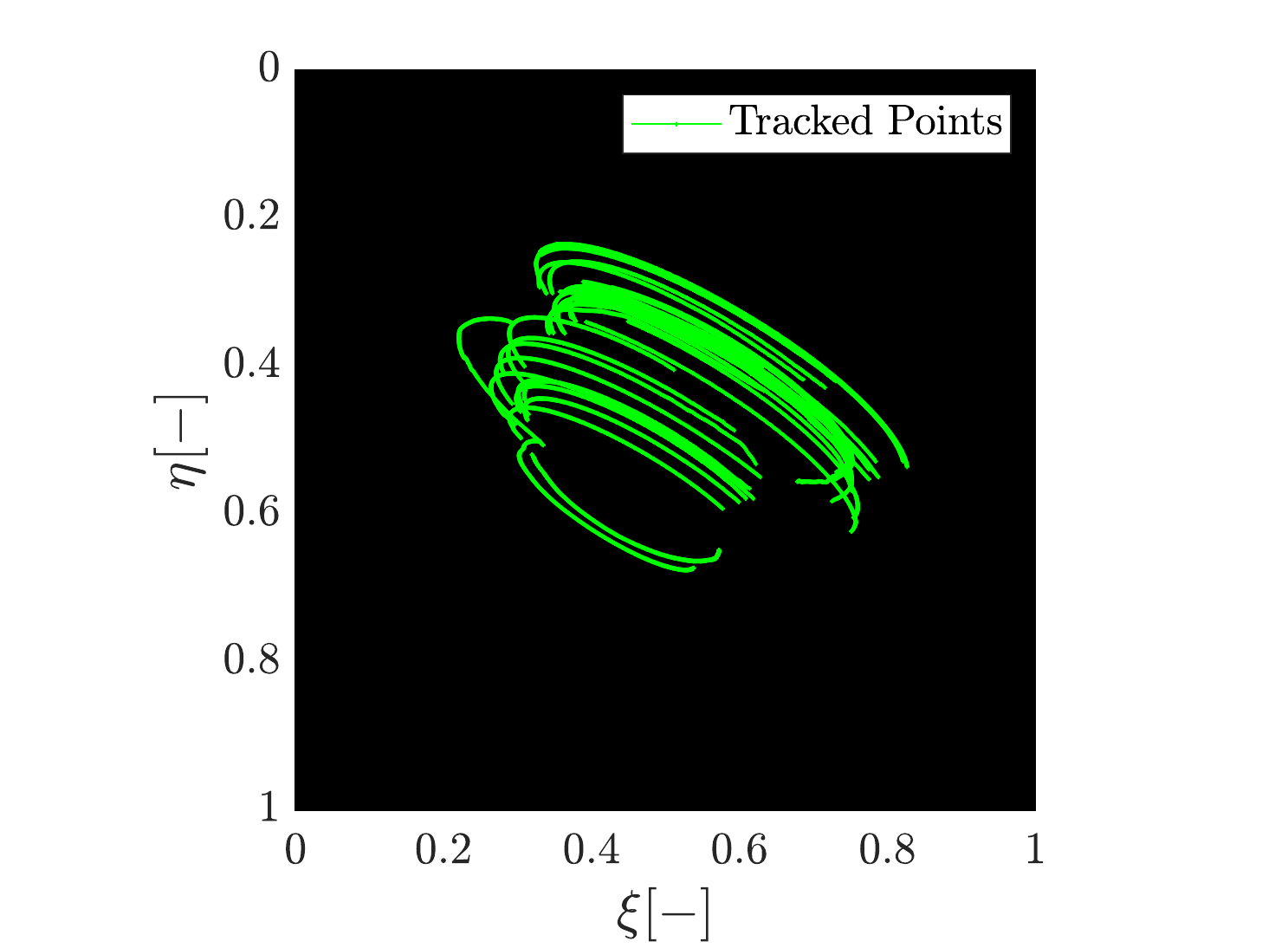}
		\caption{The tracked points}
		\label{fig:tiltedTracking}
	\end{subfigure}
	\hspace{1cm}
	\begin{subfigure}{0.4\textwidth}
		\includegraphics[width=\textwidth,trim={1.5cm 0cm 2.5cm 0cm},clip]{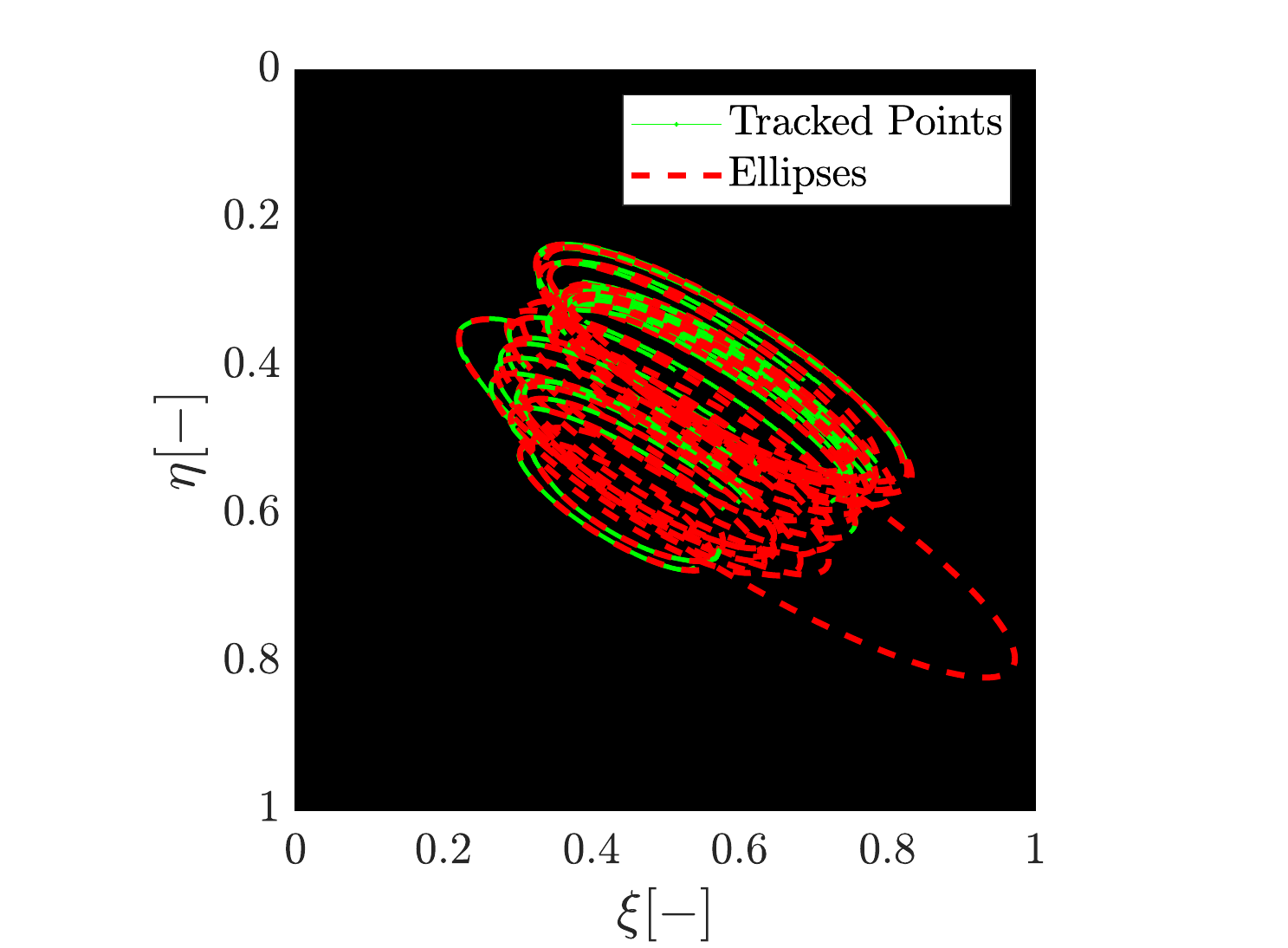}
		\caption{The ellipse reconstruction with same orientation $\phi$}
		\label{fig:tiltedEllipses_pt1}
	\end{subfigure}
	\caption{Example of the ellipse fitting from tracked features.}
\end{figure}

If the rotation axis is tilted with respect to the camera boresight, the projection of the concentric circles is a family of ellipses. As the 3D circle origins belong to the same axis, the ellipse semi-major axes are all oriented in the same direction \cite{bissonnette2015vision}. The optimization to gather the ellipses family exploits this property to compute a series of ellipses with the same orientation. This optimization technique is outlined in \citet{bissonnette2015vision} and reviewed hereafter.\newline
Let $M$ be the number of ellipses to be found. The orientation $\phi_j$ of the $j$th ellipse  according to its Cartesian representation (see Eq.~\ref{eq:conic}) is:
\begin{equation}\label{eq:orientationconstraint}
	\tan2\phi_j = \frac{B_j}{C_j-A_j}
\end{equation}
Eq.~\ref{eq:orientationconstraint} shows that to obtain a family of ellipses with the same orientation, the fraction of the three Cartesian parameters must be the same for all the $M$ ellipses. By introducing $k_j = C_j-A_j$, it is straightforward to impose that $B_j=B$ and $k_j=k\;\forall\,j\in\left[1,\,M\right]$ to implicitly verify the constraint in Eq.~\ref{eq:orientationconstraint}. Thus, the equation of the $j$th ellipses becomes \cite{bissonnette2015vision}: 
\begin{equation}
	A_j \left(\xi^2 + \eta^2\right) + B\xi\eta + k\eta^2 + D_j\,\xi + E_j\,\eta + F_j=0
\end{equation}
By defining the vectors $\bm{a}_1 = \left(B,\, k,\, A_1,\,\cdots,\,A_M\right)^T$ and $\bm{a}_2 = \left(D_1,\,E_1,\,F_1,\, \cdots,\,D_M,\,E_M,\,F_M\right)^T$ and the matrices
\begin{equation}
	\left[D_2\right] = \begin{bmatrix}
		\xi_{1,1}\eta_{1,1} & \eta_{1,1}^2 & \xi_{1,1}^2+\eta_{1,1}^2& 0 & 0 & 0 \\
		\vdots& \vdots& \vdots& \vdots & \vdots & \vdots \\
		\xi_{N_1,1}\eta_{N_1,1} & \eta_{N_1,1}^2 & \xi_{N_1,1}^2+\eta_{N_1,1}^2&0 & 0 & 0 \\
		\xi_{1,2}\eta_{1,2} & \eta_{1,2}^2 & 0& \xi_{1,2}^2+\eta_{1,2}^2& 0 &0 \\	
		\vdots& \vdots& \vdots& \vdots & \vdots & \vdots \\
		\xi_{N_2,2}\eta_{N_2,2} & \eta_{N_2,2}^2 &0& \xi_{N_2,2}^2+\eta_{N_2,2}^2 & 0 & 0 \\
		\vdots& \vdots& \vdots& \vdots & \ddots & \vdots \\
		\xi_{1,M}\eta_{1,M} & \eta_{1,M}^2 & 0& 0 &0& \xi_{1,M}^2+\eta_{1,M}^2 \\	
		\vdots& \vdots& \vdots& \vdots & \vdots & \vdots \\
		\xi_{N_M,M}\eta_{N_M,M} & \eta_{N_M,M}^2 & 0& 0 &0& \xi_{N_M,M}^2+\eta_{N_M,M}^2		\end{bmatrix}
\end{equation}
\begin{equation}
	\left[D_1\right] = \begin{bmatrix}
		\xi_{1,1} & \eta_{1,1} & 1& 0 & 0 & 0 & 0 & 0 &0&0 \\	
		\vdots& \vdots& \vdots& \vdots & \vdots & \vdots& \vdots &\vdots &\vdots & \vdots \\
		\xi_{N_1,1} & \eta_{N_1,1} & 1&0 & 0 & 0 & 0 &0 &0&0 \\	
		
		0&0 & 0& \xi_{1,2} & \eta_{1,2}& 1& 0 &0 &0&0 \\	
		\vdots& \vdots& \vdots& \vdots & \vdots & \vdots& \vdots &\vdots &\vdots & \vdots \\
		0&0 & 0& \xi_{N_2,2} & \eta_{N_2,2}& 1& 0 &0 &0&0 \\	
		\vdots& \vdots& \vdots& \vdots & \vdots & \vdots& \ddots &\vdots &\vdots & \vdots \\
		0 & 0 & 0& 0 &0& 0 &0 & \xi_{1,M} & \eta_{1,M}& 1 \\	
		\vdots& \vdots& \vdots& \vdots &\vdots & \vdots & \vdots& \vdots & \vdots & \vdots \\
		0 & 0 & 0& 0 &0& 0 &0 & \xi_{N_M,M} & \eta_{N_M,M}& 1 \\		\end{bmatrix} 
\end{equation}
where $\left(\xi_{i,j},\,\eta_{i,j}\right)$ is the $i$th point of the $j$th ellipse in normalized coordinates, the problem is to be rewritten as:
\begin{equation}
	\Big[\left[D_2\right]\;\left[D_1\right]\Big]\begin{pmatrix}
		\bm{a}_1\\\bm{a}_2
	\end{pmatrix} = \left[D\right] \bm{a} = \left[\mymathbb{0}_{\left(4M+2\right)\times1}\right]
\end{equation}
The solution of this homogeneous equation is found by minimizing $J = \bm{a}^T\left[D\right]^T\left[D\right]\bm{a}$ subject to $\bm{a}^T\bm{a}=1$.\newline
This optimization can be solved by solving the rank-deficient generalized eigenvalue system associated with the problem \cite{fitzgibbon1999direct}. To have a numerically-stable computation the algorithm proposed by  
\citet{halir1998numerically} is used to compute the solution. By defining $\left[S_1\right] = \left[D_1\right]^T\left[D_1\right]$, $\left[S_2\right] = \left[D_2\right]^T\left[D_1\right]$ and 	$\left[S_3\right] = \left[D_2\right]^T\left[D_2\right]$, the solution for $\bm{a}_1$ is the eigenvector associated with the smallest positive eigenvalue of $\left(\left[S_1\right]-\left[S_2\right]\left[S_3\right]^{-1}\left[S_2\right]^T\right)$. The solution for $\bm{a}_2$ is the computed as $\bm{a}_2 = -\left[S_3\right]^{-1}\left[S_2\right]^T\bm{a}_1$.\newline
This provides a family of conics all with the same orientation. Note that the conics are not necessary ellipses as no constraint has been imposed on the Cartesian coefficients to verify it. Thus, for each conic, the eccentricity is computed and, if greater or equal to 1, the conic is discarded. In Fig.~\ref{fig:tiltedTracking} and \ref{fig:tiltedEllipses_pt1}  the different optimization outputs  for the main steps are shown.

\subsubsection{Computation of the Rotation Axis Candidates}\label{sec:rotationaxestilted}
\begin{figure}[!t]
	\centering
	\includegraphics[width=0.7\textwidth]{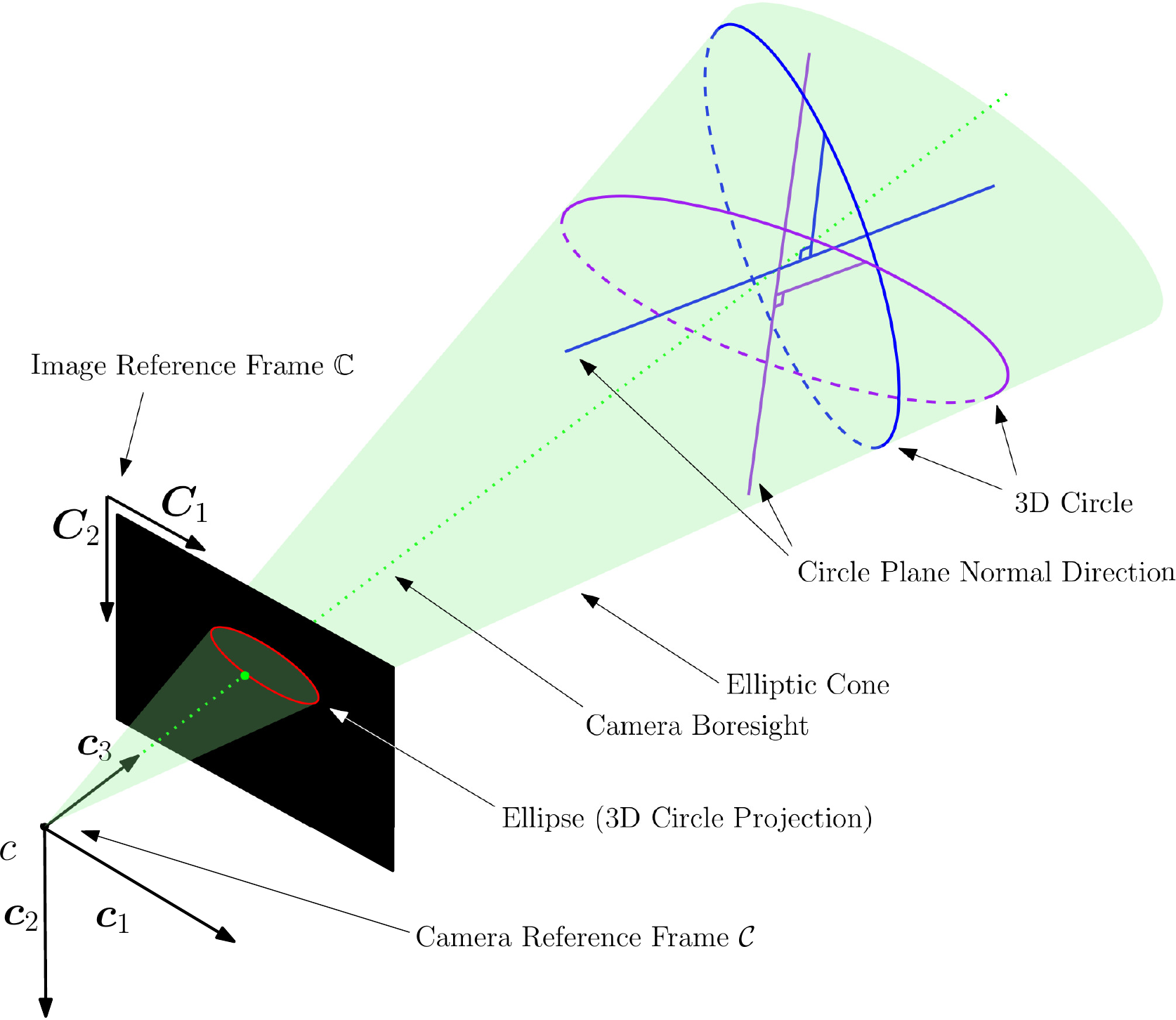}
	\caption{The geometrical construction generating the two circles.}
	\label{fig:tiltedOmega_ConeCircleIntersection}
	\includegraphics[width=0.7\textwidth]{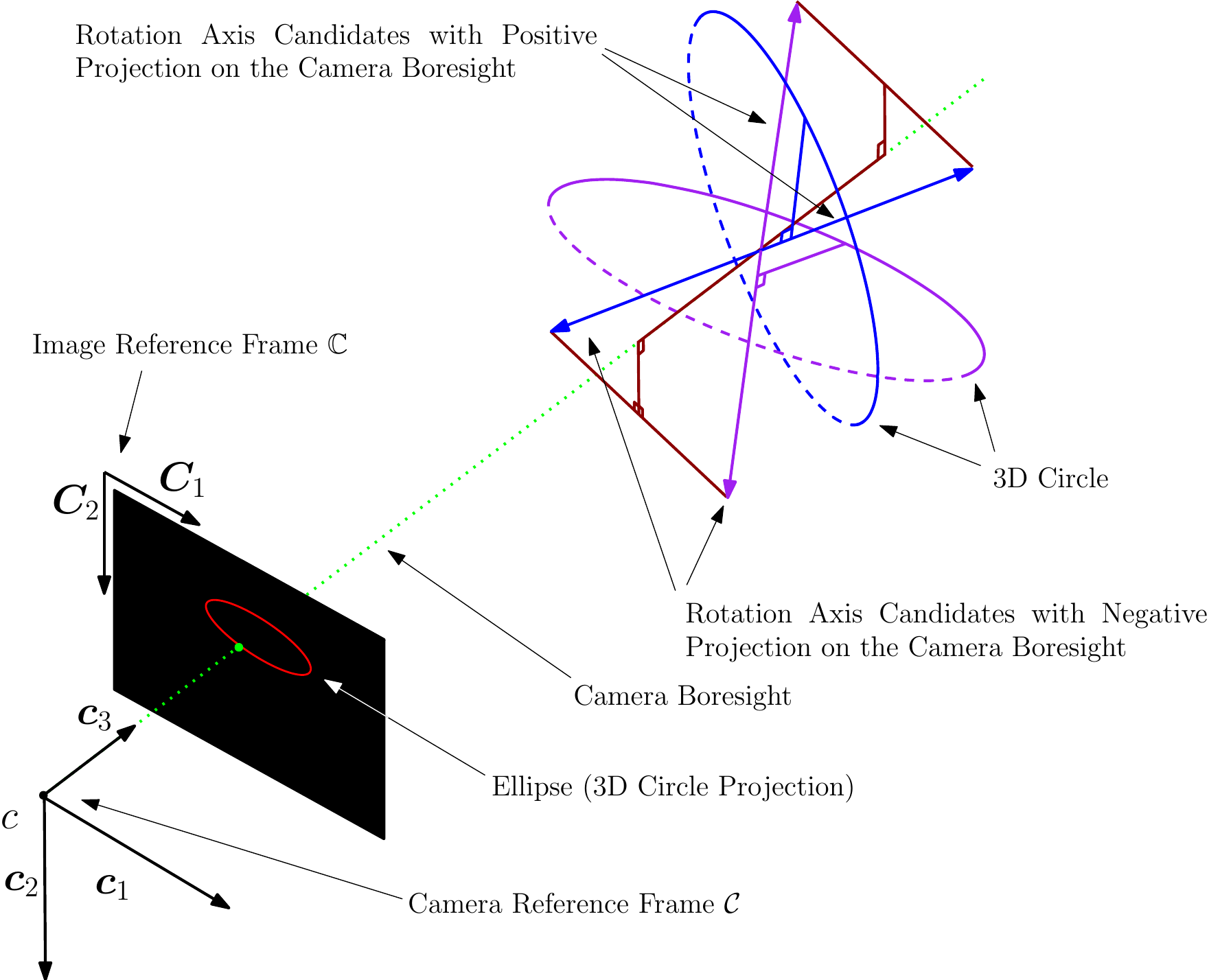}
	\caption{The four normal vector solutions from the circle computation. }
	\label{fig:tiltedOmega_Directions}
\end{figure}

Once the ellipses family is found, the rotation axis can be reconstructed by knowing that a 2D ellipse in the image is the projection of a 3D circle \cite{kanatani2004automatic}. From a single ellipse, an elliptic cone, i.e. a cone with an elliptical cross-section, is generated from backprojection. If a sheaf of planes is intersected with the cone, only two planes generate circles when intersected \cite{kanatani2004automatic}. This geometrical construction is shown in Fig.~\ref{fig:tiltedOmega_ConeCircleIntersection}. In the figure, the red ellipse is backprojected in the 3D space generating the green opaque cone. Among all the possible directions only two of them are generating a circle on the cone when intersecting it. The two circle solutions are shown in purple and blue. The solid part of the circle is the one visible from the reader's perspective and not covered by the elliptic cone, while the dashed part is the one hidden by the cone to the reader. The blue and purple lines represent the perpendicular directions to the circles. The green dotted line represents the camera boresight which intersects the image plane in the green dot.
\newline
From these 2 planes, 4 unit normal vectors can be defined. It is important to note that both the positive and the negative normal vectors define the same plane. But, if the goal is to gather the rotation axis, the two normal vectors are associated with opposite rotation directions on the same plane. Therefore, the correct normal vector must be identified.\newline
The first step is to compute the four unit normal vectors. By knowing that the matrix $\left[E\right]$ (see Eq.~\ref{eq:ellipsematrix}) represents an ellipse which is projection of a 3D circle, all the unit normal vectors generating the conics in the image can be extracted  \cite{kanatani2004automatic}. Note that the matrix $\left[E\right]$ is composed of the ellipse Cartesian parameters in pixel units. Thus the parameters found in the previous step must be rescaled according to the normalization procedure for consistency.\newline
Let $\lambda_i$ and $\bm{u}_i$ $\forall i \in \left[1,\,3\right]$ be the eigenvalues and eigenvectors of $\left[E\right]$. If $\det\left[E\right]<0$, the eigenvalues are ordered in such a way that $\lambda_3<0<\lambda_1\leq\lambda_2$. The four solutions for the circle plane normal are computed as \cite{kanatani2004automatic}:
\begin{equation}
	\hat{\bm{\nu}} = \pm\frac{\bm{\nu}}{\left|\left|\bm{\nu}\right|\right|} \quad\text{where}\quad \bm{\nu} = \sqrt{\frac{\lambda_2-\lambda_1}{\lambda_2-\lambda_3}}\bm{u}_2 \pm \sqrt{\frac{\lambda_1-\lambda_3}{\lambda_2-\lambda_3}}\bm{u}_3
\end{equation}
To estimate the four solutions, the two vectors $\bm{\nu}$ are computed for each ellipse and their mean is calculated to minimize numerical errors.
It is worth noting that the proposed method is not robust with respect to outliers. RANSAC could be exploited to identify the correct orientation among a given orientation set as in \citet{andreis2022robust}.\newline
In Fig.~\ref{fig:tiltedOmega_Directions} the geometry of the normal vectors reconstruction is shown. The four solutions are depicted in two different colors according to the generating circle plane. Moreover, their projections on the $\mathcal{C}_1$ reference frame are shown in brown. Note that they are paired in two different ways. First, each pair has the same projection on the camera boresight direction. Second, each pair has the same module in the camera plane, but opposite direction. Recall that the solutions with opposite directions but associated with the same plane denote the same circle, but swept in opposite directions.

\subsubsection{Rotation Axis Pruning and Selection}\label{sec:tiltedpruningandselection}
As the correct rotation axis must be found among the four solutions, a procedure is outlined hereafter to identify the correct normal vector among the 4 different possibilities. The main idea is to remove two spurious solutions among the four by exploiting how features move in time. The estimation of the correct rotation axis between the two remaining solutions is a more complex task as they are both acceptable with respect to the camera projection and both respect the features' motion. The idea behind the correct identification of the solution is the selection based on a heuristic, using statistical facts or a priori information. Two heuristics are investigated:
\begin{enumerate}
	\item A coarse guess of the rotation axis direction is available on board from ground-based observation. The correct solution is identified as the closest one to the onboard guess.
	\item It is remarked that, even though the observed small body can have concavities, the majority of the initialized 3D points during tracking lie between the camera and the estimated rotation center. To have statistical relevance, this reasoning must be applied to a high number of 3D points. As a consequence all 3D points that are tracked more than a given percentage of frame $\gamma_{\text{heur}}$ over one rotational period are considered. In this work, $\gamma_{\text{heur}}$ is set to be 0.2. Once it is understood which is the solution that initializes the majority of the point closer to the camera, that solution is the correct one \cite{bissonnette2015vision}. Note that increasing $\gamma_{\text{heur}}$ would imply to consider longer tracks, thus less point. On the contrary, decreasing $\gamma_{\text{heur}}$ would lead to more statistical relevance at the cost of more computational burden. The selected value is selected as a trade off between these two factors, leading to satisfying results.
\end{enumerate}
The first step is to exploit the feature rotation as seen in the image to remove two spurious solutions. Except tracking errors, matched points rotate all in the same direction. This rotation identifies uniquely the projection of the rotation axis on the camera boresight. Thus, by computing the feature rotation, it is possible to identify the two correct rotation axis candidates.\newline
To avoid selecting the wrong solutions due to tracking errors, the rotation direction of each track is computed and the rotation axis candidates not consistent with the average feature rotation are discarded. The rotation direction  $\bm{\overline{\nu}}_{{i, j}}$ between two successive frames is computed as:
\begin{equation}
{}^{\mathcal{C}}\bm{\overline{\nu}}_{{i, j}} = \left({}_h\bm{\Xi}_{i,j}\;-\; {}_h\bm{\Xi}_{c_j}\right)\times \left({}_h\bm{\Xi}_{i+1,j}\;-\;
{}_h\bm{\Xi}_{c_j}\right)
\end{equation} 
where ${}_h\bm{\Xi}_{i,j} = \left(\xi_{i,j},\,\eta_{i,j}\right)^T$ is the $j$th  tracked feature location in the $i$th image in homogeneous normalized coordinates and ${}_h\bm{\Xi}_{c_j} = \left(\xi_{c_j},\,\eta_{c_j}\right)^T$ is the $j$th ellipse center in homogeneous normalized coordinates. The vector $\bm{\overline{\nu}}_{{i, j}}$ provides information about the rotation direction of the $j$th feature between the $i$th image and the $\left(i+1\right)$th image. By counting the features moving clockwise and the ones moving counterclockwise, it is possible to understand which is the average feature motion. It is thus useful to define the clockwise index $J_{\text{cw}}$:
\begin{equation}
J_{\text{cw}}=\sum_{j=1}^{M}\sum_{i=1}^{N_j-1} J_{\text{cw}_{i,j}}\quad\text{where}\quad J_{\text{cw}_{i,j}} = \left\{\begin{aligned}
& 1 \quad \text{if}\quad {}^{\mathcal{C}}\bm{\nu}_{\text{calc}_{i, j}}^T\bm{c}_3>0\\
-&1 \quad \text{otherwise}
\end{aligned}\right.
\end{equation}
where $\bm{c}_3$ is the camera boresight unit vector. According to  $J_{\text{cw}}$ definition, features move clockwise if $J_{\text{cw}}>0$ and counterclockwise otherwise. The two correct solutions are selected as the ones providing the same rotation direction when projected on the camera boresight.\newline
At this stage, two solutions are still feasible and the correct one must be determined. To do so, two heuristic approaches are investigated. If a coarse first guess is available on board, the correct solution is selected as the one with the smallest angular error between the two. Otherwise, the second heuristic is exploited: the correct axis is the one initializing the majority of the points between the camera and the rotation center estimated in Sec.~\ref{sec:origin}. It is worth noting that the circle plane has not been estimated yet as the direction of the circle plane normal defines a sheaf of parallel planes where the circle could lie. It is thus necessary to select a point to uniquely define where the circle plane is. 
\begin{figure}[!t]
	\centering
	\includegraphics[width=0.7\textwidth]{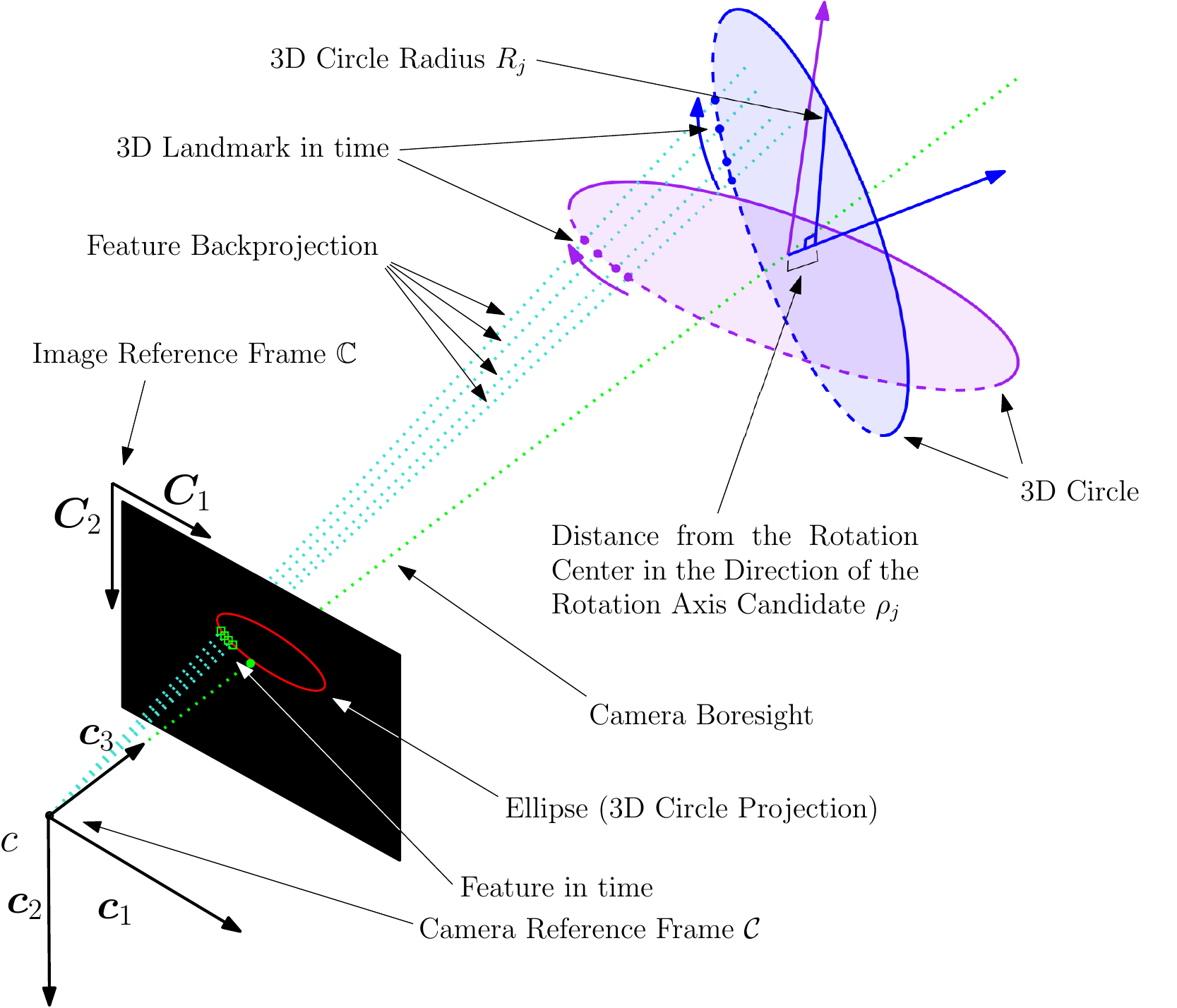}
	\caption{The estimation of the circles from the two possible rotation axis candidates.}
	\label{fig:tiltedOmega_Points}
\end{figure}
By knowing the axis direction and a point on the axis, i.e., the previously-calculated rotation center $\bm{r}_{\text{RC}}$, the circle can be fully reconstructed \cite{fremont2004turntable}. The equation of the $j$th 3D circle is written as:
\begin{equation}
U\left(u,v\right)\rho_{j}^2 + V\left(u,v\right)\rho_{j} + W\left(u,v\right) - R^2_{j}=0
\end{equation}
where $\rho_{j}$ is the distance between the rotation center $\bm{r}_{\text{RC}}$ and $j$th circle center in the direction of the rotation axis, $R_{j}$ is the $j$th circle radius, and $U\left(u,v\right)$, $V\left(u,v\right)$ and $W\left(u,v\right)$ are computed as in Appendix~\ref{app:UVW}. The parameter $\rho_{j}$ uniquely defines the plane in the sheaf, whereas $R_{j}$ uniquely identifies the circle size on that plane. Note that the pixel coordinates $\left(u,v\right)$ are used for this optimization step.\newline
This geometrical situation is depicted in Fig.~\ref{fig:tiltedOmega_Points} for two possible circle solutions colored according to the respective rotation axis. In the figure, empty green squares in the image are the tracked features and the turquoise dotted lines represent their backprojections. These backprojections intersect the two 3D circles in the full dots colored according to the intersected circle color code. The rotation direction is shown on each circle to provide information about the beginning and the end of the 3D landmark movement. Note that the rotation center is the intersection of the two possible rotation axes. Moreover, 3D circles parameters  $R_{j}$ and $\rho_{j}$ are shown only for the blue circle for the sake of completeness. Furthermore, it is worth noting that all the colored full dots on the circle are associated with one 3D point at different times during tracking. \newline
To determine the parameters $R_{j}$ and $\rho_{j}$, the following cost function is minimized:
\begin{equation}
J_{\circ,j} = \sum_{i=1}^{N_j}\left(U\left(u_i,v_i\right)\rho_{j}^2 + V\left(u_i,v_i\right)\rho_{j} + W\left(u_i,v_i\right) - R^2_{j}\right)^2
\end{equation} 
Once this procedure is performed for all tracks, an estimate of the parameters $R_{j}$ and $\rho_{j}$ $\forall j$ is determined. 
This enables the determination of the plane where the 3D circle lies and the 3D circle radius. It is now necessary to understand where the landmark has been detected at feature initialization. This is performed by computing the initialization angular position $\beta_{j}$ of the 3D landmark on the $j$th circle at the beginning of the observational period.\newline
Let $\mathcal{F}_{\hat{\bm{\nu}}_i}$ be the reference frame centered in the rotation center $\bm{r}_{\text{RC}}$ and with the vertical axis directed as $\hat{\bm{\nu}}_i$, i.e., one of the remaining solution for the rotation axis. The other two unit vectors are arbitrary. The position of the $j$th 3D landmark $^{\mathcal{F}_{\hat{\bm{\nu}}_i}}\bm{r}_{\text{LM}_j}$ in the $\mathcal{F}_{\hat{\bm{\nu}}_i}$ frame at time $t$ is:
\begin{equation}
^{\mathcal{F}_{\hat{\bm{\nu}}_i}}\bm{r}_{\text{LM}_j} = \begin{pmatrix}
R_{j}\cos\left(\beta_{j} + \dfrac{2\pi}{T_{\text{rot}}}\Delta t\right)\\
R_{j}\sin\left(\beta_{j} + \dfrac{2\pi}{T_{\text{rot}}}\Delta t\right)\\
\rho_{j}
\end{pmatrix}
\end{equation}
where $\Delta t = t-t_{\text{init}}$ is the interval between $t$ and the initialization time $t_{\text{init}}$ and $T_{\text{rot}}$ is the small body rotational period.\newline
By starting from $\beta_{j}  = 0$, the $j$th 3D landmark is projected in the image for all the available tracking times and the reprojection error is minimized to find $\beta_{j}$. Note that this optimization step could be easily removed by simple backprojection of the initial feature location and the intersection of the feature ray with the 3D circle plane, as proposed by \citet{bissonnette2015vision}. This solution has not been investigated in the present paper, but it could be a valuable option to reduce computational costs.\newline
These last two steps, i.e., the determination of the 3D circle and the angular position at the beginning of the tracking, are computed for all ellipses and the two remaining feasible orientation axis solutions. Then, by counting how many landmarks are initialized between the camera and the rotation center, it is possible to select the rotation axis solution as the one providing the higher number of point initialization closer to the camera. 

\subsection{Rotation Axis Perpendicular to Camera Boresight}\label{sec:perpaxis}
\subsubsection{Lines Fitting}
If the rotation axis is detected to be perpendicular to the camera boresight, the 3D circle projections degenerate into a sheaf of parallel lines as outlined in Sec.~\ref{sec:circleprojection}. It is worth noting that the tracked features do not form perfect lines because of tracking errors. Moreover, since the projection of the rotation axis on the camera boresight is null, only two rotation axis candidates are possible.\newline
In this section, the same notation of Sec.~\ref{sec:tiltedcase} is used for sake of simplicity. Let $M$ be the number of lines to be estimated. A sheaf of parallel lines is characterized by having the same orientation among all the lines. The equation of the $j$th line is given by Eq.~\ref{eq:lineEq}. Note that the parameters $D$ and $E$ are constant for all the lines in the sheaf by construction.\newline
Thus, $\left(\xi_{i,j},\eta_{i,j}\right)$, i.e., the $j$th feature location in the $i$th image, must verify the line equation in Eq.~\ref{eq:lineEq}. By defining the parameter vector $\bm{a} = \left(D,\,E,F_1,\,\cdots,\,F_M\right)^T$ and the matrix $\left[D_{\text{sheaf}}\right]$:
\begin{equation}
\left[D_{\text{sheaf}}\right] = \begin{bmatrix}
\xi_{1,1} & \eta_{1,1} & 1 & 0 & 0 &0 \\
\vdots & \vdots & \vdots & \vdots & \vdots & \vdots  \\ 
\xi_{N_1,1} & \eta_{N_1,1} & 1 & 0 & 0 &0 \\
\xi_{1,2} & \eta_{1,2} &0 & 1 & 0 &0 \\
\vdots & \vdots & \vdots & \vdots & \vdots & \vdots  \\ 
\xi_{N_2,2} & \eta_{N_2,2} & 0 & 1 & 0 &0 \\
\vdots & \vdots & \vdots & \vdots & \ddots & \vdots  \\ 
\xi_{1,M} & \eta_{1,M} & 0 & 0 & 0 &1 \\
\vdots & \vdots & \vdots & \vdots & \vdots & \vdots  \\ 
\xi_{N_M,M} & \eta_{N_M,M} & 0 & 0 & 0 &1\\
\end{bmatrix}
\end{equation}
the sheaf fitting problem is rewritten as 
\begin{equation}
\left[D_{\text{sheaf}}\right]\bm{a} = \left[\mymathbb{0}_{M\times1}\right]
\end{equation}
It is solved by solving through Singular Value Decomposition \cite{fitzgibbon1999direct}. An example of this procedure is reported in  Fig.~\ref{fig:perpTracking} and \ref{fig:perpEllipses}

\subsubsection{Computation of the Rotation Axis Candidates}\label{sec:rotationaxesperp}
\begin{figure}[!t]
	\centering
	\begin{subfigure}{0.4\textwidth}
		\includegraphics[width=\textwidth,trim={1.5cm 0cm 2.5cm 0cm},clip]{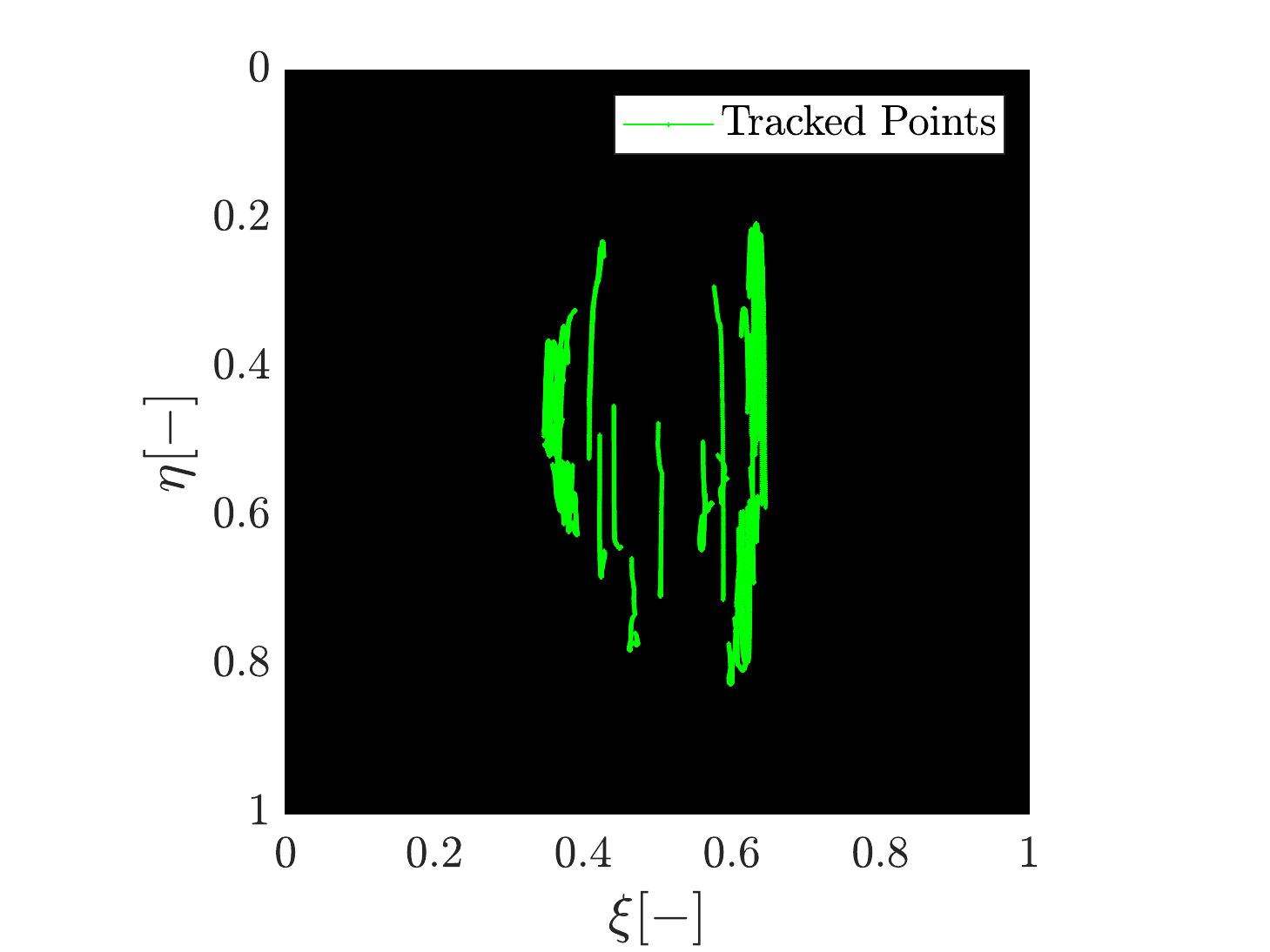}
		\caption{The tracked points}
		\label{fig:perpTracking}
	\end{subfigure}
	\hspace{0.5cm}
	\begin{subfigure}{0.4\textwidth}
		\includegraphics[width=\textwidth,trim={1.5cm 0cm 2.5cm 0cm},clip]{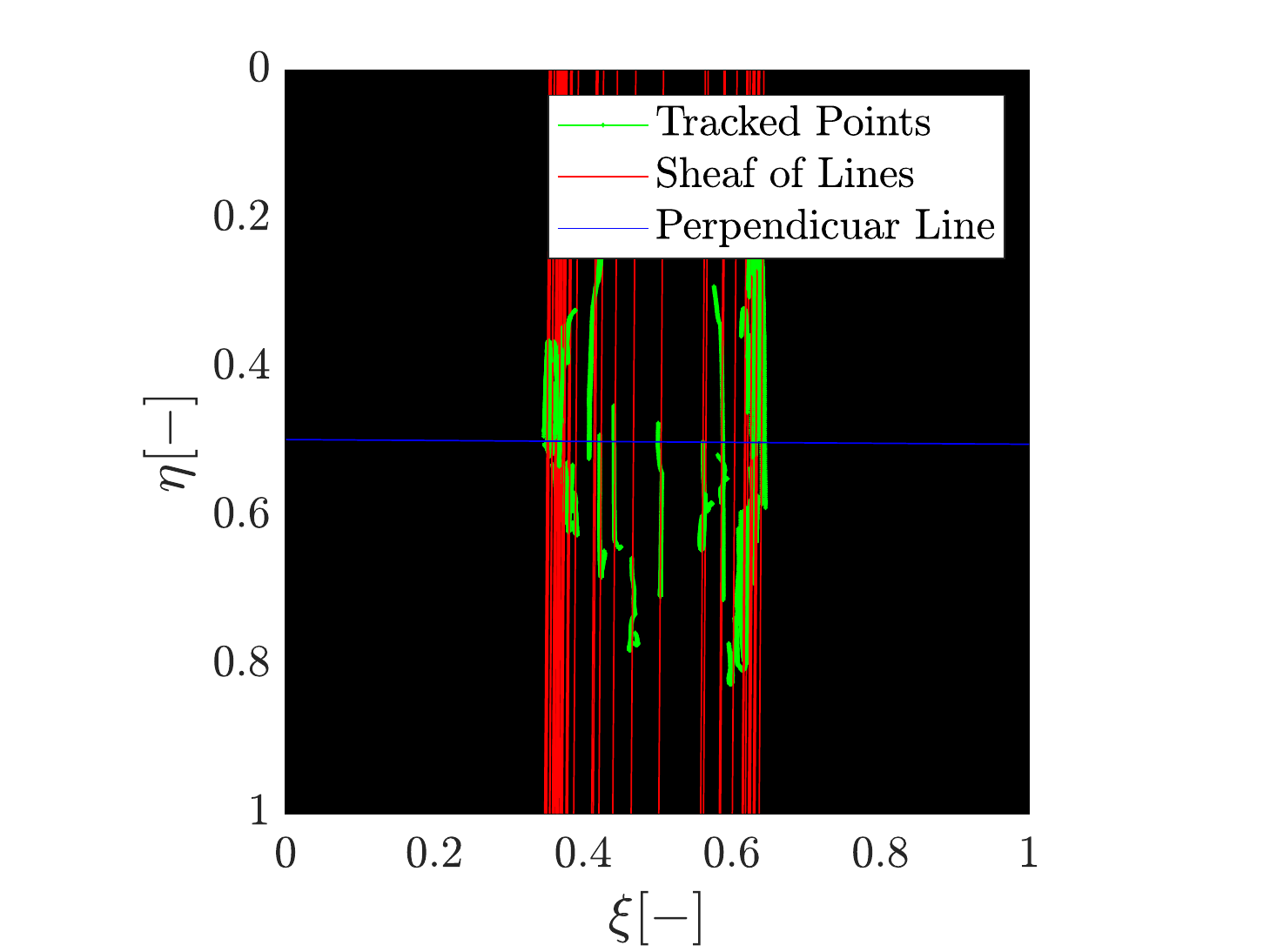}
		\caption{The lines reconstruction}
		\label{fig:perpEllipses}
	\end{subfigure}
	\caption{The perpendicular-axis case reconstruction}
	\label{fig:perpCase}
\end{figure}

As lines are the degenerate projection of the planes where the 3D circles lie, the rotation axis projection on the image is given by the line perpendicular to the sheaf of parallel lines and passing through the rotation center. This geometrical configuration is shown in Fig.~\ref{fig:perpEllipses}.
\begin{figure}[!htb]
	\centering
	\includegraphics[width=0.65\textwidth]{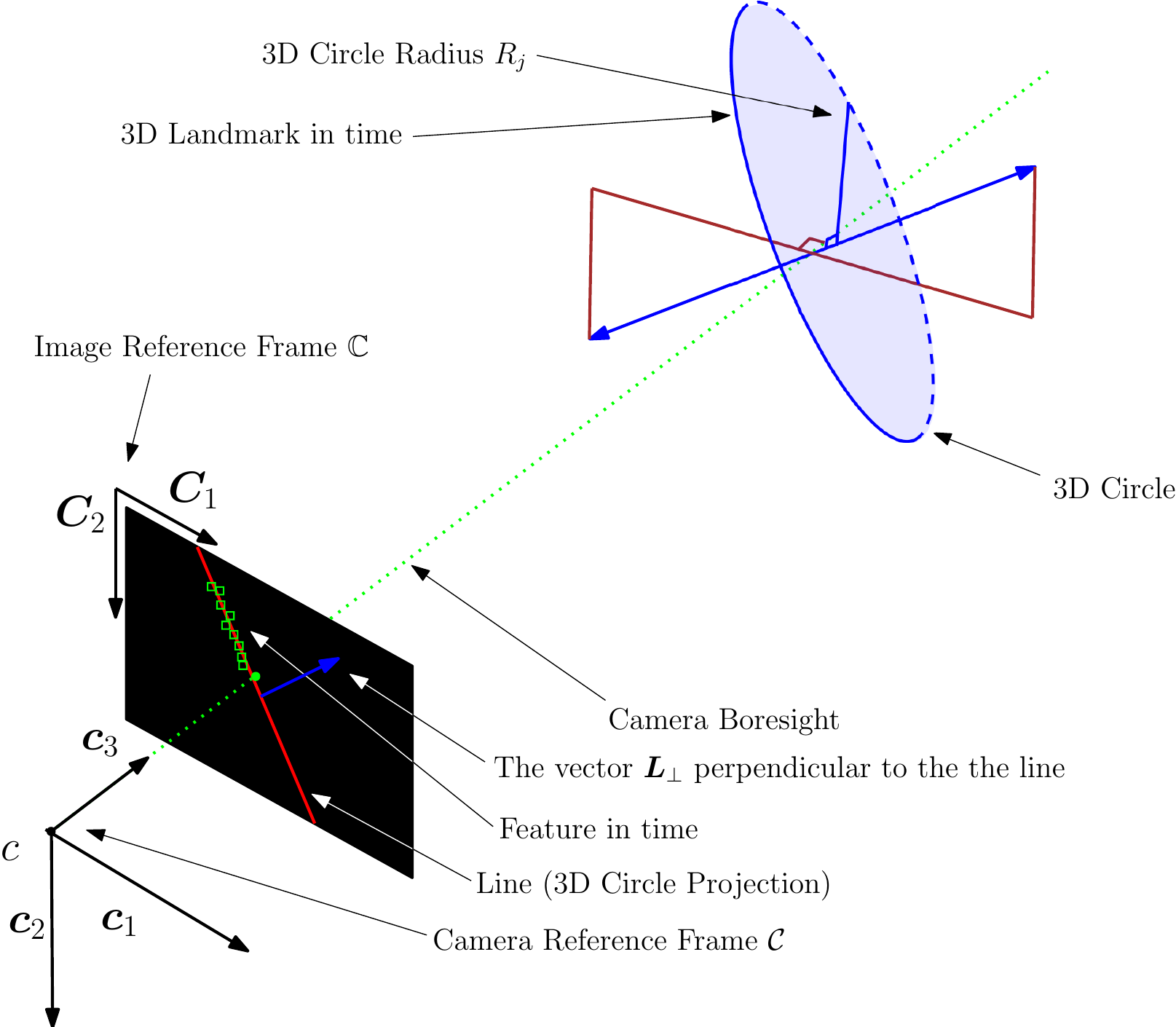}
	\caption{The estimation of the rotation axis for the perpendicular-axis case}
	\label{fig:perpOmega}
\end{figure}
Therefore, the 2D vector perpendicular to the sheaf must be backprojected in the 3D space as shown in Fig.~\ref{fig:perpOmega}. Let the vectors $\bm{L}_{\perp}$ and $\bm{L}_{\parallel}$ be respectively the normal and parallel vectors to the sheaf. These vectors can be computed from the line equation after proper rescaling of the equation in pixel coordinates. Therefore, the two solutions for the rotation axis are:
\begin{equation}
\hat{\bm{\nu}} = \pm\frac{\bm{\nu}}{\left|\left|\bm{\nu}\right|\right|} \quad\text{where}\quad \bm{\nu} = \bm{r}_{\text{SC}_1} \times \left(\left[K\right]^{-1}
{}_h\bm{L}_{\parallel} \right)
\end{equation} 

\subsubsection{Rotation Axis Pruning and Selection}
To exclude the spurious solution, a heuristic similar to the one in the previous section is introduced. If a coarse guess exists, the solution is chosen to be the closest one to the initial guess. Otherwise, it is possible to determine which is the correct solution by determining how features move in the image on average.\newline
Let $\mathcal{L} = \left\{l,\bm{l}_1,\bm{l}_2,\bm{l}_3\right\}$ be a reference frame centered in the image center and with unit vectors $\bm{l}_1= \left(\bm{l}_{\perp},0\right)^T$, $\bm{l}_2=\left(\bm{l}_{\parallel},0\right)^T$, and the perpendicular to define a right-handed reference frame. Let "in front" and "behind" be the subsets of the image defined by an observer looking in the direction of $\bm{l}_{\parallel}$ while standing on the image. If the rotation axis is projected as $\bm{l}_{\perp}$, the majority of the 3D points are moving from "behind" to "in front";  from "in front" to "behind" if projected as $-\bm{l}_{\perp}$.  Indeed, as the majority of the 3D landmarks are initialized between the camera and the rotation center, the feature mean movement respects this motion.  As in the non-degenerate conic case, shorter tracks are considered with a parameter $\gamma_{\text{heur}}=0.2$ to increase the statistical relevance of the heuristic.\newline
The movement vector $\bm{m}_{i,j}$ in the direction of $\bm{l}_2$ of $j$th feature in the $i$th image is computed as 
\begin{equation}
\bm{m}_{i,j} = \left({}_h\bm{\Xi}_{i+1,j} \;-\; {}_h \bm{\Xi}_{i,j}\right)^T\bm{l}_{2}\,\bm{l}_{2}
\end{equation}
Finally, the mean movement direction index $J_{\text{dir}}$ is computed as:
\begin{equation}
J_{\text{dir}}=\sum_{j=1}^{M}\sum_{i=1}^{N_j-1} J_{\text{dir}_{i,j}}\quad\text{where}\quad J_{\text{dir}_{i,j}} = \left\{\begin{aligned}
& 1 \quad \text{if}\quad \bm{l}_3^T\left(
\bm{l}_{1}\times\bm{m}_{i, j}\right)>0\\
-&1 \quad \text{otherwise}
\end{aligned}\right.
\end{equation}
The value of $J_{\text{dir}}$ gives information about whether the landmarks move from "in front" to "behind" or the other way around. If $ J_{\text{dir}}>0$, the rotation axis is projected in the image in the same direction of $\bm{l}_{\perp}$; the opposite otherwise. Thus, the sign of $J_{\text{dir}}$ uniquely defines the rotation axis orientation in the inertial reference frame. 

\section{Numerical Results}\label{sec:results}
\subsection{Test Cases Description}\label{sec:simsetup}
\begin{figure}[!t]
	\centering
	\begin{subfigure}{0.48\textwidth}
		\centering
		\includegraphics[width=\textwidth]{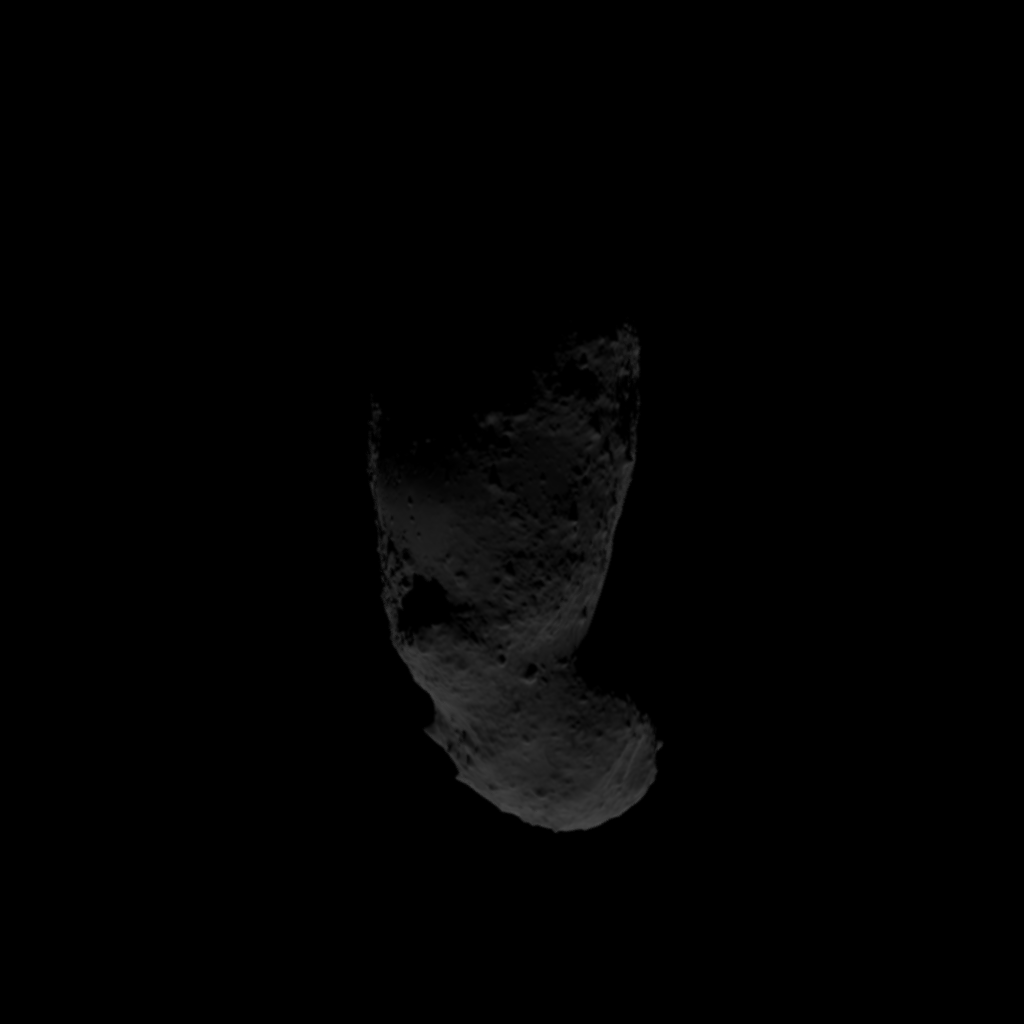}
		\caption{Itokawa with illumination angle of $75^{\circ}$}
		\label{fig:sinthImIto}
	\end{subfigure}
	\begin{subfigure}{0.48\textwidth}
		\centering
		\includegraphics[width=\textwidth]{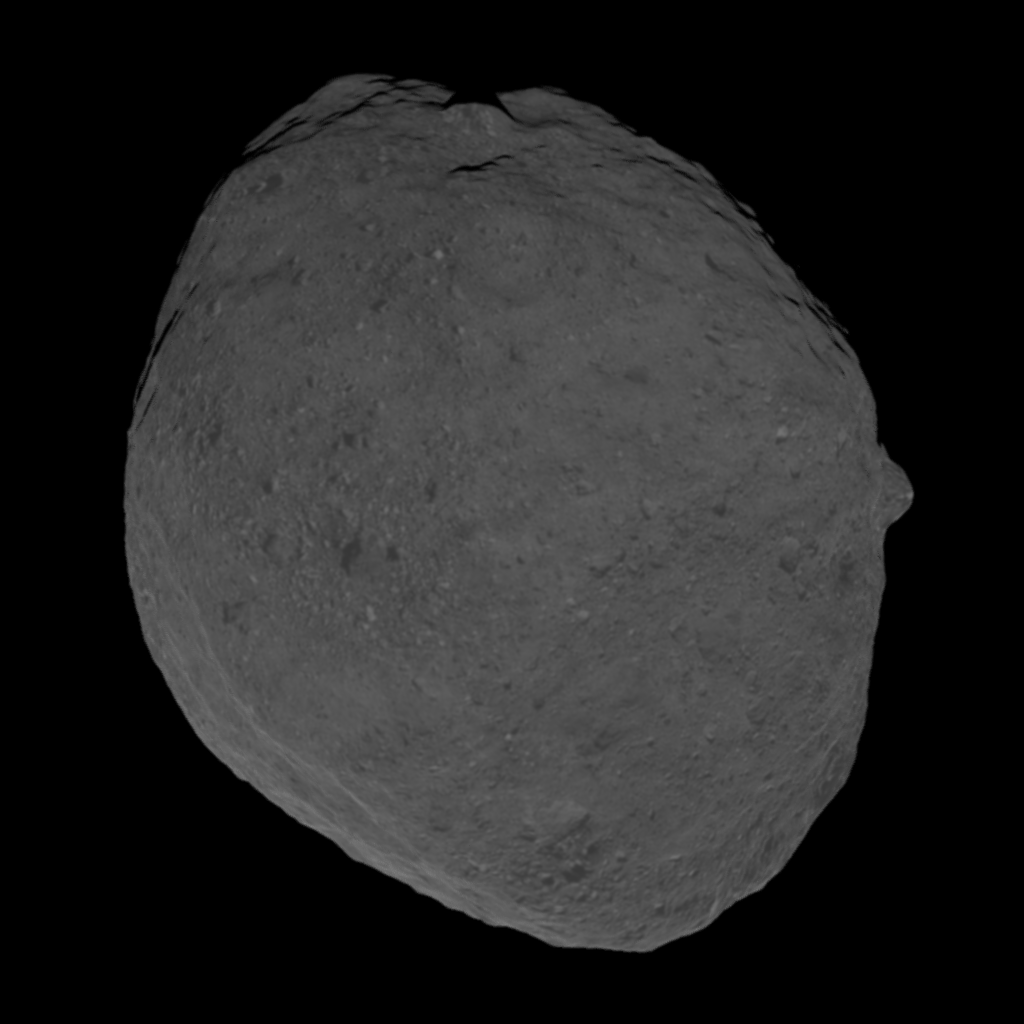}
		\caption{Bennu with illumination angle of $10^{\circ}$}
		\label{fig:sinthImBennu}
	\end{subfigure}
	\caption{Examples of synthetic images for Bennu and Itokawa with different illumination angles} 
\end{figure}

\begin{table}[!b]
	\centering
	\caption{Numerical values for the camera parameters.}
	\label{tab:numValCamera}
	\begin{threeparttable}
		\begin{tabular}{c|ccccc}
			Camera& FoV \tnote{1} $\;\left[^{\circ}\right]$ & $f$ \tnote{2} $\;\left[\text{mm}\right]$ & $s_x=s_y$ \tnote{3} $\;\left[\mu\text{m}\right]$ & Image Size [px]& $\bm{C}$ \tnote{4} $\;\left[\text{px}\right]$ \B \\
			\hline
			OSIRIS-REx's MapCam & 3.99 & 125 & 8.5 & $1024\times1024$ & $\left(512,\,512\right)^T$ \T\\
			Hayabusa's AMICA & 5.8 & 121 & 12 & $1024\times1024$ & $\left(512,\,512\right)^T$ \T\\
		\end{tabular}
		\begin{tablenotes}
			\item[1] FoV is the camera field of view
			\item[2] $f$ is the camera focal length
			\item[3] $s_x$ and $s_y$ are the pixel physical size in $x$ and $y$ components
			\item[4] $\bm{C}$ is the camera center
		\end{tablenotes}
	\end{threeparttable}
\end{table}
To test the proposed algorithm, numerical simulations have been performed. In the simulations, the spacecraft is in close approach to the small body. The considered scenarios foresee a spacecraft trajectory to simulate the approch to asteroids Bennu or Itokawa. The spacecraft starts observing the small body from about 9 km (i.e., about 60 asteroid radii for Itokawa and 36 asteroid radii for Bennu), implying that the asteroid gravity is negligible. The spacecraft velocity points towards the asteroid and its magnitude are selected to be consistent with approach velocity from SPICE kernels. As the probe is not under the influence of the small body gravity field, the trajectory has been simulated with a dynamics dominated by Solar Radiation Pressure (SRP) in the Sun-small-body rotating frame. Note that the spacecraft is orbiting the Sun but the swept arc of the conic is so short that it can be approximated by a straight line, so as to ignore the Sun gravity in the equations of motion.
The small body rotates around an inertially-fixed axis with constant angular velocity. The spacecraft orientation in the inertial frame $\mathcal{N}$ is considered known and the camera points to the small body center of mass. More details about the dynamical model and the simulation set-up can be found in Section 3.6 of \citet{panicucci2021autonomous}.\newline The mapping camera, like Hayabusa's AMICA \cite{ishiguro2010hayabusa} or OSIRIS-REx's MapCam \cite{rizk2018ocams} (see Tab~\ref{tab:numValCamera}), is constantly observing the small body. \EPP{The two cameras are selected among the OSIRIS-Rex and Hayabusa's instruments because their characteristics ensure to observe respectively Bennu and Itokawa letting the asteroid occupy hundreds of pixels at the approach distance.} Images are simulated with the SurRender software \cite{lebreton2021image}. Small-body shapes,  Bennu and  Itokawa in this study, are taken from the Planetary Data System - Small Body Node by downloading the highest resolution polyhedral shape model. The shape model is input to SurRender toghter with the small-body bulk optical properties, like albedo, reflectivity, and diffusivity, and assuming the Hapke BRDF (Bidirectional Reflectance Distribution Function) \cite{hapke2012theory} to obtain high-fidelity simulated images. A texture is added to the small body to take into account the small-body surface albedo. The camera optics is represented with a Gaussian Point Spread Function (PSF) and a pin-hole projection model. The KLT tracker Themis processes images every 60 seconds for Bennu and 180 seconds for Itokawa to account for realistic image processing operational constrains.
Numerical values, including 3D meshes, are taken from operational missions scenario  \cite{rizk2018ocams,ishiguro2010hayabusa,demura2006pole,barnouin2019shape,lauretta2019unexpected}. Note that in this study no real images are considered because of unavailability of finding frequent and successive images to provide to the KLT algorithm.\newline
Two main test cases are presented hereafter: Bennu and Itokawa. To study the performance of the proposed algorithm the approach direction is kept constant and 3 different illumination angles are considered: $10^{\circ}$, $45^{\circ}$, and $75^{\circ}$. After having defined the Sun-small-body and the approach directions, the rotation axis must be defined to fully determine the observational geometry during the approach (see Fig.~\ref{fig:RotEst_anglesapproach}). To characterize the algorithm performance, the rotation axis is varied over the entire celestial sphere by changing the right ascension $\alpha$ and the declination $\delta$ with respect to the approach direction. Note that $\alpha = 0^{\circ}$  and $\delta = 0^{\circ}$ means that the rotation axis is pointing towards the spacecraft during the approach. Moreover, all cases with $\alpha = 90^{\circ}$ or $\delta = 90^{\circ}$ imply that the rotation axis is perpendicular to the approach direction. The discretization of the $\left(\alpha,\,\delta\right)$ grid is $30^{\circ}$ as each point of the grid implies the rendering of 250 images. By considering that only one case is simulated for $\delta =\pm 90$ as $\alpha$ is not defined and that $\delta=\pm 180$ are the same orientation for all $\alpha$, 134 simulations are performed per illumination angle and asteroid leading to 804 simulations and more than 200,000 images. This rendering and simulation effort has been performed to push towards the validation and the performance assessment of the algorithm when illumination changes, shape varies, or the rotation axis orientation is unforeseeable. Examples of synthetic images for both asteroids with different illumination angles are shown in Fig.~\ref{fig:sinthImIto} and \ref{fig:sinthImBennu} for the sake of completeness.

\subsection{Origin Estimation}
\begin{figure}[t]
	\centering
	\begin{subfigure}[t]{0.47\textwidth}
		\includegraphics[width=\textwidth,trim={0cm 0cm 0cm 0cm},clip]{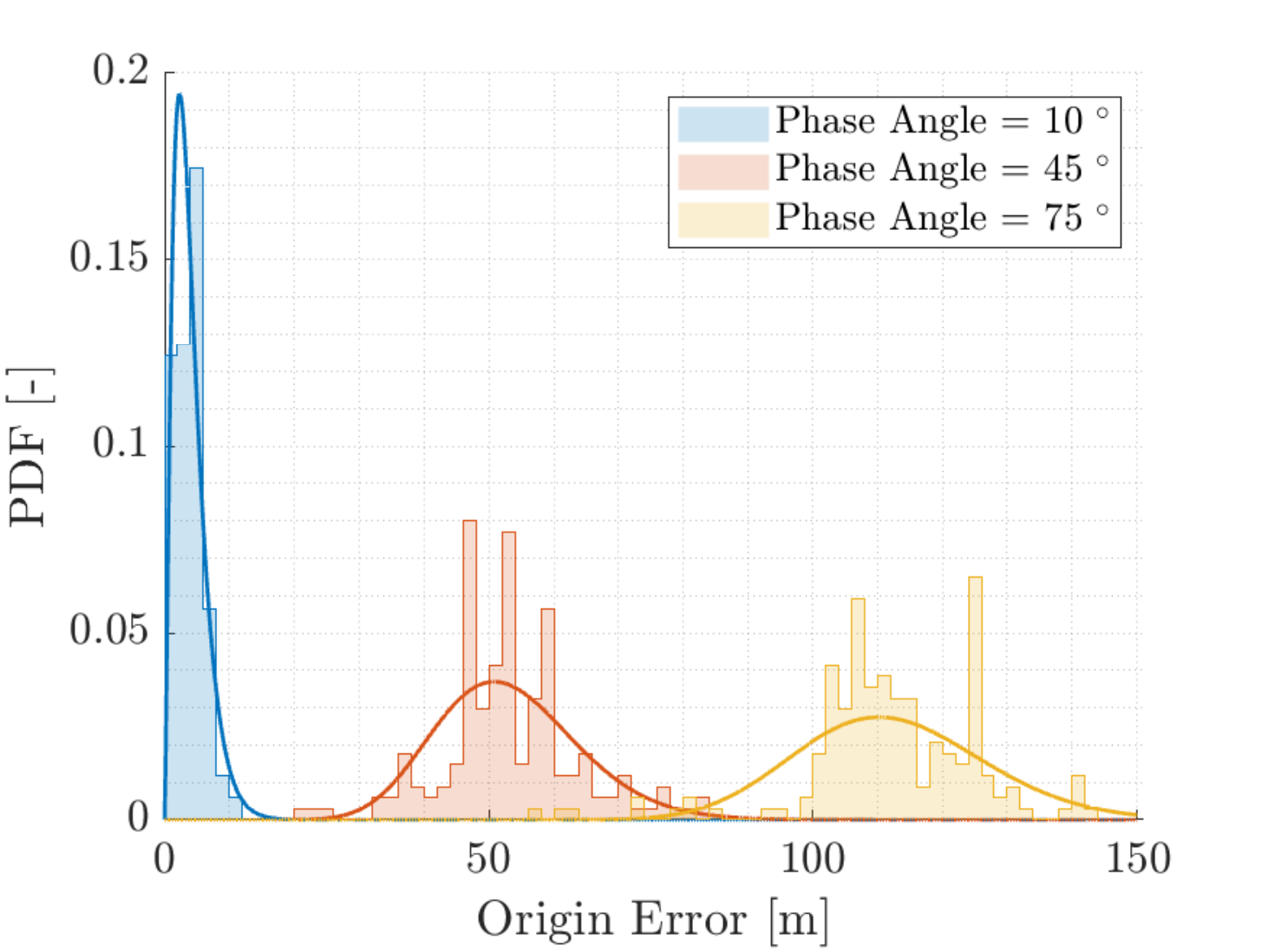}
		\caption{Bennu}
		\label{fig:PDFOrigin_Bennu_Initilization}
	\end{subfigure}
	\hspace{0.5cm}
	\begin{subfigure}[t]{0.47
			\textwidth}
		\centering
		\includegraphics[width=\textwidth,trim={0cm 0cm 0cm 0cm},clip]{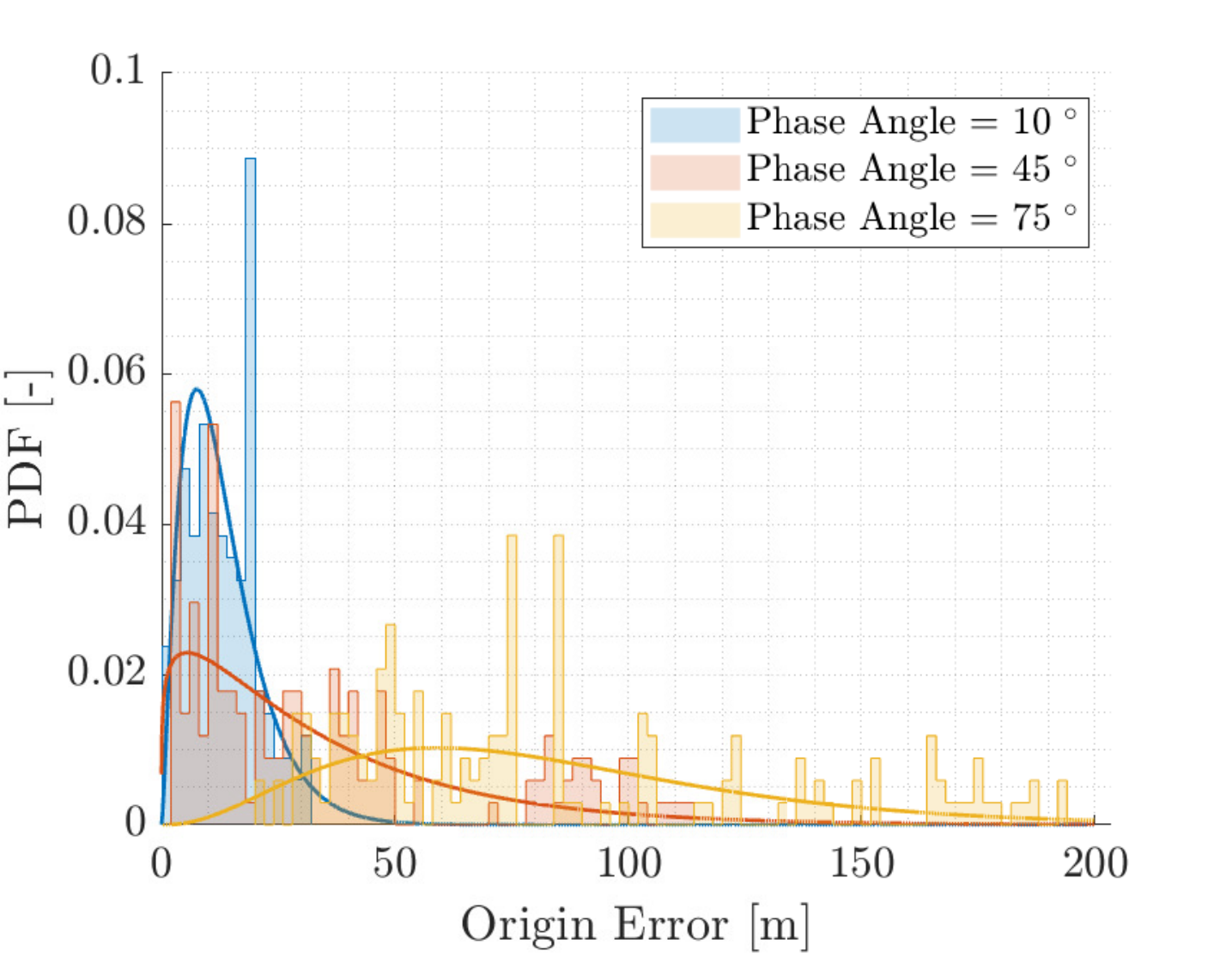}
		\caption{Itokawa}
		\label{fig:PDFOrigin_Itokawa_Initilization}
	\end{subfigure}
	\caption{Probability Density Function (PDF) of the origin estimation error. The distributions in the figures are the best-fit Gamma distribution}
\end{figure}

The first analyzed results deal with the small-body-fixed reference frame estimation. To assess the performance in estimating the origin, the origin error is defined as the norm between the estimated origin and the true one.\newline
Note that the origin error is mainly influenced by the center of britghtness computation, as the distance to the small body is considered know (see Sec.~\ref{sec:probstat}). The probability density functions (PDFs) of the origin error are reported in Fig.~\ref{fig:PDFOrigin_Bennu_Initilization} and \ref{fig:PDFOrigin_Itokawa_Initilization} for different values of the illumination angle. Recall that Bennu and Itokawa have a best-fitting-ellipse semiaxes of 252 m $\times$ 246 m $\times$ 228.3 m and 267.5 m $\times$ 147 m $\times$ 104.5 m, respectively.  As known from the literature \cite{pugliatti2022data, bhaskaran1998autonomous, wright2018optical}, the center of brightness deviates from its true value when the illumination angle increases. Thus its backprojection deviates from the true origin accordingly. For very high illumination angle the performance degrades  and the origin is estimated to be far from the true value. This is even more important for an highly concave and oblate body as Itokawa. It is worth noting that this error can influence the rotation axis estimation as the origin determination is necessary for the determination of the 3D circle parameter. Despite this behavior, the rotation axis estimation shows good performance for the majority of the simulated scenarios, implying that the origin estimation play a minor role. For the sake of completeness, Tab.~\ref{tab:meanVarOriginError} shows the mean and the standard deviation of the best Gamma distribution fitting the origin estimation error data. The Gamma distribution is preferred with respect to other distribution as the PDF support is limited to the positive real numbers. The means and the standard deviations confirm the previous interpretation of the error. The performance degrades when increasing the illumination angle as shown by the errors means, but this effect is more relevant for the Itokawa test-case where the variability of the results due to selfshadowing implies higher standard deviations with respect to the Bennu test-case. As stated in Sec.~\ref{sec:origin}, algorithms were designed in the past to compensate for this effect implying a possible reduction of the origin error and its influence on the rotation axis estimation.

\begin{table}[t]
	\centering%
	\caption{Mean and standard deviation of the best-fit Gamma distribution of the origin estimation error.}
	\label{tab:meanVarOriginError}
		\begin{tabular}{c|ccc|ccc}
		Test-case & \multicolumn{3}{c|}{Bennu} & \multicolumn{3}{c}{Itokawa}\\
		\hline
		Illumination Angle $\left[{}^{\circ}\right]$ & 10 & 45 & 75 & 10 & 45 & 75 \\
		\hline
		Mean [m] &  6.38 &  121.05 & 214.22  &  13.12 & 34.07  &  82.98 \\
		Variance [m] &  3.93 &  53.09 & 112.18  &  73.05 & 975.18  & 1998.88 
	\end{tabular}
\end{table}

\subsection{Detection of Degenerate Solutions}\label{sec:detectionresults}
\begin{figure}[p]
	\centering
	\begin{minipage}{1cm}
		\begin{tabular}{c m{2.4cm}}
			\colorbox{PerpendicularTrueDetection}{\hspace{\lineskip}} & Perpendicular\par Axis True\par Detection\\
			&\\
			\colorbox{TiltedTrueDetection}{\hspace{\lineskip}}& Tilted Axis\par True Detection\\
			&\\
			\colorbox{PerpendicularFalseDetection}{\hspace{\lineskip}}
			& Perpendicular\par Axis False\par Detection\\
			&\\
			\colorbox{TiltedFalseDetection}{\hspace{\lineskip}}& Tilted Axis\par False Detection
		\end{tabular}
	\end{minipage}
	\hfill
	\begin{minipage}{14cm}
		\centering
		\begin{subfigure}[t]{0.26\paperwidth}
			
			\includegraphics[width=\textwidth]{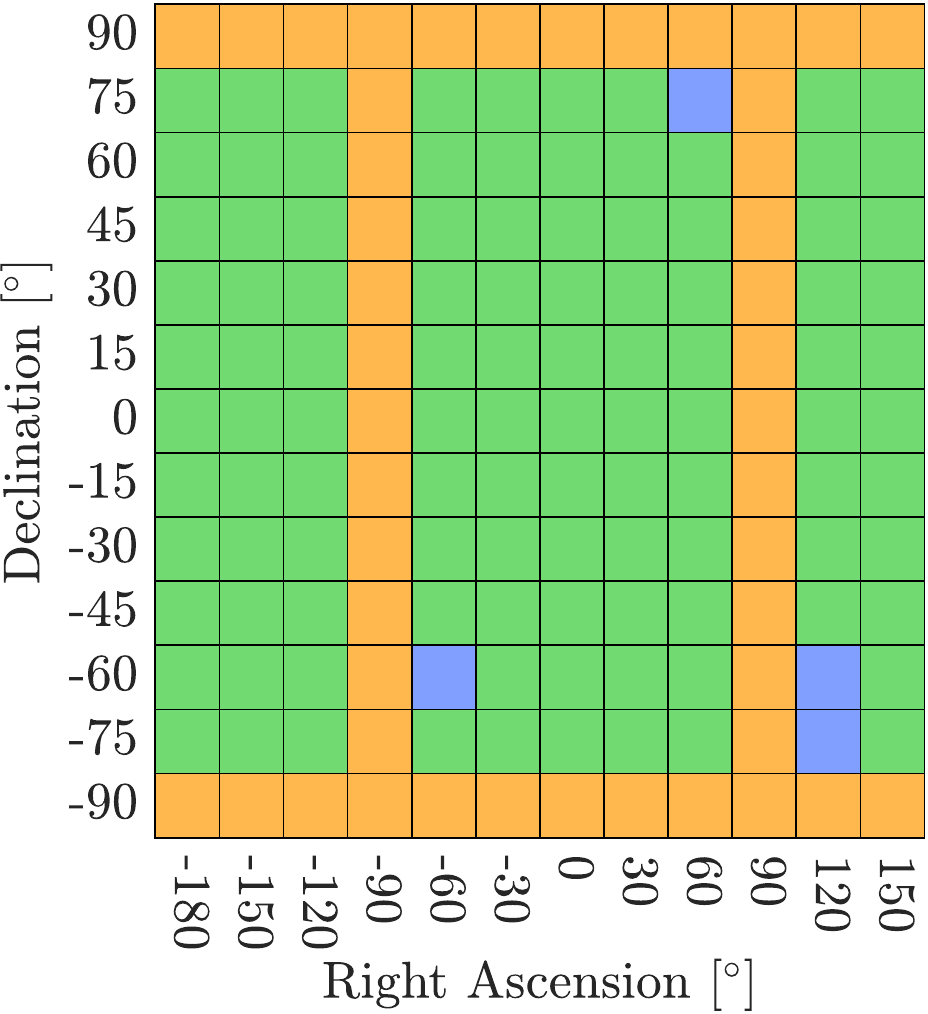}
			\caption{Bennu with illumination angle of $10^{\circ}$}
			\label{fig:Detection_Bennu_PA10_OnboardGuess}
		\end{subfigure}
		\hspace{1cm}
		\begin{subfigure}[t]{0.26\paperwidth}
			\centering
			\includegraphics[width=\textwidth]{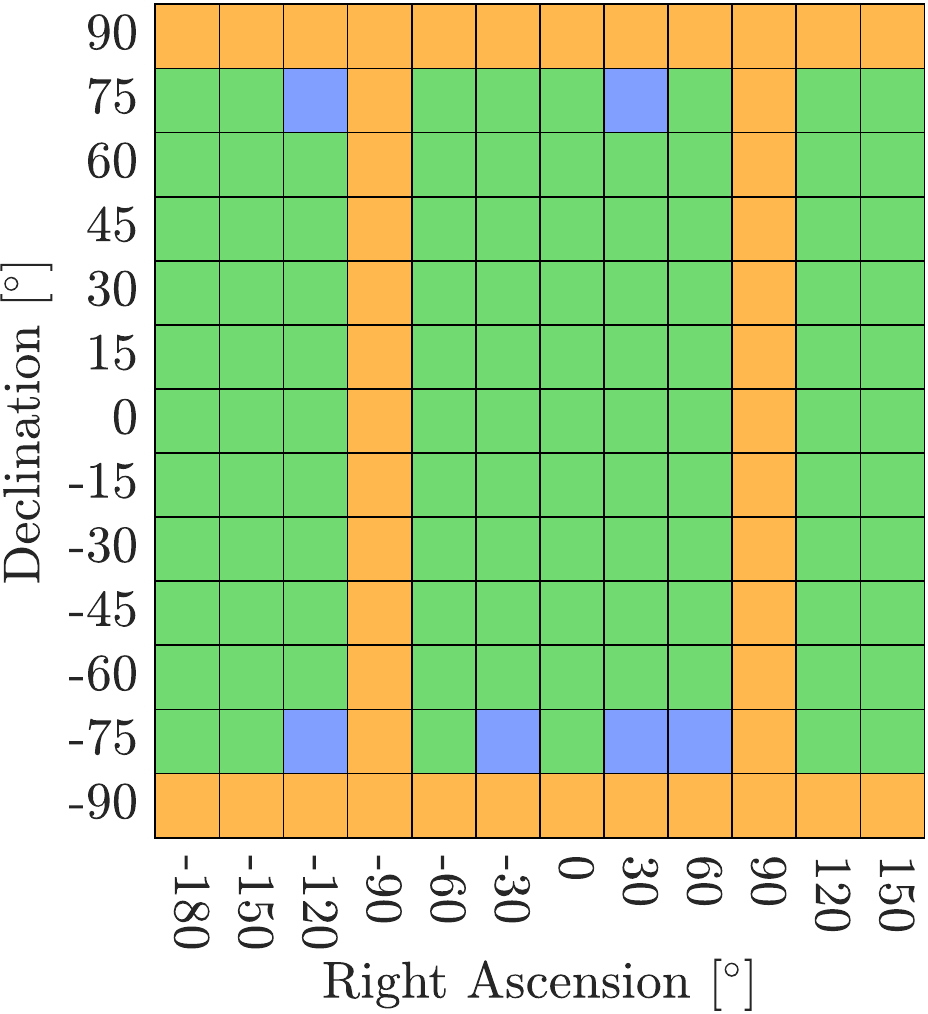}
			\caption{Itokawa with illumination angle of $10^{\circ}$}
			\label{fig:Detection_Itokawa_PA10_OnboardGuess}
		\end{subfigure}\\[0.3cm]
		\begin{subfigure}[t]{0.26\paperwidth}
			\centering
			\includegraphics[width=\textwidth,trim={0cm 0cm 0cm 0cm},clip]{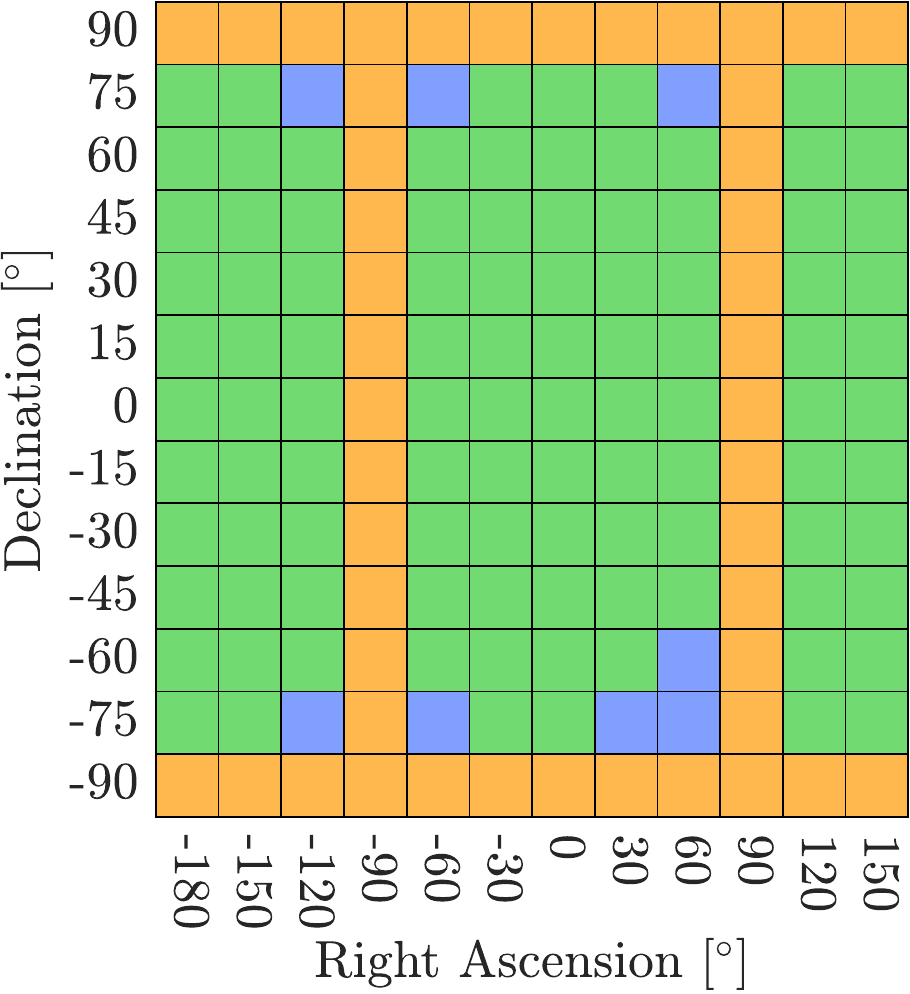}
			\caption{Bennu with illumination angle of $45^{\circ}$}
			\label{fig:Detection_Bennu_PA45_OnboardGuess}
		\end{subfigure}
		\hspace{1cm}
		\begin{subfigure}[t]{0.26\paperwidth}
			\centering
			\includegraphics[width=\textwidth]{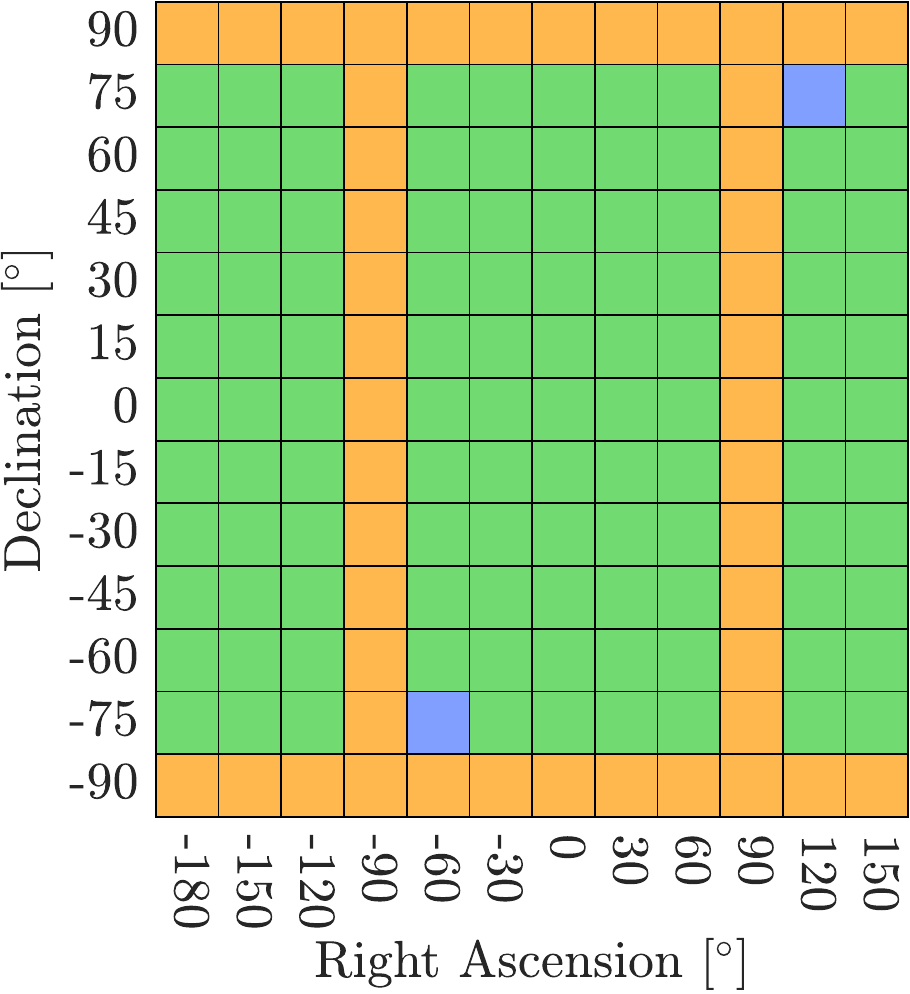}
			\caption{Itokawa with illumination angle of $45^{\circ}$}
			\label{fig:Detection_Itokawa_PA45_OnboardGuess}
		\end{subfigure}\\[0.3cm]
		\begin{subfigure}[t]{0.26\paperwidth}
			\centering
			\includegraphics[width=\textwidth]{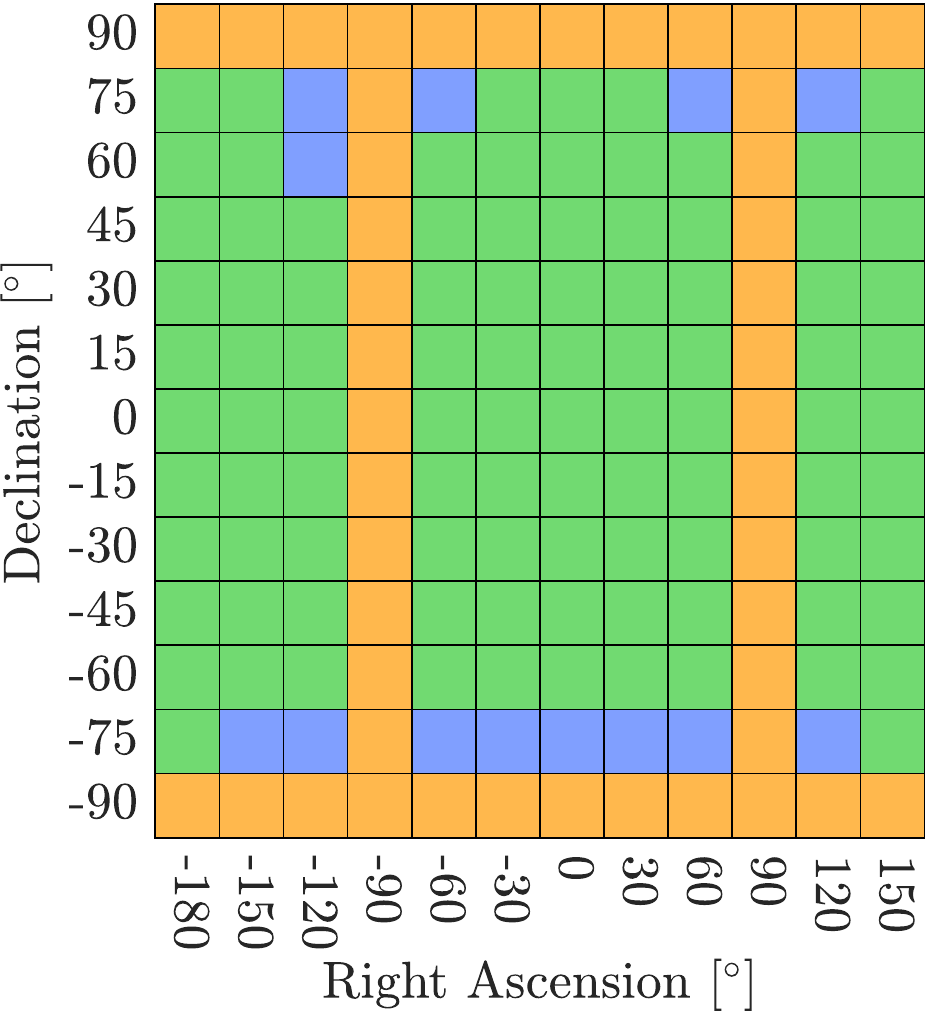}
			\caption{Bennu with illumination angle of $75^{\circ}$}
			\label{fig:Detection_Bennu_PA75_OnboardGuess}
		\end{subfigure}
		\hspace{1cm}
		\begin{subfigure}[t]{0.26\paperwidth}
			\centering
			\includegraphics[width=\textwidth]{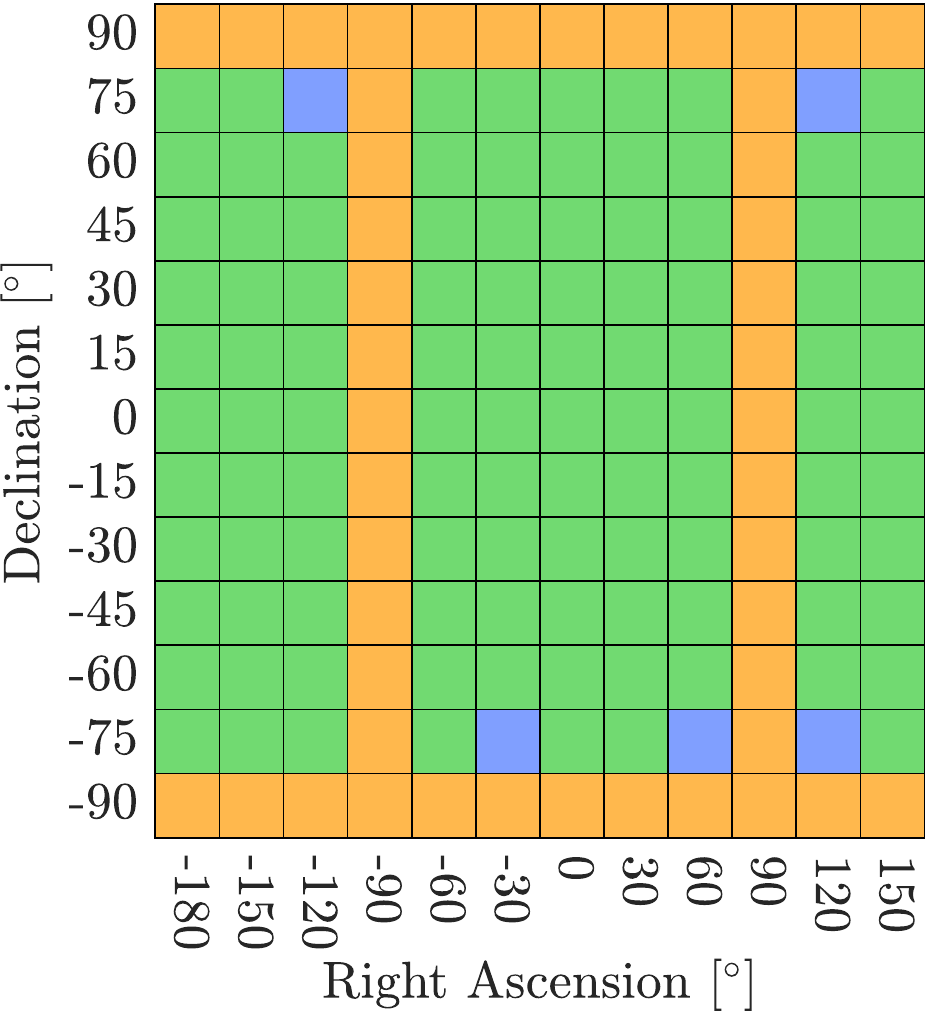}
			\caption{Itokawa with illumination angle of $75^{\circ}$}
			\label{fig:Detection_Itokawa_PA75_OnboardGuess}
		\end{subfigure}	
	\end{minipage}
	\caption{Degenerate solutions detection performance varying the rotation axis orientation with respect to the camera boresight}
\end{figure}

In this section, the results of the degenerate solutions detection are presented. To evaluate the performance of the detection, note that two classes are present: "Tilted Rotational Axis" and "Perpendicular Rotational Axis". For each class it is possible to define precision $Pr$, recall $Re$, and the F-score $F_1$:
\begin{equation}
	Pr = \frac{TP}{TP + FP}
\end{equation}
\begin{equation}
	Re = \frac{TP}{TP + FN}
\end{equation}
\begin{equation}
	F_1 = \frac{2 Pr\, Re}{Re + Pr}
\end{equation}
where $TP$ is the true positive number, $FP$ is the false positive number, and $FN$ is the false negative number. These metrics are selected as precision $Pr$ provides information about how many detections within a class are correct, recall $Re$ states how many samples are correctly detected in the considered class, and the F-score $F_1$ is a mix of precision and recall which is high when both are high. For the two considered classes, the considered metrics are reported in Tab.~\ref{tab:metricsDetection}. The precision of the "Tilted Rotation Axis" class is 1 which implies that all the solutions labeled within the class belong to this class. Its recall is always above $0.85$ which implies that more than $85\%$ of the tilted axis cases are identified correctly.  Fig.~\ref{fig:Detection_Bennu_PA10_OnboardGuess} - \ref{fig:Detection_Itokawa_PA75_OnboardGuess} show the detection results by varying the right ascension and declination. It is worth noitng that no false detection of the tilted rotation axis is present. Finally, the F-score for the "Tilted Rotation Axis" is high implying that a good balance between precision and recall is achieved.
\begin{table}[t]
	\centering
	\caption{Metrics for the Degenerate Solution Detection.}
	\label{tab:metricsDetection}
	\begin{tabular}{c|ccc|ccc}
		Test-case & \multicolumn{3}{c|}{Bennu} & \multicolumn{3}{c}{Itokawa}\\
		\hline
		Illumination Angle $\left[{}^{\circ}\right]$ & 10 & 45 & 75 & 10 & 45 & 75 \\
		\hline
		Tilted Rotation Axis Precision &  1 &  1 & 1  &  1 & 1  &  1 \\
		Tilted Rotation Axis Recall &  0.9636 &  0.9272 & 0.8818  &  0.9454 & 0.9818  & 0.9545  \\
		Tilted Rotation Axis F-score &  0.9814 & 0.9622  & 0.9372  & 0.9719 &  0.9908 &  0.9767 \\
		\hline
		Perpendicular Rotation Axis Precision &  0.92 & 0.8518  & 0.7796 & 0.8846 &  0.9583 &  0.9019 \\
		Perpendicular Rotation Axis Recall & 1  & 1  &  1 &  1 & 1  & 1  \\	
		Perpendicular Rotation Axis F-score & 0.9583  & 0.92  & 0.876  & 0.9387  & 0.9787  & 0.9484  \\
	\end{tabular}
\end{table}
\newline
Regarding the "Perpendicular Rotation Axis" class, its recall is 1 which means that all the perpendicular rotation axis are correctly identified. The precision is lower implying that some of the found solutions do not belong to the class. By looking at Fig.~\ref{fig:Detection_Bennu_PA10_OnboardGuess} - \ref{fig:Detection_Itokawa_PA75_OnboardGuess}, all the false detections are close to the perpendicular case in the right-ascension-declination plane. This is because tracks are almost linear and the algorithm classifies them as lines. This problem could be avoided by increasing the conditioning number parameter $\gamma_{\text{CN}}$, but this could cause the presence of false detection of perpendicular rotation axis as tracks could be detected as ellipses due to tracking errors. Otherwise, $\gamma_{\text{sheaf}}$ could be increased to have a more stringent selection .when labeling a curve as a line. Nevertheless, the proposed numerical values show good performance in terms of precision and recall, implying that the detection of the degenerate solution is performed correctly. The F-score is high also for this class which implies a good balance between precision and recall.\newline
Note that results are similar for both asteroids meaning that performance is similar despite the different onboard cameras. It is worth noting that the Bennu test case has more false detection of the perpendicular axis which is probably due to the low concavity of the asteroid (see Fig.~\ref{fig:Detection_Bennu_PA10_OnboardGuess} - \ref{fig:Detection_Itokawa_PA75_OnboardGuess} and Tab.~\ref{tab:metricsDetection}). Indeed, when the body is convex, it is harder to detect tracks that deviate from a line for high approach angles. Moreover, the performance is not affected by the illumination angle variation for the Itokawa test case. This is not true for the Bennu test case where the performance metrics decrease with the illumination angle increase. This is mainly due to the difference in shadowing. The performance of the Itokawa test case is mainly influenced by self-shadowing, whereas the tracking performance is governed by the terminator line shadows for the Bennu test case. As self-shadowing is present even for low illumination angles, the tracks have generally the same length for all the simulations in the Itokawa test case. This is not happening in the Bennu test case where features are initiated close to the terminator line and are tracked for more time when the illumination angle is low.

\subsection{Rotation Axis Estimation}
The last result to be analyzed is the performance of the rotation axis estimation. As explained in Sec.~\ref{sec:simsetup}, the rotation axis is varied by defining a spaced grid in the right-ascension-declination plane to cover all the geometrical configurations that the probe could encounter during the approach. To assess the algorithm performance, the angular error with respect to the true rotation axis is used as the performance index. Moreover, two different simulations are reported hereafter in accord with the two heuristics proposed in Sec.~\ref{sec:rotstate}: the final selection based on onboard guess or the one based on the point initialization.
\begin{figure}[!b]
	\centering
	\begin{subfigure}[t]{0.47\textwidth}
		\includegraphics[width=\textwidth,trim={0cm 0cm 0cm 0cm},clip]{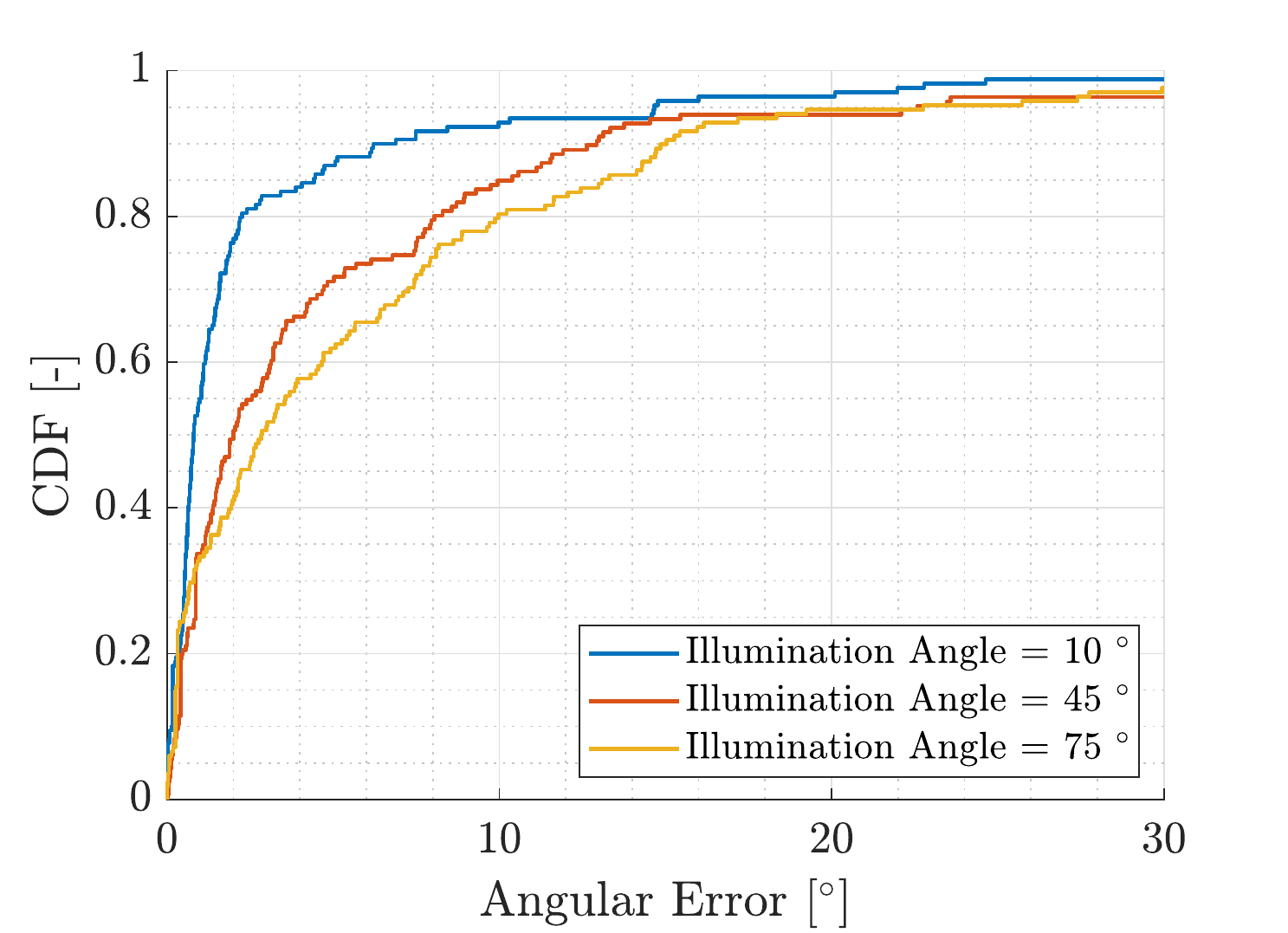}
		\caption{Bennu}
		\label{fig:CDFAxis_Bennu_OnboardGuess}
	\end{subfigure}
	\hspace{0.5cm}
	\begin{subfigure}[t]{0.47\textwidth}
		\centering
		\includegraphics[width=\textwidth,trim={0cm 0cm 0cm 0cm},clip]{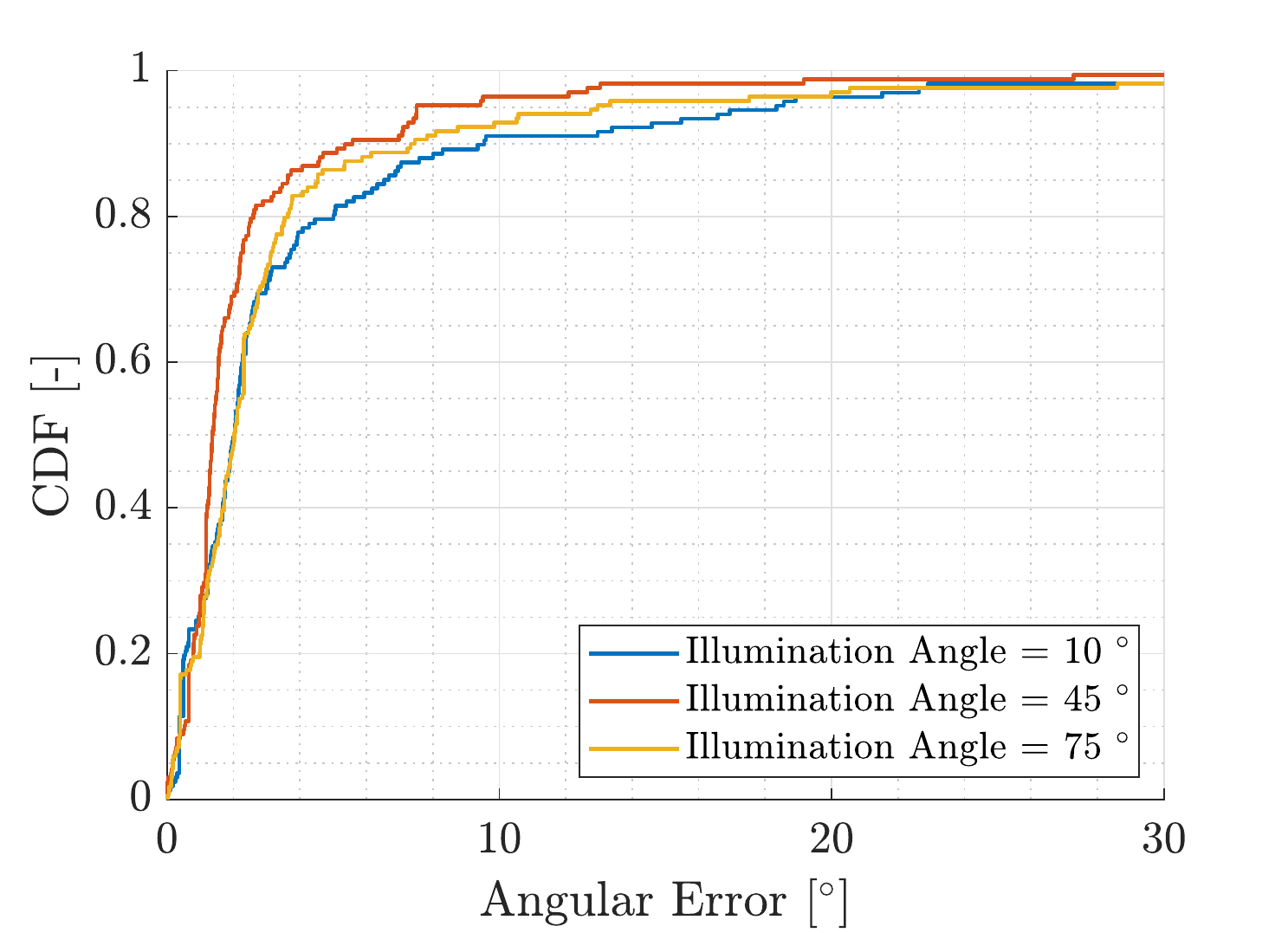}
		\caption{Itokawa}
		\label{fig:CDFAxis_Itokawa_OnboardGuess}
	\end{subfigure}
	\caption{Zoom of the CDFs of the rotation axis estimation error with the heuristic based on onboard guess.}
\end{figure}
\begin{figure}[p]
	\centering
	\begin{subfigure}[t]{0.45\textwidth}
		\includegraphics[width=\textwidth,trim={1cm 0cm 1cm 0cm},clip]{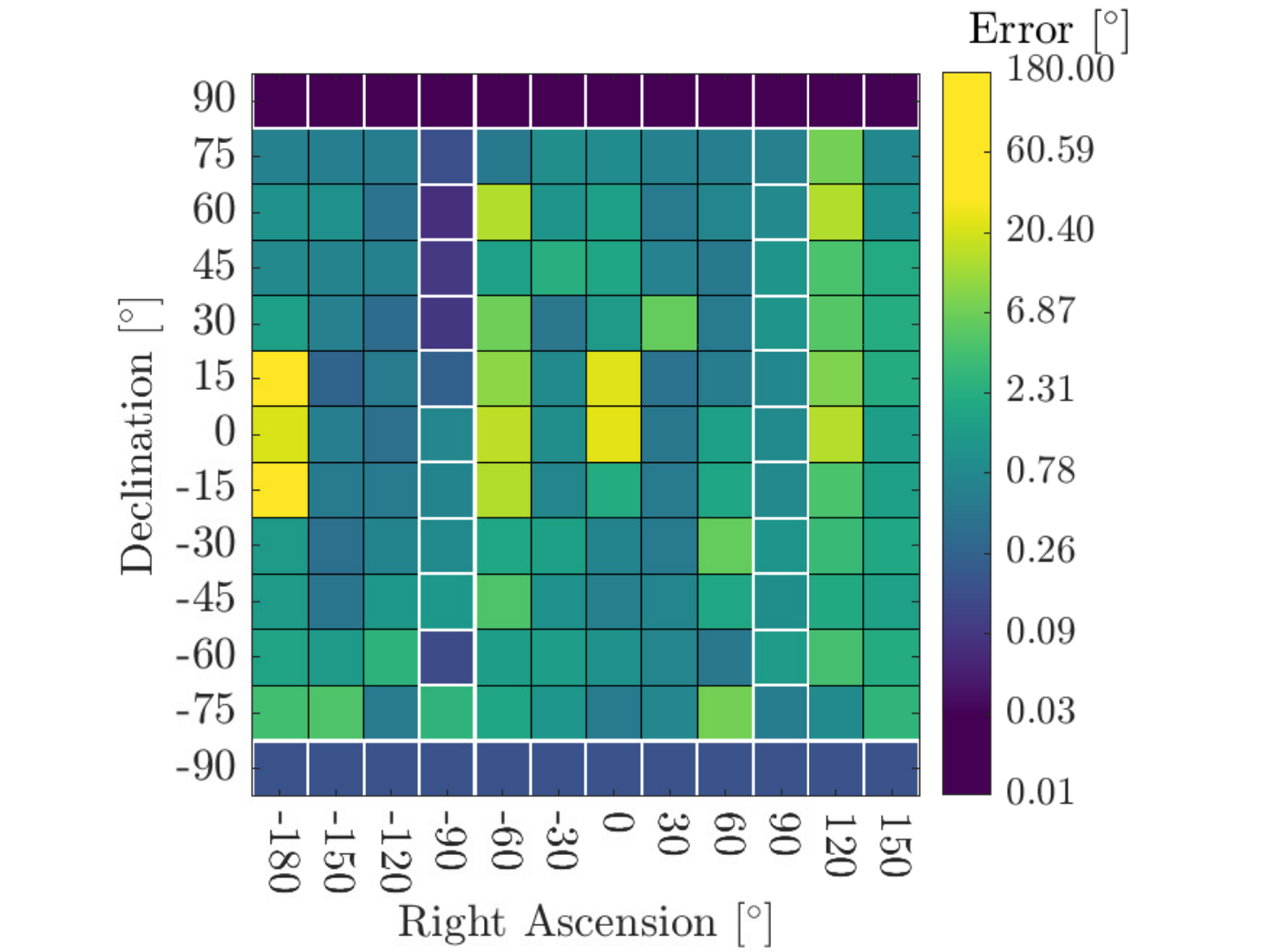}
		\caption{Bennu with illumination angle of $10^{\circ}$}
		\label{fig:AngularError_Bennu_PA10_OnboardGuess}
	\end{subfigure}
	\hspace{0.5cm}
	\begin{subfigure}[t]{0.45\textwidth}
		\centering
		\includegraphics[width=\textwidth,trim={1cm 0cm 1cm 0cm},clip]{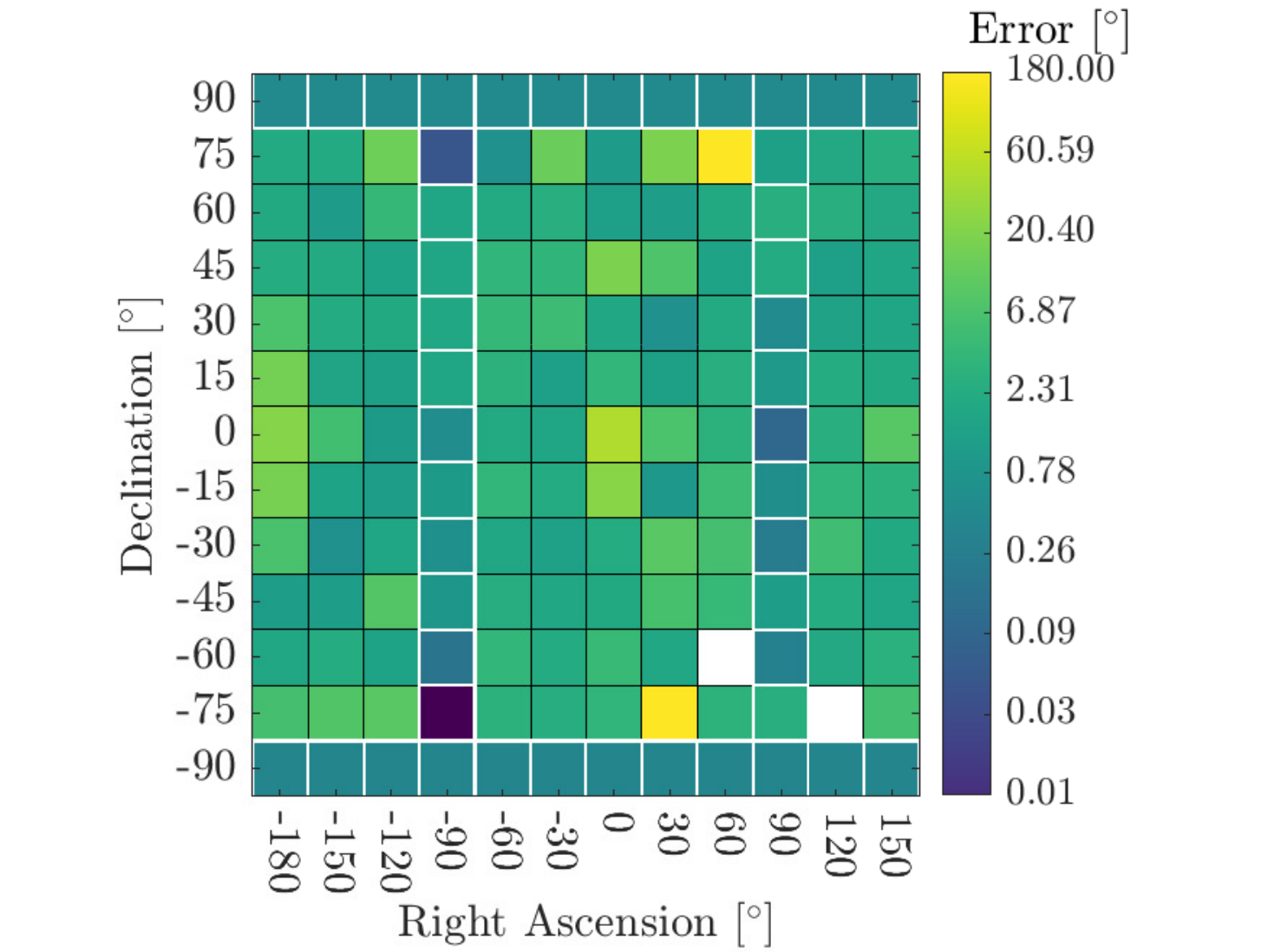}
		\caption{Itokawa with illumination angle of $10^{\circ}$}
		\label{fig:AngularError_Itokawa_PA10_OnboardGuess}
	\end{subfigure}\\[0.3cm]
	\begin{subfigure}[t]{0.45\textwidth}
		\centering
		\includegraphics[width=\textwidth,trim={1cm 0cm 1cm 0cm},clip]{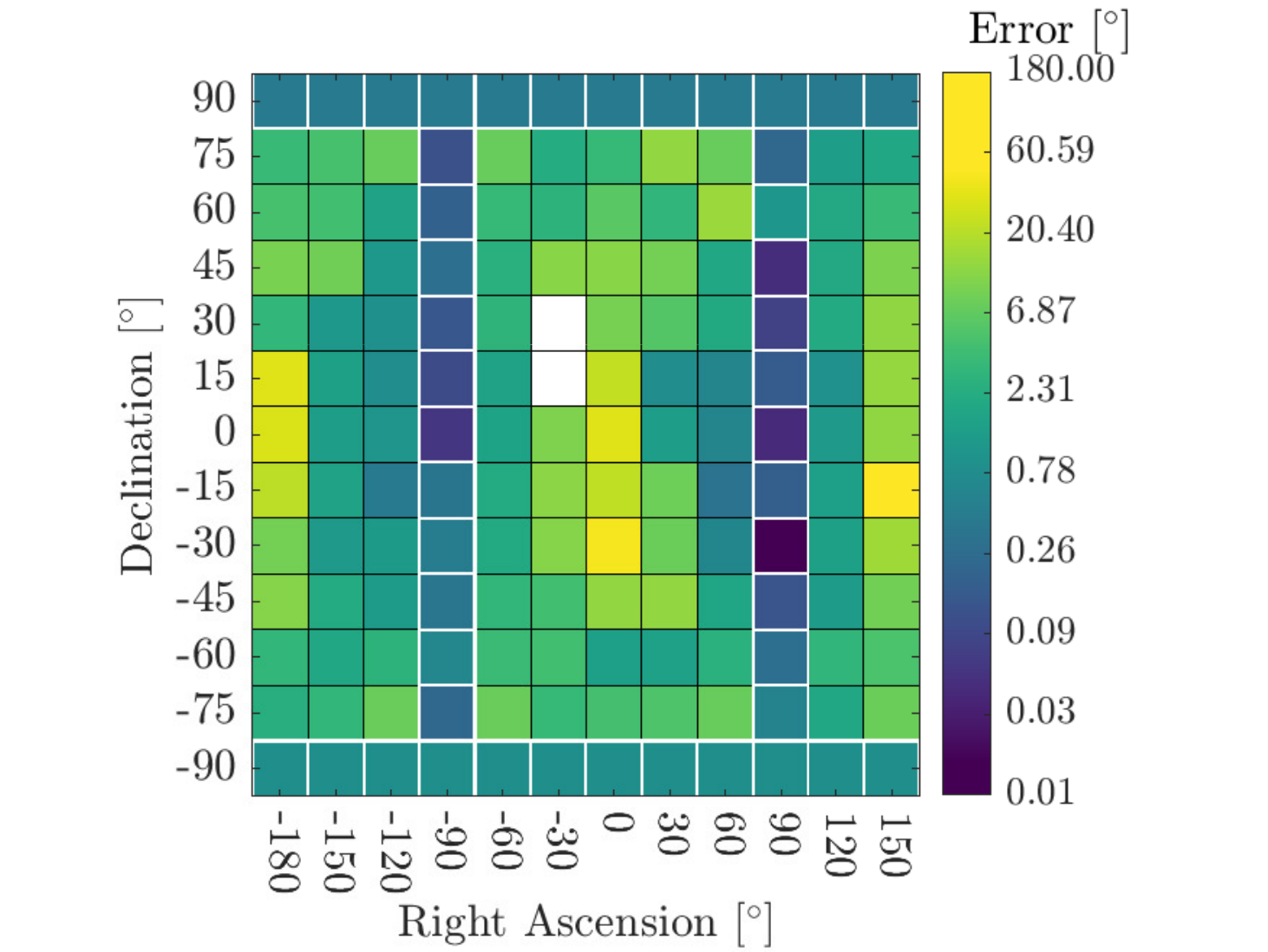}
		\caption{Bennu with illumination angle of $45^{\circ}$}
		\label{fig:AngularError_Bennu_PA45_OnboardGuess}
	\end{subfigure}
	\hspace{0.5cm}
	\begin{subfigure}[t]{0.45\textwidth}
		\centering
		\includegraphics[width=\textwidth,trim={1cm 0cm 1cm 0cm},clip]{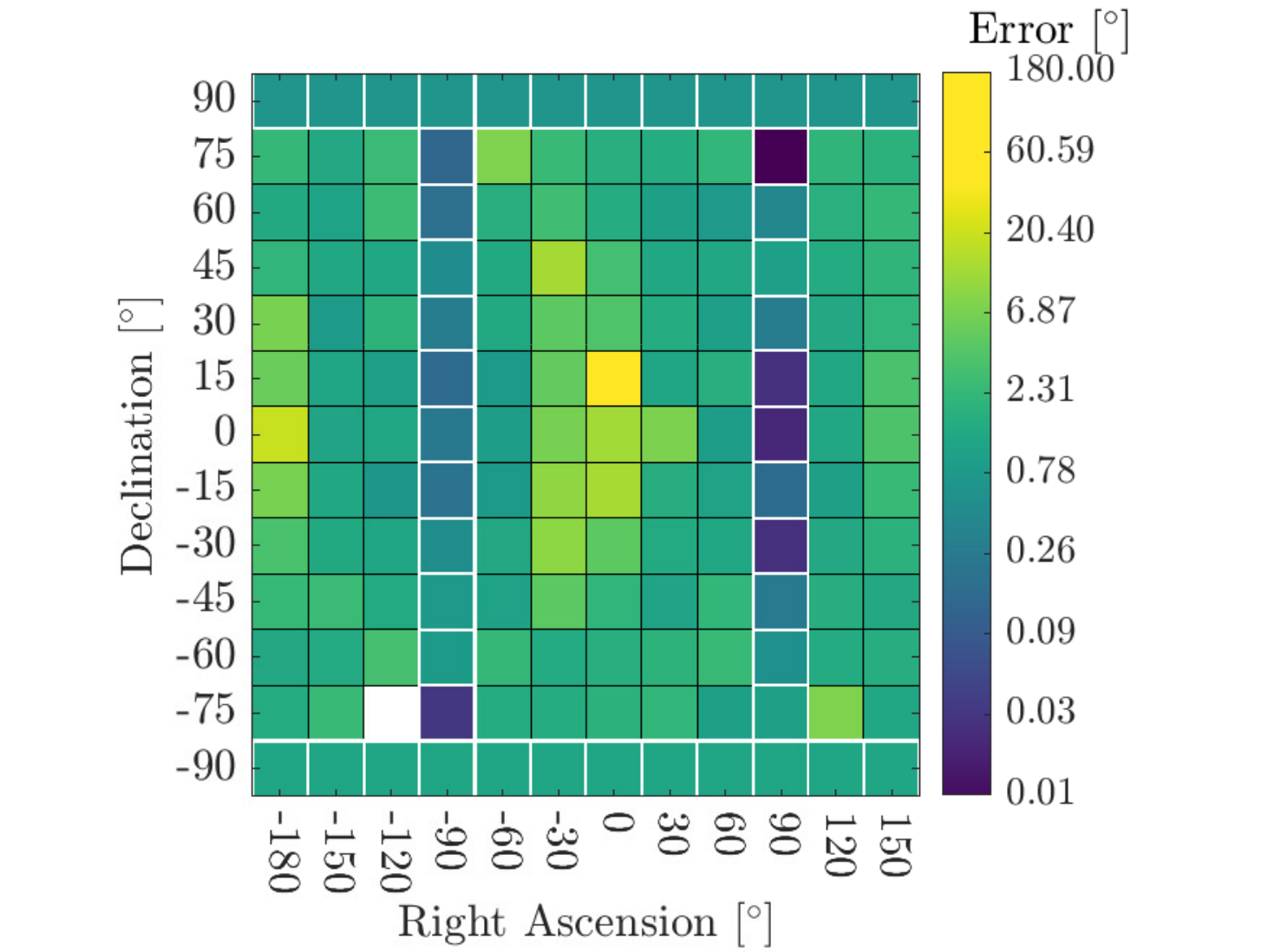}
		\caption{Itokawa with illumination angle of $45^{\circ}$}
		\label{fig:AngularError_Itokawa_PA45_OnboardGuess}
	\end{subfigure}\\[0.3cm]
	\begin{subfigure}[t]{0.45\textwidth}
		\centering
		\includegraphics[width=\textwidth,trim={1cm 0cm 1cm 0cm},clip]{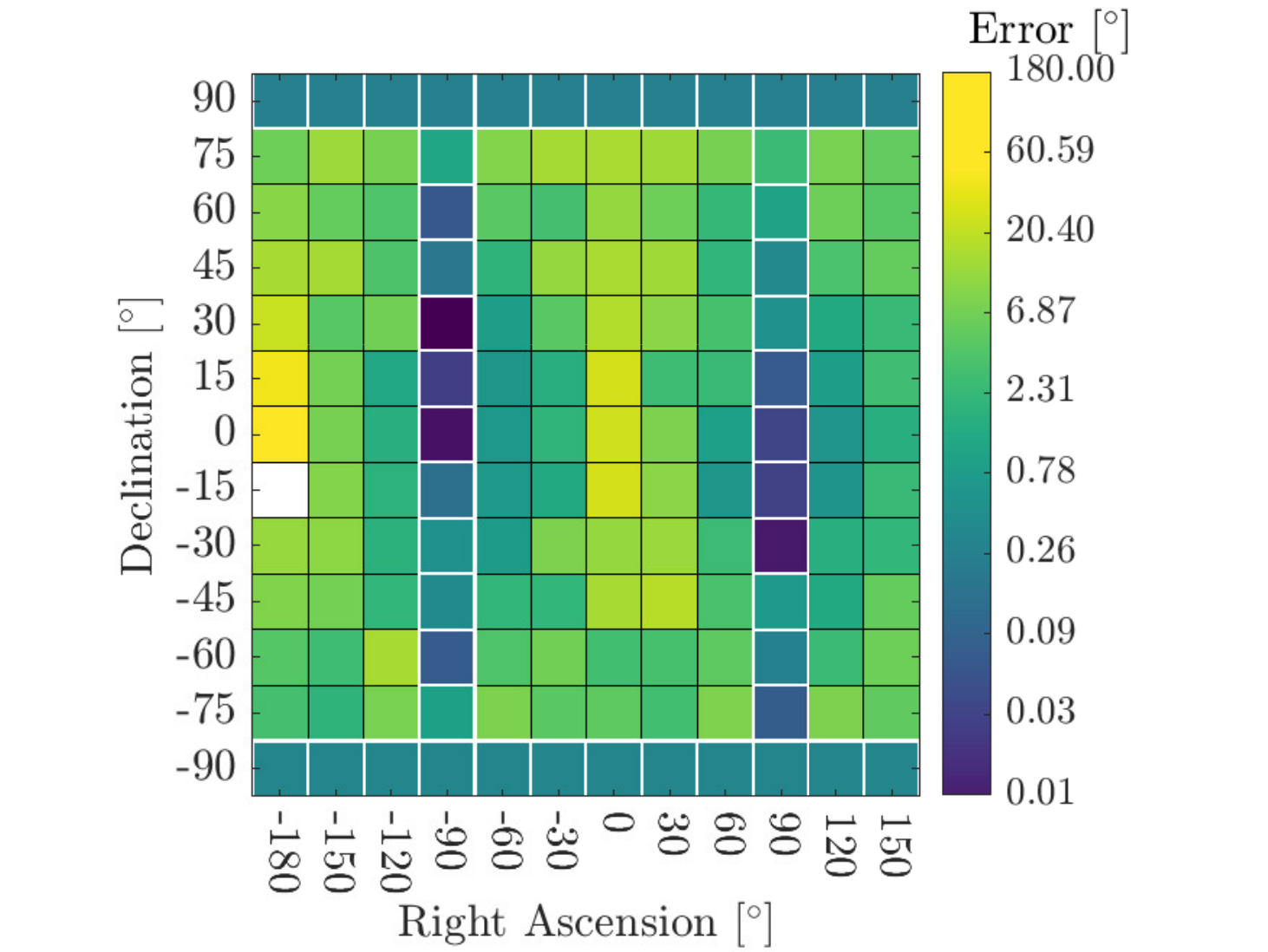}
		\caption{Bennu with illumination angle of $75^{\circ}$}
		\label{fig:AngularError_Bennu_PA75_OnboardGuess}
	\end{subfigure}
	\hspace{0.5cm}
	\begin{subfigure}[t]{0.45\textwidth}
		\centering
		\includegraphics[width=\textwidth,trim={1cm 0cm 1cm 0cm},clip]{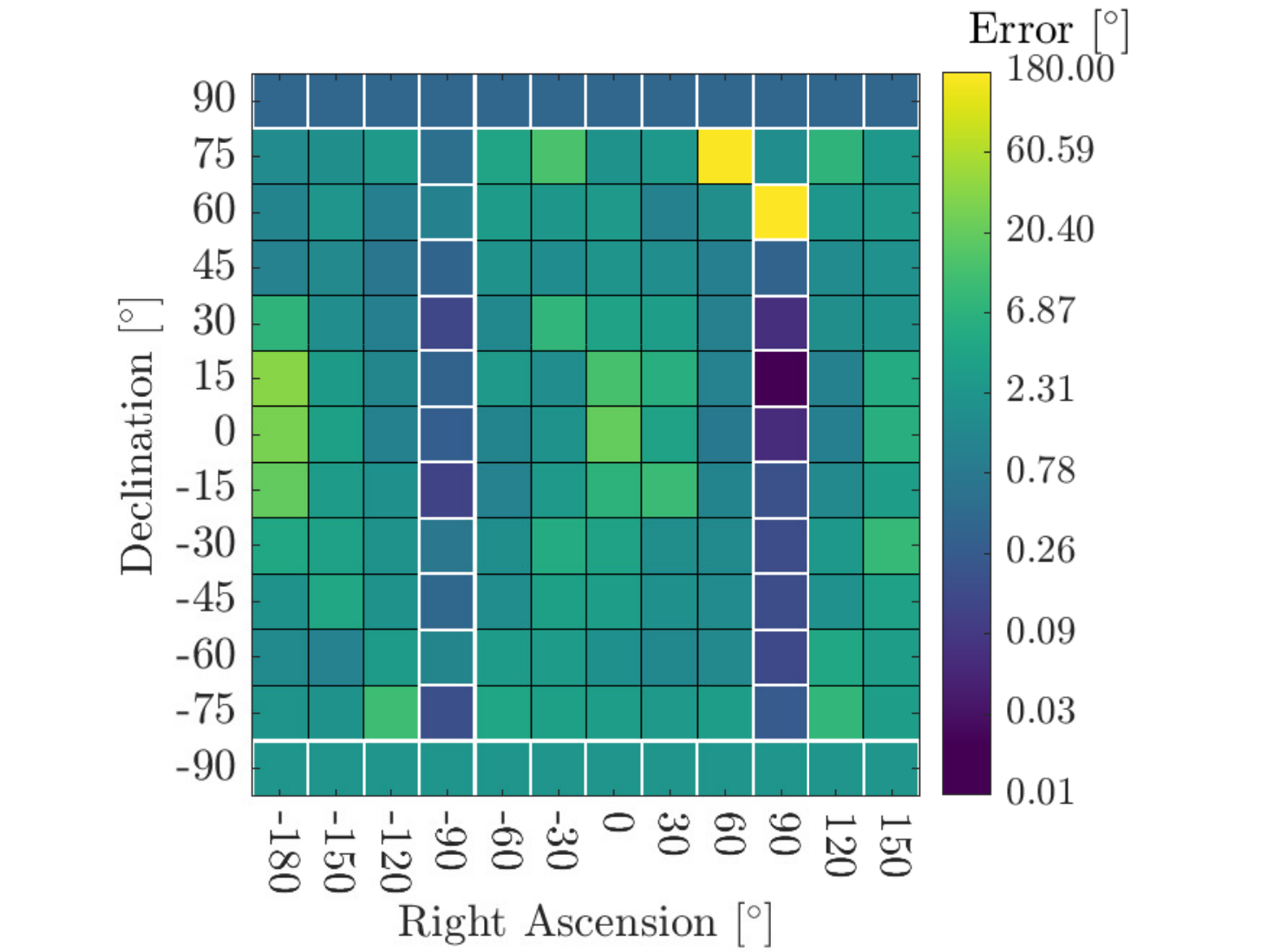}
		\caption{Itokawa with illumination angle of $75^{\circ}$}
		\label{fig:AngularError_Itokawa_PA75_OnboardGuess}
	\end{subfigure}	
	\caption{Rotation axis estimation performance with the heuristic based on onboard guess.}
\end{figure}
\newline
First, the results for the rotation axis estimation based on the available onboard guess are analyzed. This means that the final rotation axis is selected as the closest one to a coarse estimation of the rotation axis available on board. The coarse first guess is simulated in this study by applying a rotation of 20 degrees to the true axis in a random direction.  A compact visualization of the results is reported in the Cumulative Density Function (CDFs) in Fig.\ref{fig:CDFAxis_Bennu_OnboardGuess} and \ref{fig:CDFAxis_Itokawa_OnboardGuess} for various illumination angles and the two studied asteroids. Note that the CDFs are cropped at $30^{\circ}$ to have a detailed look close to the low error solutions. In both cases, the angular error is below $10^{\circ}$ for $80\%$ of the cases and it is below $20^{\circ}$ for $90\%$ of the cases. This implies that the proposed method estimates correctly the 4 solutions and identifies the correct one. Note that the illumination does not seem to play a major role for the Itokawa test case (see Fig.~\ref{fig:CDFAxis_Itokawa_OnboardGuess}), while the performance degrades  degrade for the Bennu test case when the illumination angle increase (see Fig.~\ref{fig:CDFAxis_Bennu_OnboardGuess}). As stated in Sec.~\ref{sec:detectionresults}, this is due to the different shadows present in the two test cases. A more detailed view of the results is reported in Fig.~\ref{fig:AngularError_Bennu_PA10_OnboardGuess} - \ref{fig:AngularError_Itokawa_PA75_OnboardGuess} where the angular error is shown in the right-ascension-declination plane. In the figures, the sectors with the white contours are associated with rotation axes perpendicular to the camera boresight. Moreover, the white sectors label the solutions where the algorithm has not converged. This is mainly due to the impossibility of detecting a series of ellipses all with the same orientation leading to the detection of a conic family with an eccentricity greater than 1. Finally, the yellow sectors are associated with very high errors which are due to wrong detection among the four solutions. It is worth noting that the perpendicular cases have generally low errors and higher errors are present when the rotation axis is parallel to the approach direction.  Note that the wrong classifications outlined in \ref{sec:detectionresults} do not show high errors because they occur when the rotation axis is almost perpendicular.
\begin{figure}[!t]
	\centering
	\begin{subfigure}[t]{0.47\textwidth}
		\includegraphics[width=\textwidth,trim={0cm 0cm 0cm 0cm},clip]{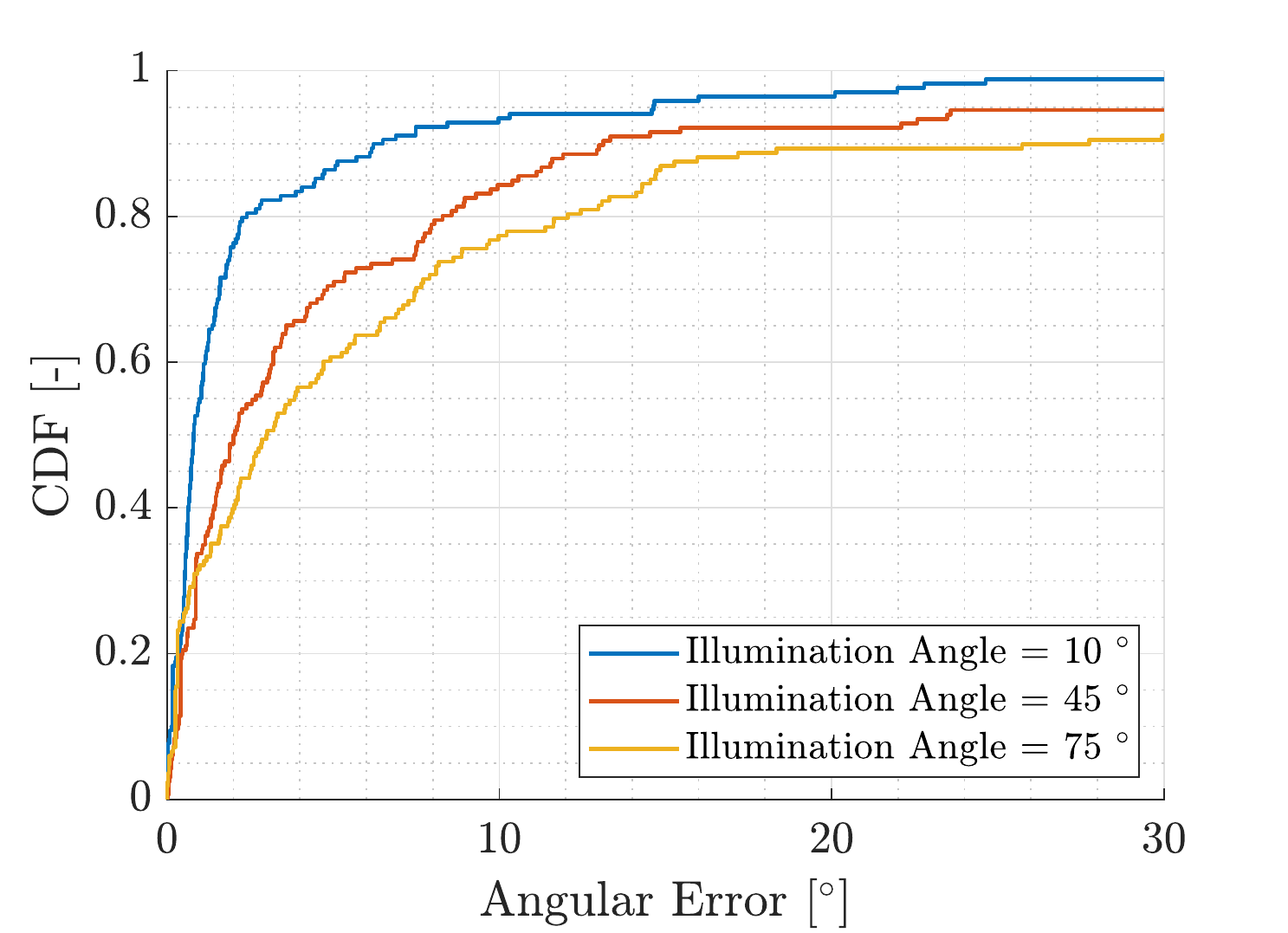}
		\caption{Bennu}
		\label{fig:CDFAxis_Bennu_Initilization}
	\end{subfigure}
	\hspace{0.5cm}
	\begin{subfigure}[t]{0.47\textwidth}
		\centering
		\includegraphics[width=\textwidth,trim={0cm 0cm 0cm 0cm},clip]{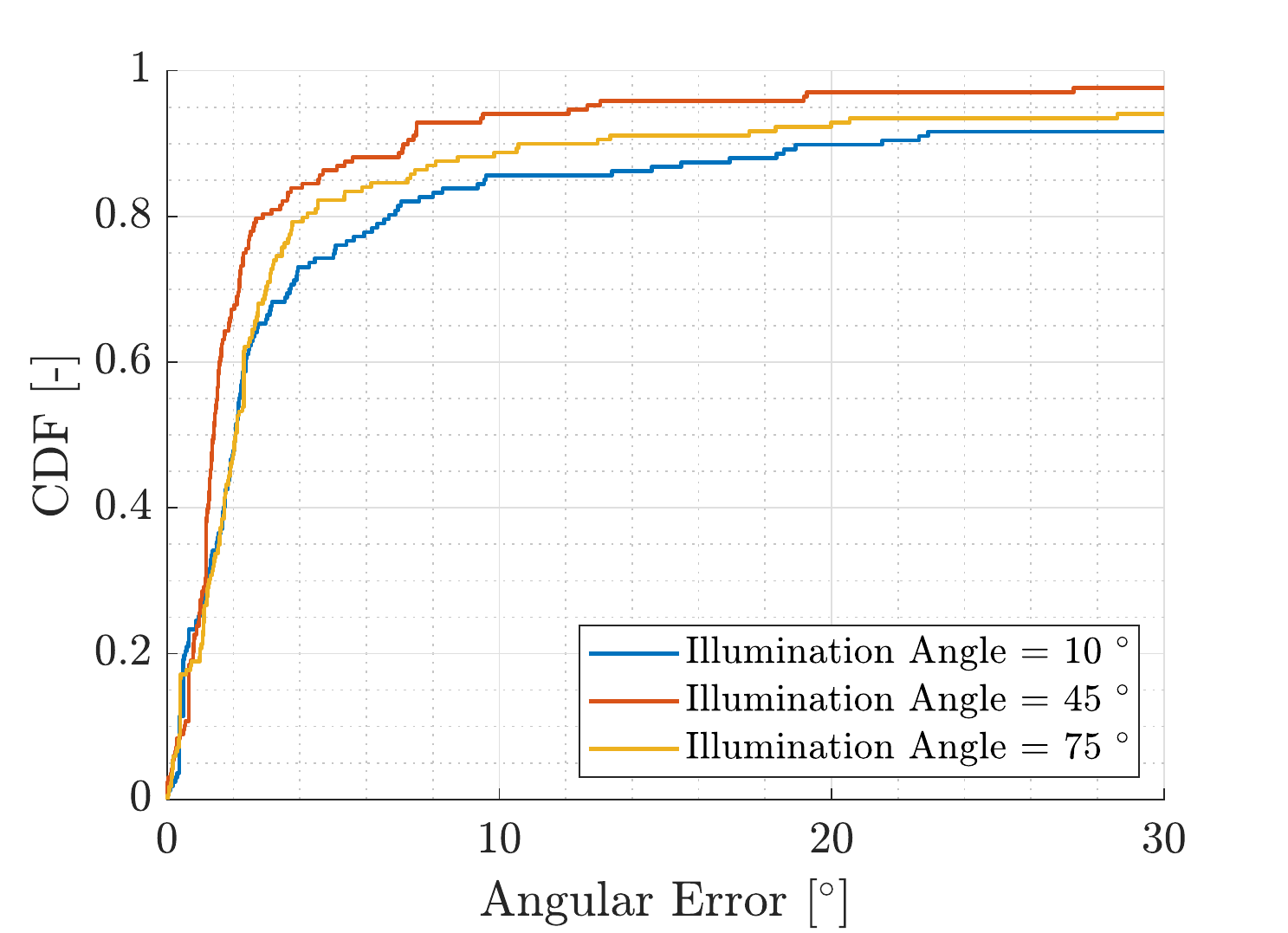}
		\caption{Itokawa}
		\label{fig:CDFAxis_Itokawa_Initilization}
	\end{subfigure}
	\caption{Zoom of the CDFs of the rotation axis estimation error with the heuristic based on landmark initialization.}
\end{figure}
\begin{figure}[p]
	\centering
	\begin{subfigure}[t]{0.45\textwidth}
		\includegraphics[width=\textwidth,trim={1cm 0cm 1cm 0cm},clip]{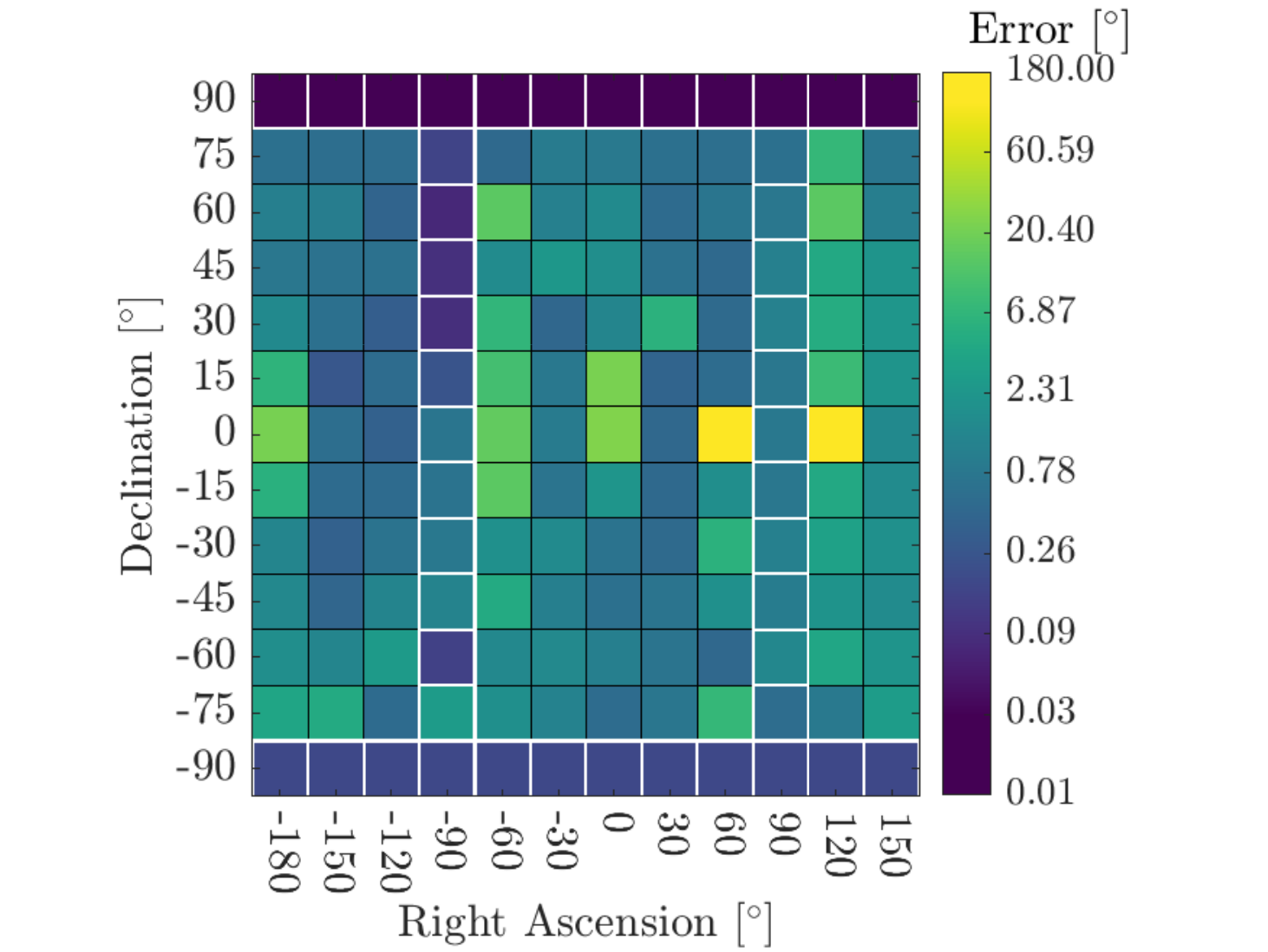}
		\caption{Bennu with illumination angle of $10^{\circ}$}
		\label{fig:AngularError_Bennu_PA10_Initilization}
	\end{subfigure}
	\hspace{0.5cm}
	\begin{subfigure}[t]{0.45\textwidth}
		\centering
		\includegraphics[width=\textwidth,trim={1cm 0cm 1cm 0cm},clip]{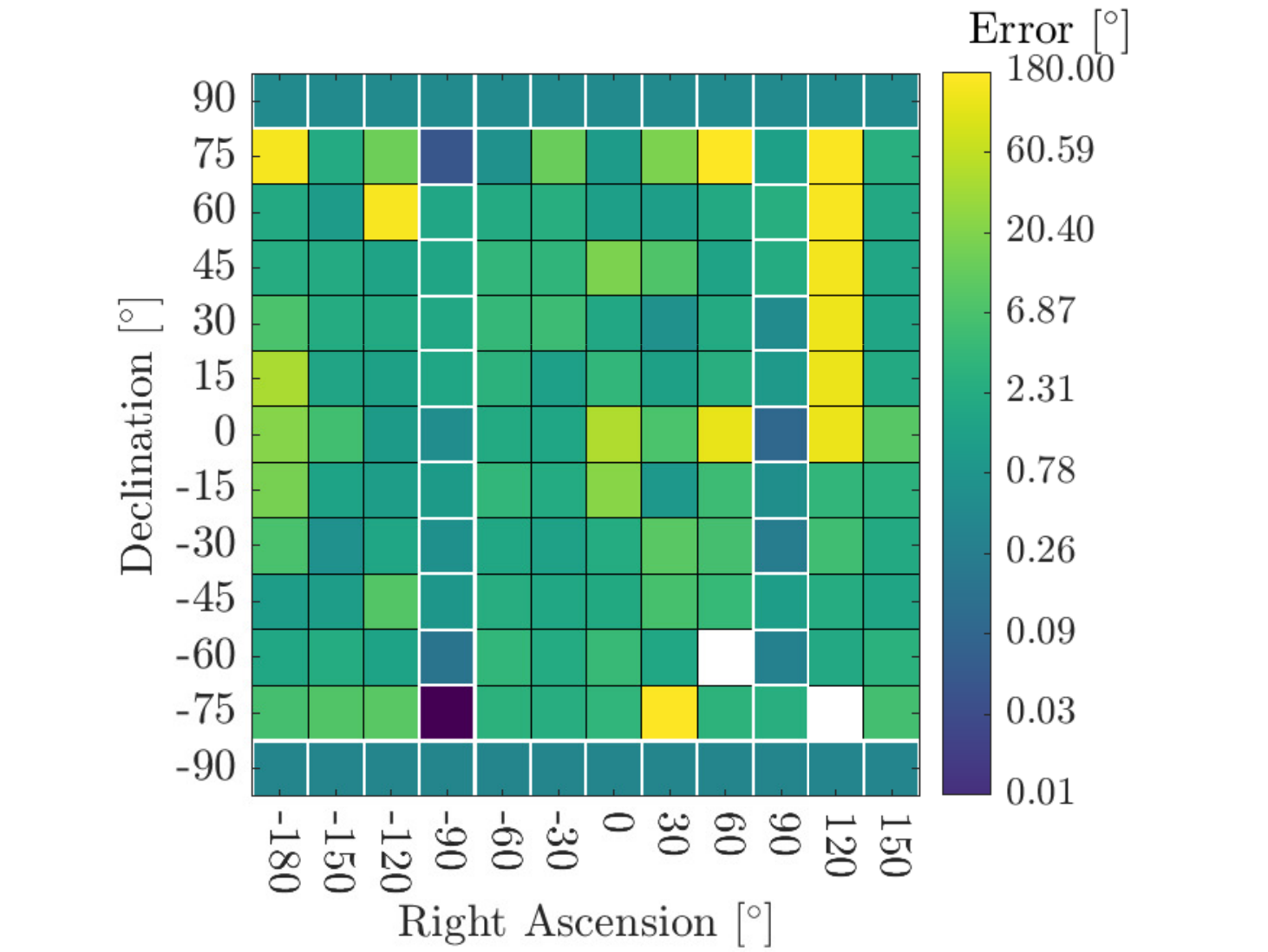}
		\caption{Itokawa with illumination angle of $10^{\circ}$}
		\label{fig:AngularError_Itokawa_PA10_Initilization}
	\end{subfigure}\\[0.5cm]
	\begin{subfigure}[t]{0.45\textwidth}
		\centering
		\includegraphics[width=\textwidth,trim={1cm 0cm 1cm 0cm},clip]{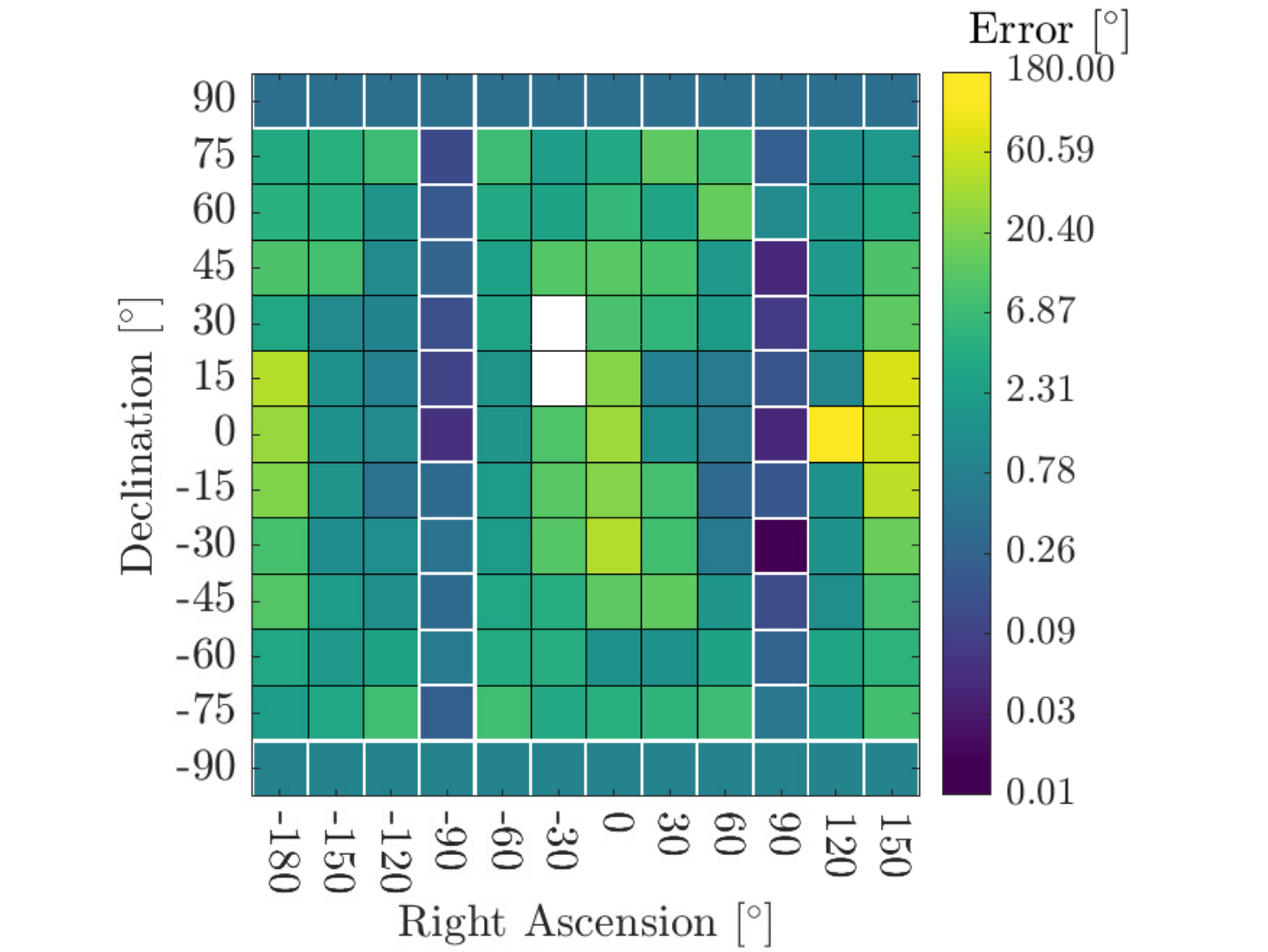}
		\caption{Bennu with illumination angle of $45^{\circ}$}
		\label{fig:AngularError_Bennu_PA45_Initilization}
	\end{subfigure}
	\hspace{0.5cm}
	\begin{subfigure}[t]{0.45\textwidth}
		\centering
		\includegraphics[width=\textwidth,trim={1cm 0cm 1cm 0cm},clip]{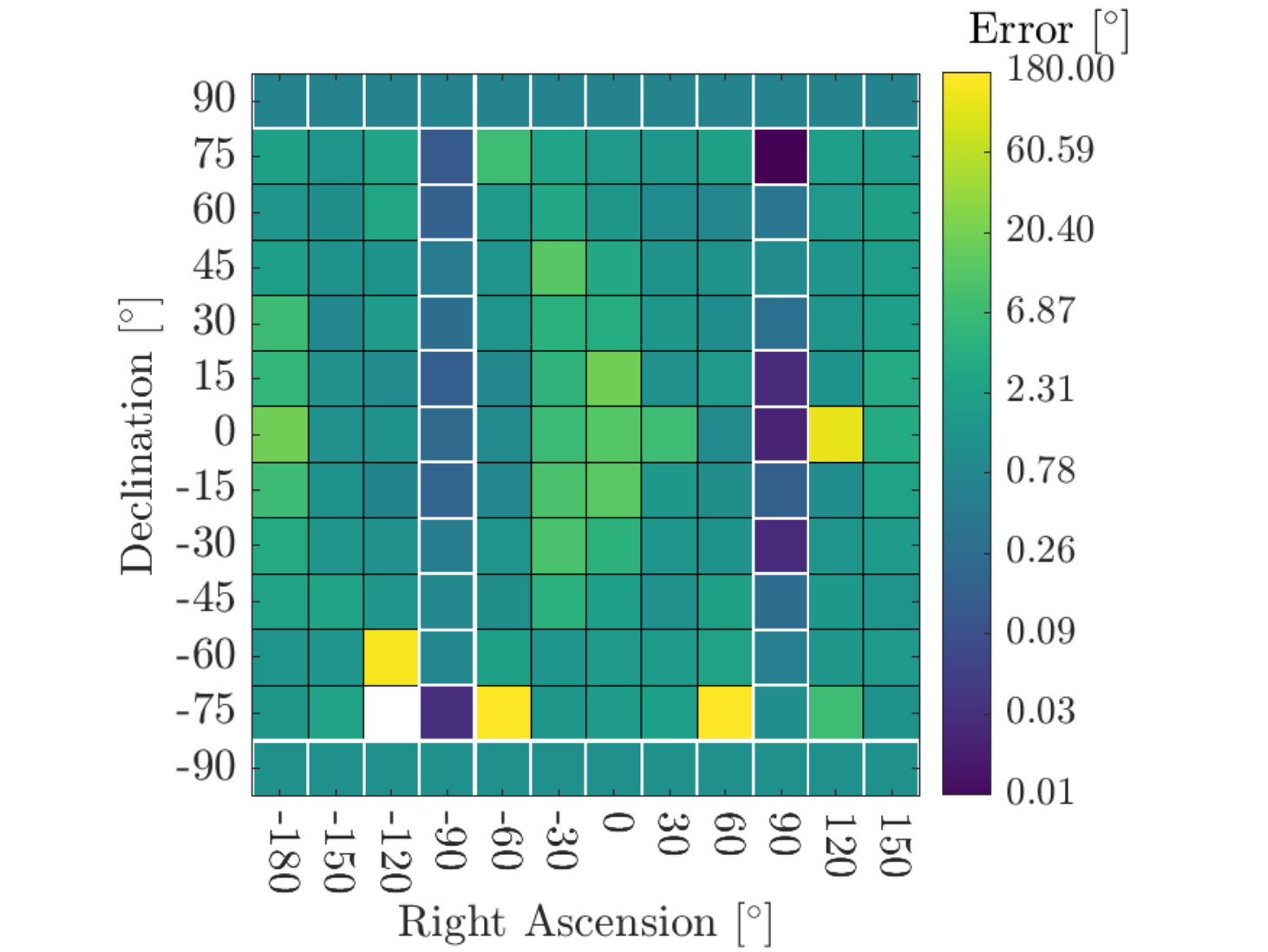}
		\caption{Itokawa with illumination angle of $45^{\circ}$}
		\label{fig:AngularError_Itokawa_PA45_Initilization}
	\end{subfigure}\\[0.3cm]
	\begin{subfigure}[t]{0.45\textwidth}
		\centering
		\includegraphics[width=\textwidth,trim={1cm 0cm 1cm 0cm},clip]{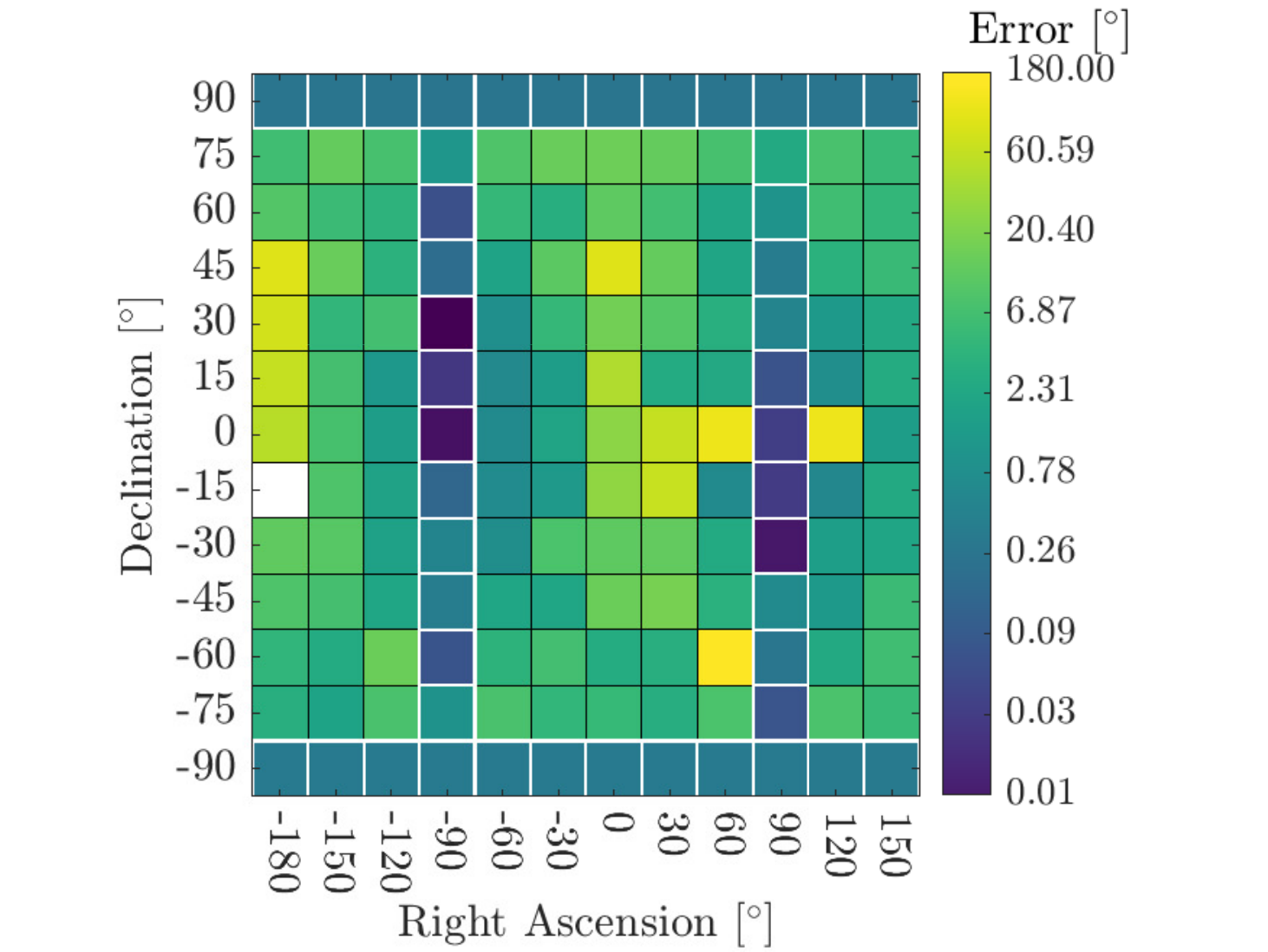}
		\caption{Bennu with illumination angle of $75^{\circ}$}
		\label{fig:AngularError_Bennu_PA75_Initilization}
	\end{subfigure}
	\hspace{0.5cm}
	\begin{subfigure}[t]{0.45\textwidth}
		\centering
		\includegraphics[width=\textwidth,trim={1cm 0cm 1cm 0cm},clip]{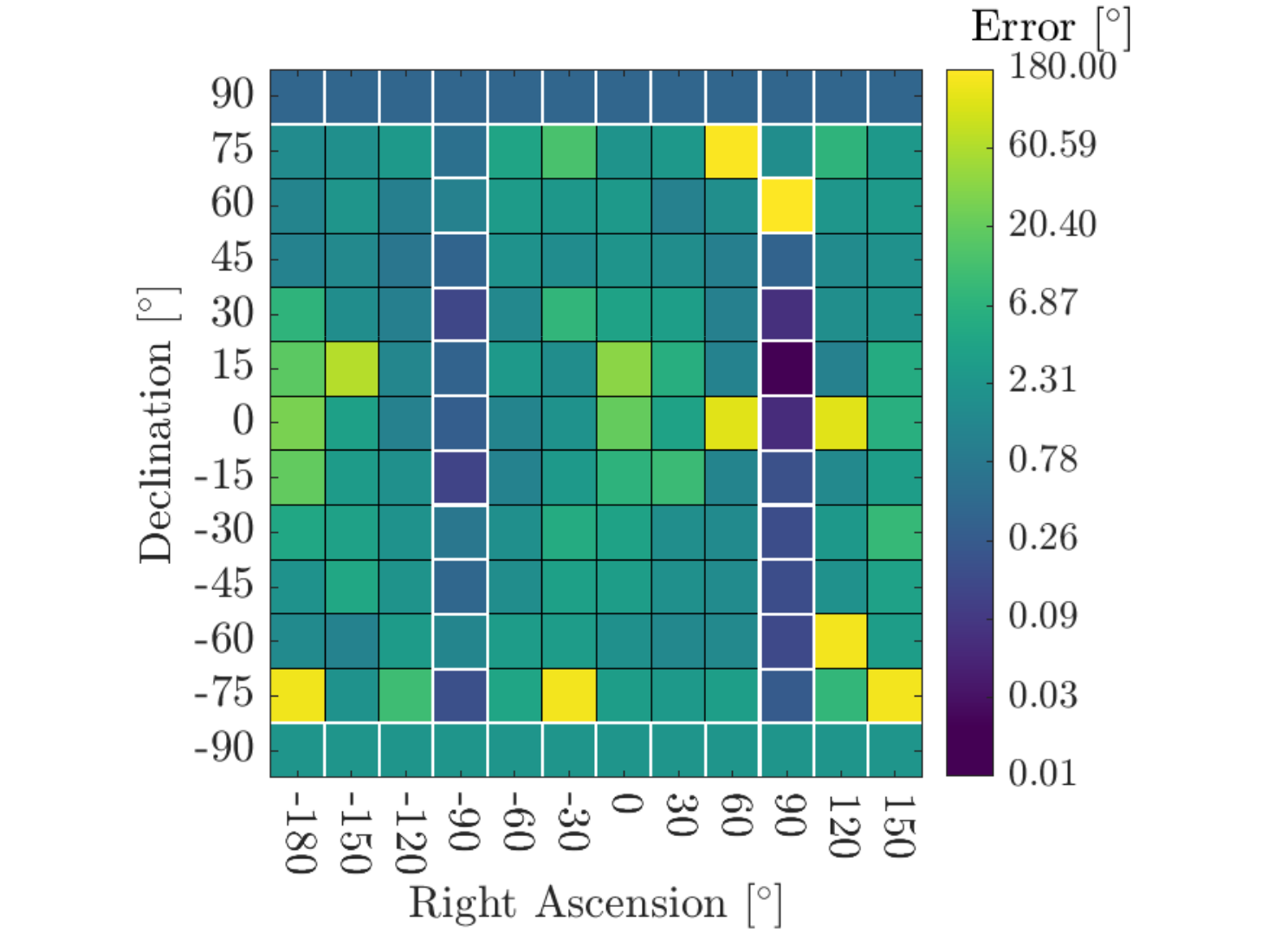}
		\caption{Itokawa with illumination angle of $75^{\circ}$}
		\label{fig:AngularError_Itokawa_PA75_Initilization}
	\end{subfigure}	
	\caption{Rotation axis estimation performance with the heuristic based on landmark initialization.}
\end{figure}
\newline
Second, the results for the rotation axis estimation based on the landmark initialization are studied. Results are similar with respect to the previous ones with small performance degradation, mainly because the correct axis is selected less often. The performance reduction is anyway low as shown in Fig.~\ref{fig:CDFAxis_Bennu_Initilization} and \ref{fig:CDFAxis_Itokawa_Initilization}. Indeed,  the angular error is below $10^{\circ}$ for $75\%$ of the samples for all asteroids and illumination angles and it is below $20^{\circ}$ for $85\%$ of the cases. Note that the performance reduction is not due to a change in the angular error, which is the same as the tracked features are the same, but to the higher number of samples in the distribution tails. This is  visible in Fig.~\ref{fig:AngularError_Bennu_PA10_Initilization} - \ref{fig:AngularError_Itokawa_PA75_Initilization}
where the angular error is shown in the right-ascension-declination plane. The number of yellow sectors increases because the heuristic fails in detecting the correct solution more often than the onboard guess heuristic. However, it is worth noting that this heuristic does not rely on a previously-computed guess and the rotation axis estimation is performed fully autonomously from images.

\section{Conclusion}\label{sec:conclusions}
This paper outlines an algorithm to autonomously determine the rotational state of a small body during the approach phase. Under the assumption of knowing the rotational period, the algorithm determines the rotation axis orientation of a small body exploiting images during the approach phase. The spacecraft camera acquires images of an unknown small body and extracts features from these images. The long-tracked features are the projection of 3D landmarks whose movement is due to the small body rotation. By exploiting the 2D feature tracks, the algorithm fits a family of conics which are the projection of the landmark 3D circle. By backprojecting the found conics in 3D, the algorithm finds all the possible solutions for the small body rotation axis. The correct one is selected by ensuring coherence with the movement of the features in the image and by exploiting a heuristic approach leading to identifying the right rotation axis. Moreover, the center of rotation is determined from the first image by backprojecting the small body center of brightness. \newline 
Numerical simulations are performed with a concave and a convex asteroid over a wide range of illumination angles and pole orientations. The simulations underline that the algorithm can efficiently detect the case in which the rotation axis is perpendicular to the spacecraft approach direction. Moreover, the rotation axis estimation is performed with limited error in most cases. Indeed the error between the true rotation axis and its estimation is below $10^{\circ}$ for $80\%$ of the considered test cases. The computed rotation axis is a valuable first guess of the rotation axis to enable spacecraft localization around an unknown small body. The proposed rotation axis estimation and the origin determination are valuable tools to define the small-body-fixed reference frame during the approach phase. This is an important and preliminary task to be fulfilled before performing insertion maneuvers, shape determination, or small-body-fixed localization. 

\appendix
\section{The computation of $U\left(u,v\right)$, $V\left(u,v\right)$ and $W\left(u,v\right)$}\label{app:UVW}
The functions $U\left(u,v\right)$, $V\left(u,v\right)$ and $W\left(u,v\right)$ introduced in Sec\ref{sec:tiltedpruningandselection} are computed as follows:
\begin{equation}
U\left(u,v\right) = U^2_1\left(u,v\right) + W^2_1\left(u,v\right)
\end{equation}
\begin{equation}
V\left(u,v\right) = 2U_1\left(u,v\right)V_1\left(u,v\right) + 2W_1\left(u,v\right)Y_1\left(u,v\right)
\end{equation}
\begin{equation}
W\left(u,v\right) = V^2_1\left(u,v\right) + Y^2_1\left(u,v\right)
\end{equation}
\begin{equation}
U_1\left(u,v\right) = U_{1,1}\left(u,v\right)W_1\left(u,v\right) + V_{1,1}\left(u,v\right)
\end{equation}
\begin{equation}
V_1\left(u,v\right) = U_{1,1}\left(u,v\right)Y_1\left(u,v\right) + W_{1,1}\left(u,v\right)
\end{equation}
where $\left(u,v\right)$ are pixel coordinates. By defining the projection matrix $\left[P\right]$ between the 3D world homogeneous coordinates to the 2D homogeneous image coordinates, the needed quantities are computed:
\begin{equation}
W_{1}\left(u,v\right) = \frac{\left(\left[P\right]_{2,1}-\left[P\right]_{3,1}v\right)V_{1,1} + \left[P\right]_{2,3}-\left[P\right]_{3,3}v}{
	\left(\left[P\right]_{3,1}v-\left[P\right]_{2,1}\right)U_{1,1} + \left[P\right]_{3,2}v-\left[P\right]_{2,2}}
\end{equation}
\begin{equation}
Y_{1}\left(u,v\right) = \frac{\left(\left[P\right]_{2,1}-\left[P\right]_{3,1}v\right)W_{1,1} + \left[P\right]_{2,4}-\left[P\right]_{3,4}v}{
	\left(\left[P\right]_{3,1}v-\left[P\right]_{2,1}\right)U_{1,1} + \left[P\right]_{3,2}v-\left[P\right]_{2,2}}
\end{equation}
\begin{equation}
U_{1,1}\left(u,v\right) = \frac{\left[P\right]_{1,2}-\left[P\right]_{3,2}u}{\left[P\right]_{3,1}u-\left[P\right]_{1,1}}
\end{equation}
\begin{equation}
V_{1,1}\left(u,v\right) = \frac{\left[P\right]_{1,3}-\left[P\right]_{3,3}u}{\left[P\right]_{3,1}u-\left[P\right]_{1,1}}
\end{equation}
\begin{equation}
W_{1,1}\left(u,v\right) = \frac{\left[P\right]_{1,4}-\left[P\right]_{3,4}u}{\left[P\right]_{3,1}u-\left[P\right]_{1,1}}
\end{equation}
where $\left[P\right]_{k,l}$ is the $k$th row and $l$th column element of the matrix.

\section*{Acknowledgment}
This research received funding from the CNES, the ISAE-SUPAERO and Airbus Defence \& Space under the doctoral contract CNES-2879. P. P. would like to thank Felice Piccolo for the discussions and advises.

\bibliographystyle{unsrtnat}
\bibliography{mybibliography.bib}

\end{document}